\documentclass{aa}  

\pdfoutput=1
\usepackage{url,twoopt,natbib}
\usepackage[varg]{txfonts}
\usepackage{amsmath}
\usepackage{graphicx}

\bibpunct{(}{)}{;}{a}{}{,} 

\def\xin{RX\,J0720.4$-$3125}
\def\zm{$\mathrm{Z^{2}_{m}}$\,}
\newcommand{\xmm}{XMM-{\it Newton}}
\newcommand{\be}{\begin{displaymath}} 
\newcommand{\ee}{\end{displaymath}} 
\newcommand{\beq}{\begin{equation}} 
\newcommand{\eeq}{\end{equation}} 

\begin{document} 
  \title{On the compactness of the isolated neutron star RX\,J0720.4$-$3125\thanks{Based 
         on observations obtained with \xmm, an ESA science mission with instruments and 
         contributions directly funded by ESA Member States and the USA (NASA)}}
 
  \author{V. Hambaryan 
          \inst{1} 
          \and
	  V. Suleimanov 
          \inst{2} 
          \and
          F. Haberl
          \inst{3} 
	  \and 
	  A.D. Schwope
          \inst{4} 
          \and
	  R. Neuh\"auser 
          \inst{1} 
          \and 
	  M. Hohle 
          \inst{1,5} 
          \and 
	  K. Werner 
          \inst{2} 
          }
 
   \institute{Astrophysikalisches Institut und Universit\"ats-Sternwarte,  
              Universit\"at Jena, Schillerg\"a\ss chen 2-3, 07745 Jena, Germany\\ 
              \email{vvh@astro.uni-jena.de} 
         \and 
           Eberhard-Karls-Universit\"at, Institut f\"ur Astronomie und Astrophysik, 72076 T\"ubingen, Germany 
         \and
           Max-Planck-Institut f\"ur extraterrestrische Physik, Giessenbachstrasse, D-85741 Garching, Germany
         \and 
	   Leibniz-Institut f\"ur Astrophysik Potsdam, An der Sternwarte 16, 14482 Potsdam, Germany 
         \and
           Gene Center of the LMU, Department of Biochemistry, 
           Feodor-Lynen-Strasse 25, 81377 M\"unchen, Germany
         } 
 
   \date{Received ... / Accepted ...} 
 
  
  \abstract 
	{} 
	{To estimate the compactness of the thermally emitting isolated neutron star
          RX\,J0720.4$-$3125, an X-ray spin phase-resolved spectroscopic study is conducted. 
          In addition, to identify the genuine spin-period, an X-ray timing analysis is performed.} 
        {The data from all observations of  RX\,J0720.4$-$3125 conducted by
          \xmm\ EPIC-pn  with the same instrumental setup 
        in 2000-2012 were reprocessed to form a homogenous data set of solar
        barycenter corrected photon arrival times registered from RX\,J0720.4$-$3125.
        A Bayesian method for the search, detection, and estimation of the
        parameters of an unknown-shaped periodic signal was employed as developed by
        Gregory \& Loredo (1992). 
        A number of complex models (single and double peaked) of light curves 
        from pulsating neutron stars were statistically analyzed. The distribution of phases 
        for the registered photons was calculated by folding the arrival times
        with the derived spin-period and the resulting
        distribution of phases -- approximated with a mixed von Mises distribution --, 
        and its parameters were estimated by using the Expected Maximization method.
        Spin phase-resolved spectra were
        extracted, and a number of highly magnetized atmosphere models of an INS were used to fit
        simultaneously, the results were verified via an MCMC approach.} 
        {The phase-folded light curves in different energy bands with high S/N ratio show a
         high complexity and variations depending on time and energy. 
        They can be parameterized with a mixed von Mises distribution, i.e. with 
        double-peaked light curve profile showing a dependence of the estimated 
         parameters (mean directions, concentrations, and proportion) upon the energy 
         band, indicating that radiation emerges from at least two emitting areas.}
        {The genuine spin-period of the isolated neutron star RX\,J0720$-$3125
          derived as more likely is twice of that reported in the literature (16.78s instead of 8.39s). 
        The gravitational redshift of RX\,J0720.4$-$3125 was
        determined to $z=0.205_{-0.003}^{+0.006}$ and the compactness was 
        estimated to $(M/M_{\sun})/(R/\mathrm{km})=0.105 \pm 0.002 $.
      }
   {}
 
 
   \titlerunning{On the compactness of  RX\,J0720.4$-$3125}

   \maketitle 
%
 
\section{Introduction}\label{intro}
 
A comprehensive understanding of neutron stars is one of the key challenges facing astrophysics 
and astronomy. In particular, the observational determination of the 
gravitational redshift of an isolated neutron star (INS) 
can be used to put constraints on the theoretical models of superdense matter. 
This goal can be achieved, for instance, by modeling and analyzing 
the observed emergent spectrum of a thermally emitting, pulsating INS via
rotational phase-resolved X-ray spectroscopy 
\citep[][also developed and successfully applied by us]{1995A&A...297..441Z,2003A&A...408..323M,2006A&A...459..175P,2006MNRAS.366..727Z,2007MNRAS.375..821H,2010A&A...522A.111S,2011A&A...534A..74H,2014JPhCS.496a2015H}.

This spectroscopy involves simultaneous fitting of 
{\em high quality spectra}\footnote{High S/N ratio (relative errors in each spectral bin are of 
the order of a few percent) spectra for different rotational phases can 
be achieved by combining observations performed at different epochs.}
from different spin phases with models of highly
magnetized atmospheres of INSs. This allows us to constrain a
number of physical properties of the X-ray emitting areas, including
their temperatures, magnetic field strengths at the poles, and their
distribution parameters. In addition, we may place some constraints on
the viewing geometry of the emerging X-ray emission and the
gravitational redshift and, hence, the compactness of the INS.
This model of spectra is based on various local models,
such as blackbody or condensed (iron or other heavy element) surface,
covered by a thin hydrogen layer. 

On the other hand, it is clear that for such kind of study the {\em true} 
spin parameter of the rotating INS must be known at different epochs, i.e.
X-ray timing analysis is also needed for the application of advanced methods (see Sect.~\ref{tim}).

At the beginning of 1990ies, 
the {\em ROSAT} X-ray observatory has discovered a small group (so
 far 7) nearby, thermally emitting and radio-quiet INSs 
\citep[e.g. see reviews by][and references therein]{2007Ap&SS.308..181H,2008A&ARv..15..225M,2009ASSL..357..141T}.
Meanwhile, multi-epoch X-ray  observations using the \xmm\  and {\em Chandra} 
telescopes allowed to detect X-ray pulsations and to uncover absorption
features in the spectra of some of them 
\citep[][]{2003A&A...403L..19H,2009A&A...497L...9H,2015ApJ...807L..20B}.

RX\,J0720.4$-$3125 is a special case among the nearby thermally emitting
INSs showing long-term variations in its timing and
spectral parameters. A clear variation in the spectrum has been detected by
the \xmm\  high spectral resolution Reflection Grating
Spectrometers \citep[RGS;][]{2004A&A...415L..31D}. This 
discovery was subsequently confirmed and analyzed by imaging-spectroscopic EPIC-pn
observations \citep{2006A&A...451L..17H,2009A&A...498..811H}.

The spin-phase-averaged X-ray spectra of \xin\ can formally be 
modeled by a blackbody (kT $\sim$84$-$94\,eV) plus a broad ($\sim$ 10$-$70\,eV wide) absorption
feature centered at $\sim$ 300\,eV \citep[][]{2004A&A...419.1077H,2009A&A...498..811H,2012MNRAS.423.1194H} 
undergoing rather sensible changes over a timescale of a few years 
\citep[however, see also][stating that the observed broad 
absorption feature, partially, might be owing to the 
X-ray spectral distortions induced by inhomogeneous temperature
distributions of the neutron star surface]{2014MNRAS.443...31V}. 

A narrow ($\sim 5$\,eV wide) absorption feature was identified in \xmm\ RGS
data at $\sim$ 570\,eV, possibly due to highly ionized oxygen of  
circumstellar origin \citep{2004cosp...35.2075M,2009A&A...497L...9H,2012MNRAS.419.1525H}.

In this paper, we performed a rotational phase-resolved spectral re-analysis 
of the \xmm\ EPIC-pn data from the avaliable multi-epoch X-ray observations of \xin.

\section{Observations and data reduction}\label{ditum}
 
RX\,J0720.4$-$3125 was observed many times by \xmm\ with different instrumental setup (Table~\ref{obslog}). 
Here we focus mainly on the high quality data collected with EPIC-pn in Full
Frame mode \citep{2001A&A...365L..18S} from the 16 \xmm\ observations with
same instrumental setup (positioned on-axis),  
distributed over 12.4 years, in total presenting about 350~ks of effective exposure time, and 
consisting of $\sim$~2.2 Million registered source photons. 

\begin{table*} 
\centering                          
\caption[]{\xmm\ EPIC-pn Full Frame mode observations of RX\,J0720.4$-$3125} 
\label{obslog}      
\begin{tabular}{cccccrc} 
\hline\noalign{\smallskip} 
Obs. ID & Date begin &  MJD & Exposure & Effective  & Number  & Group\\ 
        &            & [start] &  [ksec]    &exposure  [ksec]            & of  counts & No.  \\ 
\hline\noalign{\smallskip}
0124100101   & 2000-05-13 01:42:22  & 51677.104103  & 65.9 & 33.6 & 205200 &  I\\
0156960201   & 2002-11-06 17:51:49  & 52584.761786  & 30.2 & 19.7 & 125305 &  " \\
0156960401   & 2002-11-08 19:25:02  & 52586.826653  & 32.0 & 30.0 & 189113 &  " \\
\hline\noalign{\smallskip}
0164560501   & 2004-05-22 10:15:22  & 53147.442236  & 52.0 & 25.4 & 166528 & II \\
0300520201   & 2005-04-28 08:41:05  & 53488.377978  & 53.3 & 30.0 & 193632 &  "  \\
0300520301   & 2005-09-22 23:44:23  & 53636.004188  & 53.0 & 35.2 & 230808 &  "  \\
\hline\noalign{\smallskip}
0400140301   & 2006-05-22 04:44:47  & 53877.228785  & 21.9 & 16.2 & 103291 & III  \\
0400140401   & 2006-11-05 11:19:29  & 54044.489372  & 21.9 & 20.0 & 127835 &  "   \\
0502710201   & 2007-05-05 17:01:25  & 54225.732373  & 21.9 & 15.0 & 94214  &  "   \\
0502710301   & 2007-11-17 05:14:32  & 54421.236398  & 24.9 & 23.0 & 142855 &  "  \\
\hline\noalign{\smallskip}
0554510101   & 2009-03-21 14:14:24  & 54911.669656  & 21.9 & 14.8 & 92506 & IV    \\
0601170301   & 2009-09-22 04:27:35  & 55096.200867  & 30.8 & 13.2 & 81991 &  "     \\
0650920101   & 2011-04-10 23:59:26  & 55662.016898  & 21.9 & 16.1 & 99286 &  "     \\
\hline\noalign{\smallskip}
0670700201   & 2011-05-02 23:25:17  & 55684.067143  & 28.8 & 10.1 & 62887 & V   \\
0670700301   & 2011-10-01 03:47:26  & 55835.184239  & 26.9 & 24.1 & 150050 &  "  \\
0690070201   & 2012-09-18 08:50:53  & 56188.369444  & 25.1 & 24.9 & 159712 &  "     \\
\hline                                   
\end{tabular} 
\end{table*} 

The data were reduced using standard threads from the \xmm\ ~data analysis  
package SAS version 14.0.0. We reprocessed all observations listed in Table~\ref{obslog}
with the standard metatask {\em epchain}.  
To determine good time intervals, which were free of background flares, 
we applied a filter to the background light curves and performed a visual inspection. 
This reduced the total exposure time by $\sim$\,35\%.  
Solar barycenter corrected source and background photon events files 
were produced from the cleaned single-pixel events, using an extraction radius
of 45$-$60$\arcsec$  depending on the brightness of \xin\ in the corresponding pointed observation.

\begin{figure}[t] 
\resizebox{\hsize}{!}{ 
\includegraphics[clip=]{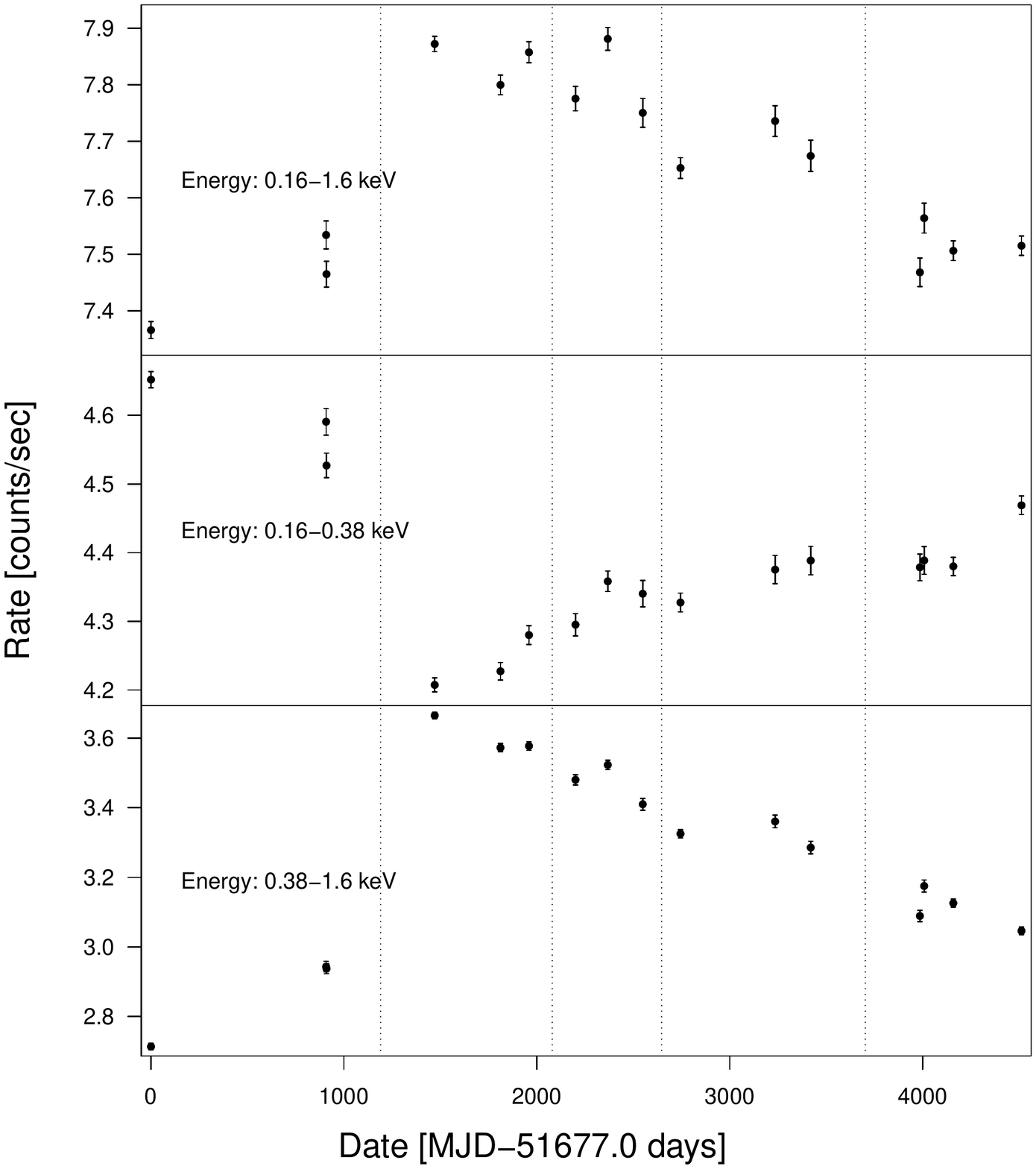}} 
\caption{Observed count rates of \xin\ in the
energy ranges 0.16$-$1.6\,keV, 0.16$-$0.38\,keV, and 0.38$-$1.6\,keV 
in different \xmm\ EPIC-pn Full Frame mode observations. 
} 
\label{obs_ctr} 
\end{figure}

We extracted light curves of \xin\  and the corresponding background from
nearby, source-free regions. We then used the SAS task {\em epiclccorr}  
to correct the observed count rates for various sorts of detector inefficiencies  
(vignetting, bad pixels, dead time, effective areas, etc.) in different energy bands  
for each pointed observation for uninterrupted good time interval exceeding
900 sec (see Fig.~\ref{obs_ctr}). 
 
\section{Data analysis}\label{da}

In order to extract spin-phase-resolved spectra,
one has to determine the rotational phase of each registered photon, i.e.
an accurate estimate of the most probable value of the spin-period is required for each epoch.

In the case of \xin, apart from the clear variation detected in the spectrum, 
a significant variation of the spin-period also was reported.
A number of published papers are devoted to the timing analysis and 
behavior of spectral variations of this enigmatic object 
\citep{2001A&A...365L.302C,2004MNRAS.351.1099C,2007ApJ...659L.149V,2009A&A...498..811H}.
To explain significant phase residuals from a steady spin-down model, different 
interpretations have been suggested;
free precession and glitch possibilities are mostly debated \citep[see, e.g.][possible glitch ($\Delta \nu \sim 4.1(12)$ nHz at  MJD=52866$\pm$73d)]{2009A&A...498..811H,2012MNRAS.419.1525H}.

However, sometimes an unambiguous identification of the {\em true} spin-period of 
a spinning compact star (where the pulsed emission is dominated from emitting areas with slightly 
different physical characteristics in comparison to the other parts of the surface or atmosphere, 
e.g. around the magnetic poles) is very difficult from simple analysis of 
a periodogram \citep{1983A&A...128..245B,1989ASIC..262.....O}\footnote{A
  commonly used statistics in X-ray astronomy: \\
$Z_{m}^{2}=\frac{2}{N}{\displaystyle\sum_{k=1}^{m}\left\{\left(\sum_{i=1}^{N}\sin\left(2k\pi\phi_{i}\right)\right)^{2}+\left(\sum_{i=1}^{N}\cos\left(2k\pi\phi_{i}\right)\right)^{2}\right\}}$  
with phases $\phi_{i}$ at trial frequency. Here, $\phi_i = \nu\,(t_i-t_0) + \dot{\nu}\, (t_i-t_0)^2/2$
is the phase, $\nu$ and $\dot{\nu}$ are the trial frequency and its derivative, 
$t_i$ ($i=1,...,N$) are the photon arrival times, and  $t_0$ is the epoch of zero phase. 
Note, $Z_{m}^{2}$ is a sufficient statistic
for detection of sinusoidal signals in comparison to the arrival times, 
they are distributed uniformly (obeying constant-rate Poissonian counting
statistic, i.e. a periodic component is absent).  
}, 
i.e. by assigning the frequency corresponding to the most significant peak in a 
periodogram to the {\em true} rotational period.

The matter is that, the {\it genuine} spin frequency in a periodogram highly depends 
on the {\it unknown true} light curve shape (e.g. single-, double-, or
multiple-peaked profile), which in turn depends on the physical
characteristics of the emitting areas (e.g. temperatures, sizes, locations,
i.e. ratios and locations of the maxima and minima in a light curve) as well from viewing geometry 
\citep[angles between rotation, magnetic axes, line of sight, gravitational
  redshift, etc., e.g., see][]{2006MNRAS.373..836P}.

On the other hand, it is clear that if there is a periodic signal in the data
set with frequency $\nu$, then there is also a less significant signal at
$\mathrm{0.5}\nu$. Therefore, the most significant peak in the 
periodogram unequivocally corresponds to the {\it genuine} spin frequency for
a {\it single-peaked} light curve shape.
The situation is not simple for the case of a signal from a {\it double- or
  multiple-peaked} light curve shape. In this case,
depending on the parameters mentioned above, the periodogram can be identical to a single-peaked one.
Moreover, the dependence of an applied statistic 
(e.g. $\mathrm{Z^{2}_{m}}$, see also  Appendix~\ref{meth}) from a trial frequency will show 
a highly significant detection {\it only} at twice of the {\it true} spin frequency ($\mathrm{2}\nu$), 
if the two emitting areas have identical physical characteristics 
and favorable viewing geometry \citep[e.g. orthogonal rotator, see expression
  for $\mathrm{Z^{2}_{1}}$ and Fig.\,2 in][]{2015AN....336..545H}.

These arguments have motivated us to perform a complete X-ray timing re-analysis of the observational 
data sets of \xin\ for the identification of the {\em true} pulsation period by using
a more advanced analysis \citep[e.g. Bayesian analysis of the search, estimate, and testing the 
hypothesis following][hereafter GL]{1992ApJ...398..146G} and to explore some
effects, such as folding into other possible periods 
(as well half of the frequency of the most significant peak in the periodogram)  and checking of 
the presence of possible asymmetries in the folded light curves 
  over time and various energy ranges, i.e. peak-to-peak and 
minimum-to-minimum ratios and phase differences 
\citep[for details of methodical issues, see][]{2015AN....336..545H}. 
Moreover, in order to solve this challenging problem, a
  study of the distribution of the spin-phases (resulting from the folding of
  the photon arrival times into above mentioned various frequencies, see
  Appendix~\ref{meth}) has to be performed.

\subsection{X-ray timing analysis}\label{tim}

\begin{figure*} 
\centering
    \vspace{-0.2cm}
    \vbox{
      \hbox{
        \includegraphics[width=8.0cm]{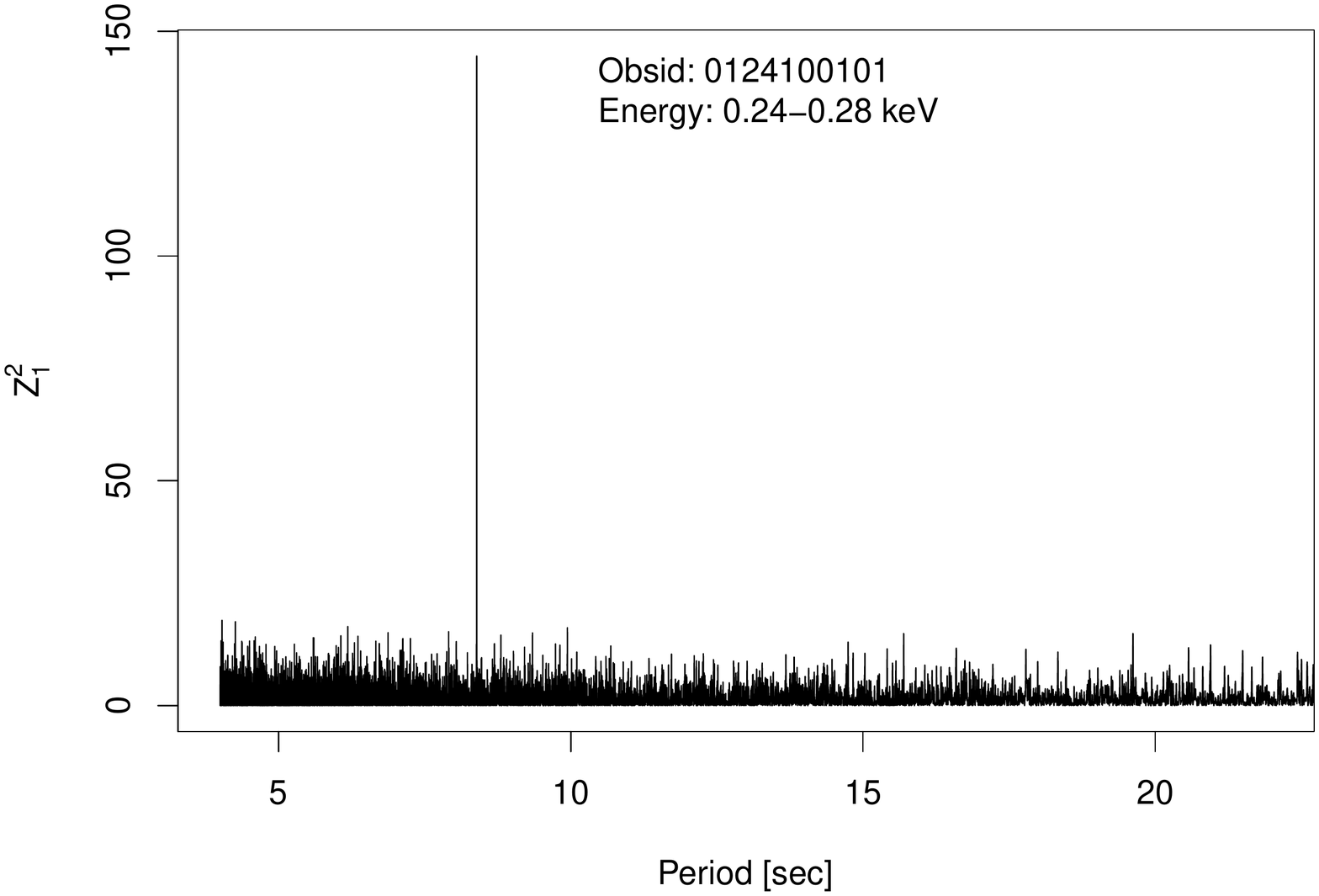}
        \includegraphics[width=8.0cm]{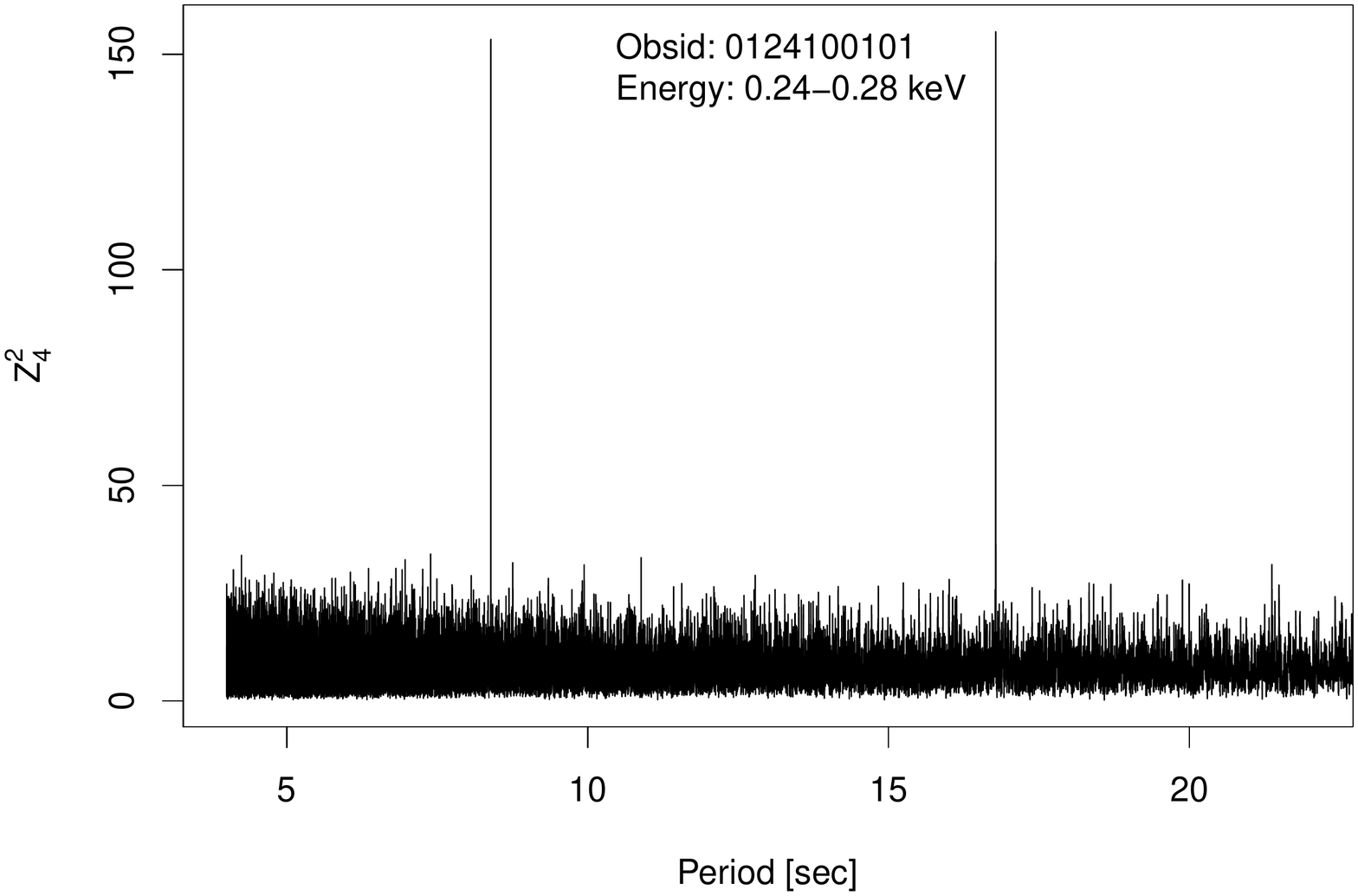}
      }
      \hbox{
        \includegraphics[width=8.0cm]{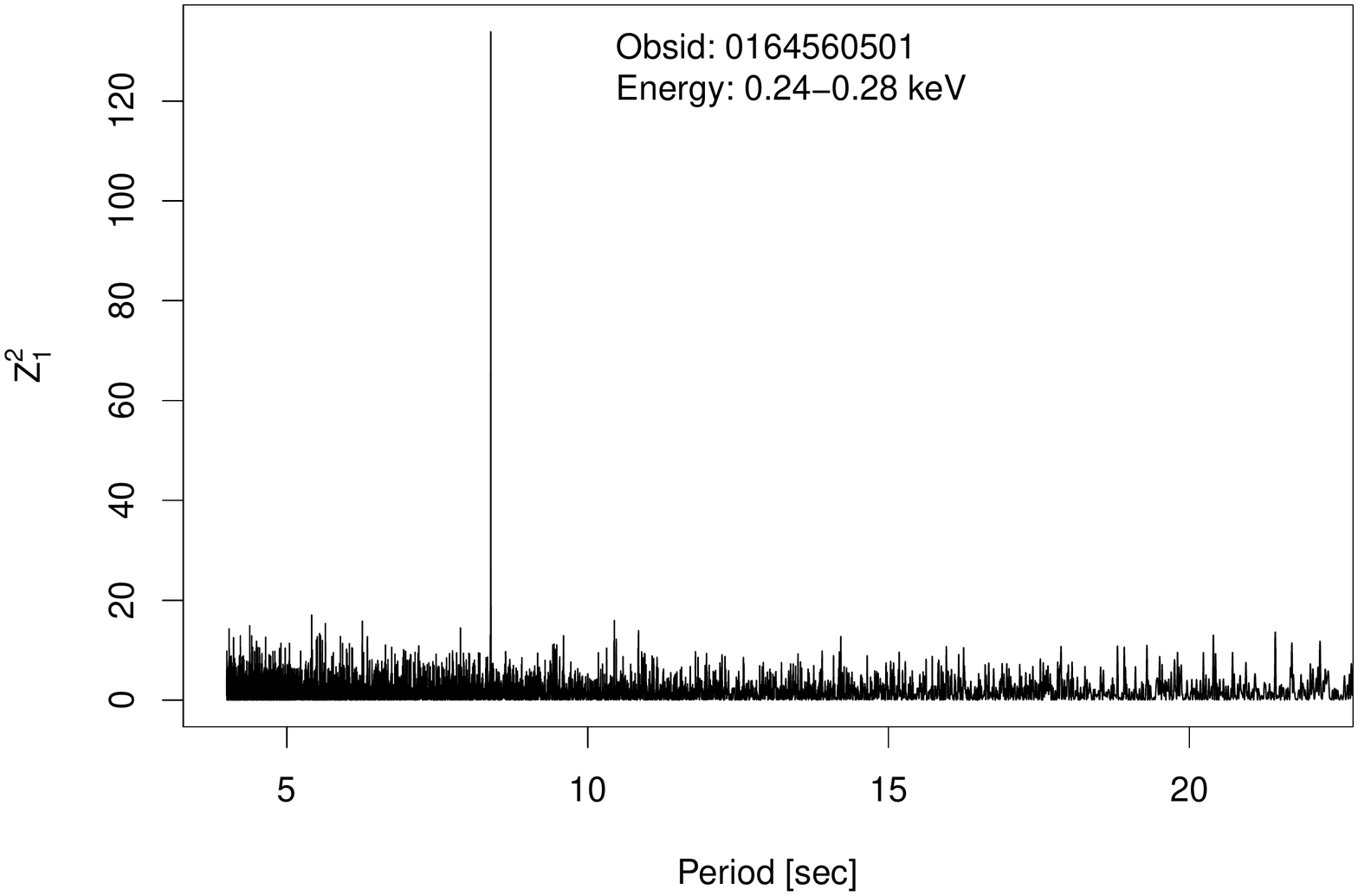}
        \includegraphics[width=8.0cm]{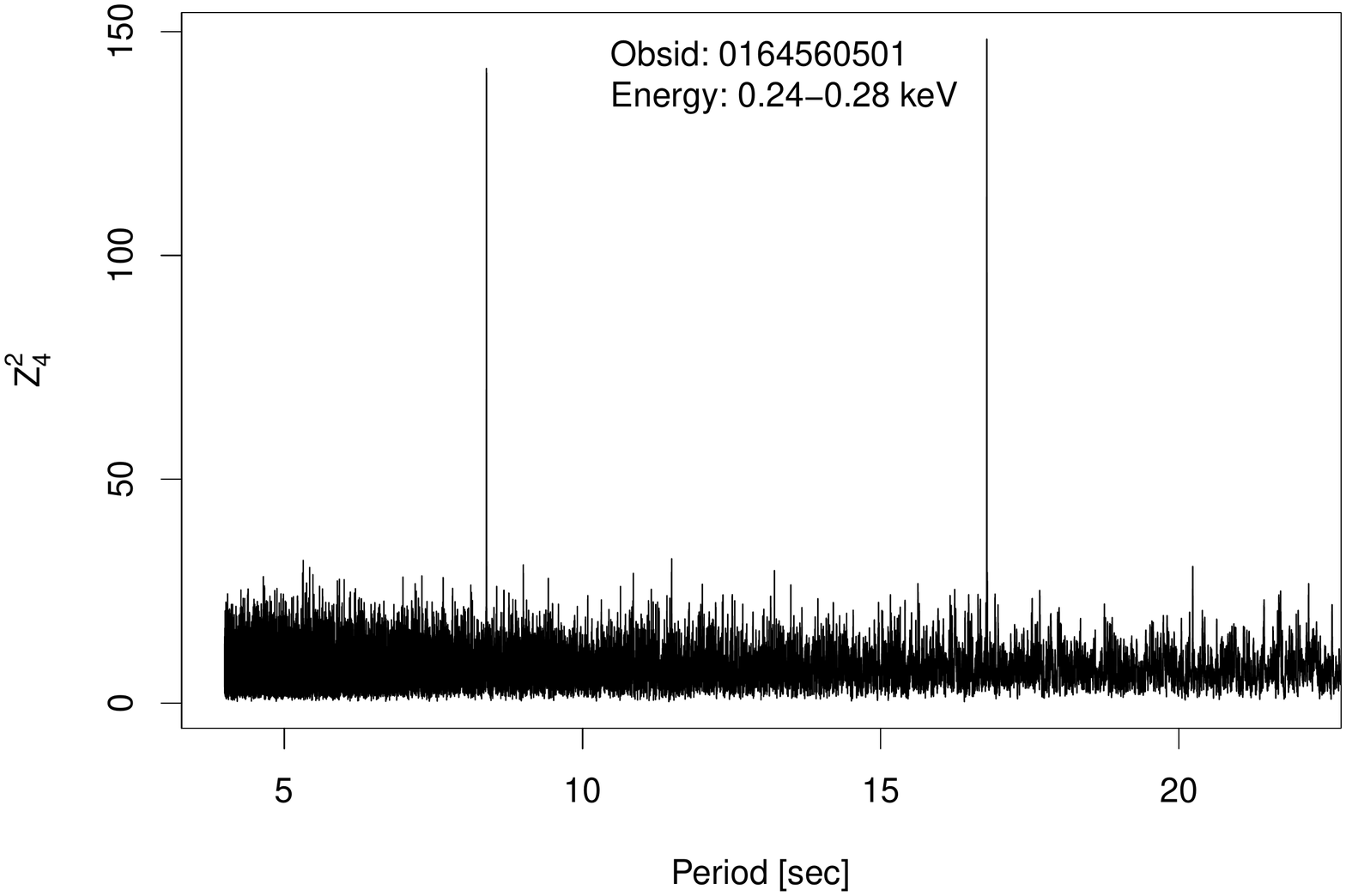}
      }
    }
    \vspace{-0.2cm}
\caption{\zm test statistic ($Z_{1}^{2}$ left and $Z_{4}^{2}$ right panel, respectively) vs trial period of two
pointed \xmm\  EPIC-pn observations (Obsids: 0124100101 and 0164560501, see Table~\ref{obslog} 
and Fig.~\ref{obs_ctr}) in the 0.24-0.28 keV narrow energy band of \xin. 
} 
\label{nkzz} 
\end{figure*}

First of all, for each pointed observation (Table~\ref{obslog}),
we used the cleaned photon arrival times (see Sect.~\ref{ditum})
to apply the $\mathrm{Z^{2}_{m}}$ test for periodicity detection in the frequency range
of 0.05$-$0.5~Hz in the broad energy band (0.16$-$1.6~keV). 
Note, that this range includes, as reported in the literature, the
spin-frequency 0.119~Hz of \xin\ as well as the second harmonic and
subharmonic frequencies. As already mentioned in Sect.~\ref{da}, the
periodograms revealed a very significant peak 
at the frequency 0.119~Hz. Apart from that they also show a less significant
second peak at the frequency of 0.0596~Hz, i.e. close to half the frequency of the 
most significant peak. In Fig.~\ref{nkzz}, we present the \zm
statistic periodogram of two
observations (Obsids: 0124100101 and 0164560501, see Table~\ref{obslog} and Fig.~\ref{obs_ctr}), 
corresponding to the first and  the brightest phase of \xin\ in the long-term light curve.

If we apply the H-test developed by 
\cite{1989A&A...221..180D,2010A&A...517L...9D}\footnote{The H-test $H\equiv\mathrm{Z^{2}_{m}-4m+4}$ 
is a modified $Z_m^2$ test including higher harmonics and finding optimal number of  them 
to be included in the search and detection of a periodic signal.} to the data sets, we identify 
a second peak in all pointed observations of \xin\ also in different energy bands. 
Moreover, for some energy bands, they do show a more significant peak corresponding to 0.0596~Hz
rather than 0.119~Hz.

These results led us to consider the frequency at 0.0596~Hz as the most probable {\em true} spin 
period of \xin\ and to perform a detailed analysis of the light curve shapes in different energy bands.

For this purpose, (together with \zm and $H$-statistics) 
we applied the GL method to the different data sets in various energy bands 
for the frequency ranges 0.05$-$0.07 Hz and 0.10$-$0.14 Hz separately. 
In general, the GL method first tests if a constant, variable or periodic
signal is present in a data set 
(given photon arrival times of events or a binned data set obeying Poissonian counting statistics)
when we have no specific prior information about the shape of the signal.

\begin{figure*}
\centering
    \vspace{-0.2cm}
      \hbox{
        \includegraphics[width=8.6cm]{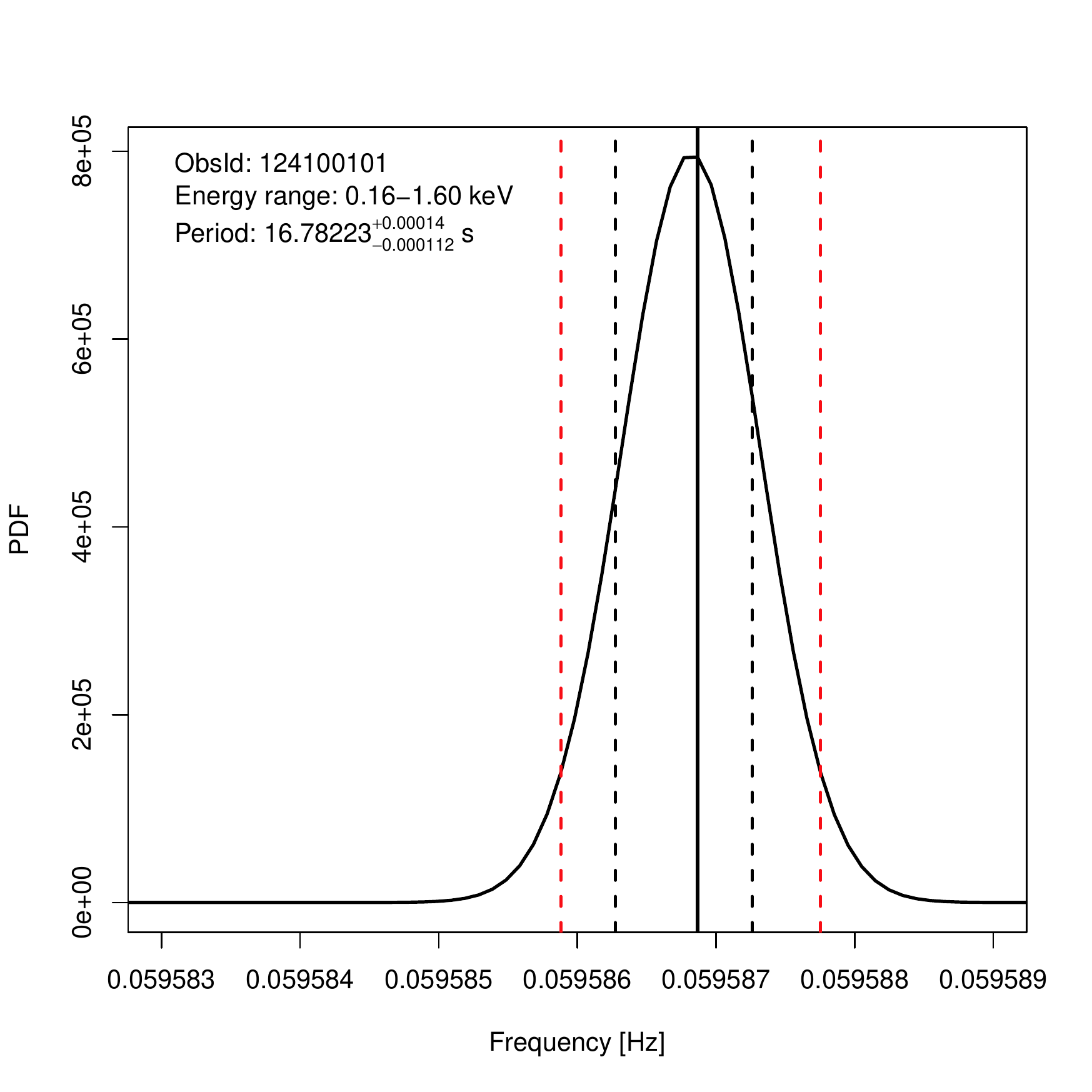}
        \includegraphics[width=8.6cm]{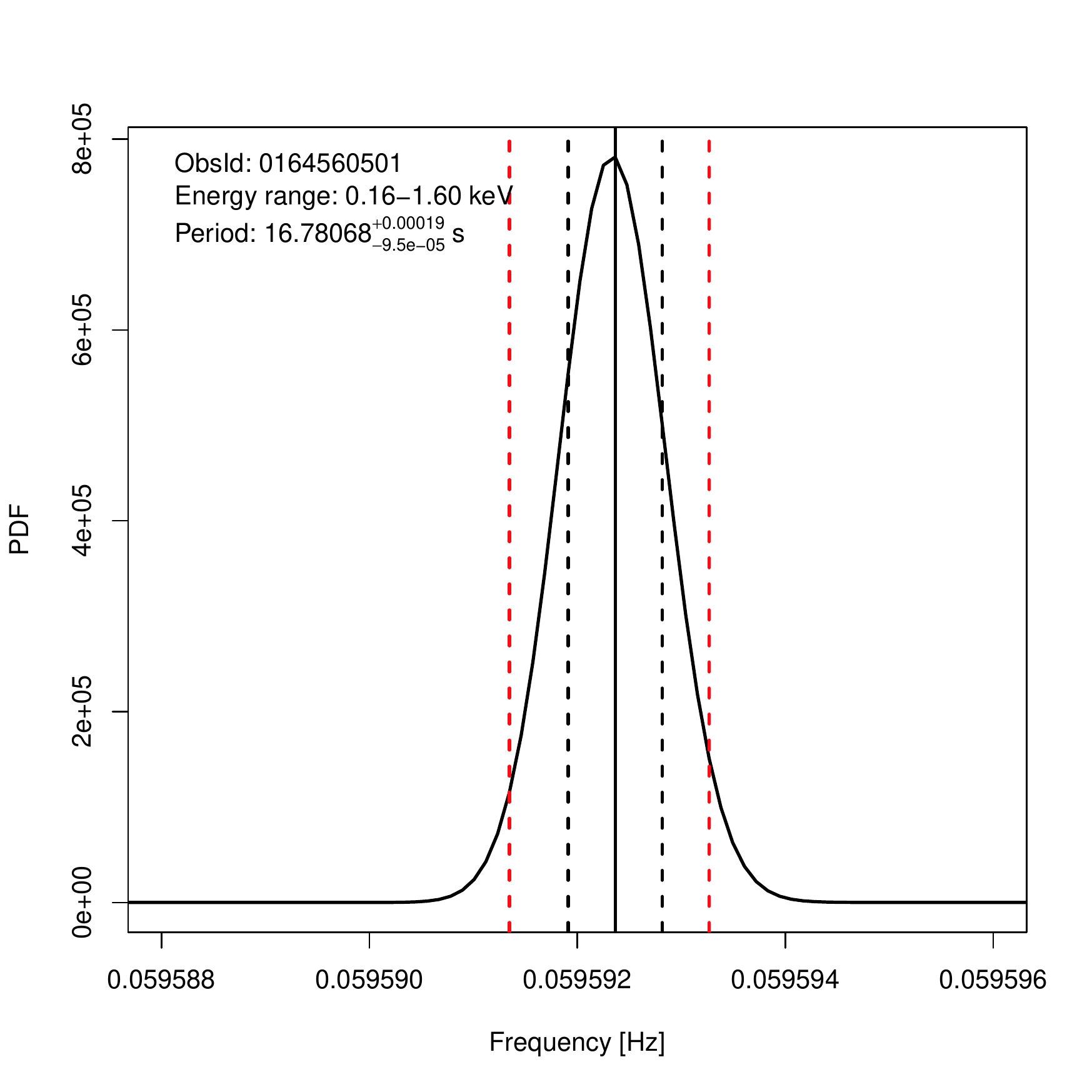}
}
    \caption{Estimated most probable spin-frequencies of \xin\ (for details, see text and Table~\ref{ax2}) 
by application of the GL method for periodicity search in the broad energy band 
for ObsIds 0124100101 and 0164560501 (\xmm\ EPIC-pn Full Frame mode). 
The highest posterior density (HPD) intervals, 68\% and 95\%, are determined as a probabilistic 
region around a posterior mode and depicted as vertical dashed-lines in black and red, respectively.}
\label{PestimGL} 
\end{figure*}

\begin{figure} 
\resizebox{\hsize}{!}{\includegraphics[bb=1 1 475 446,angle=90]{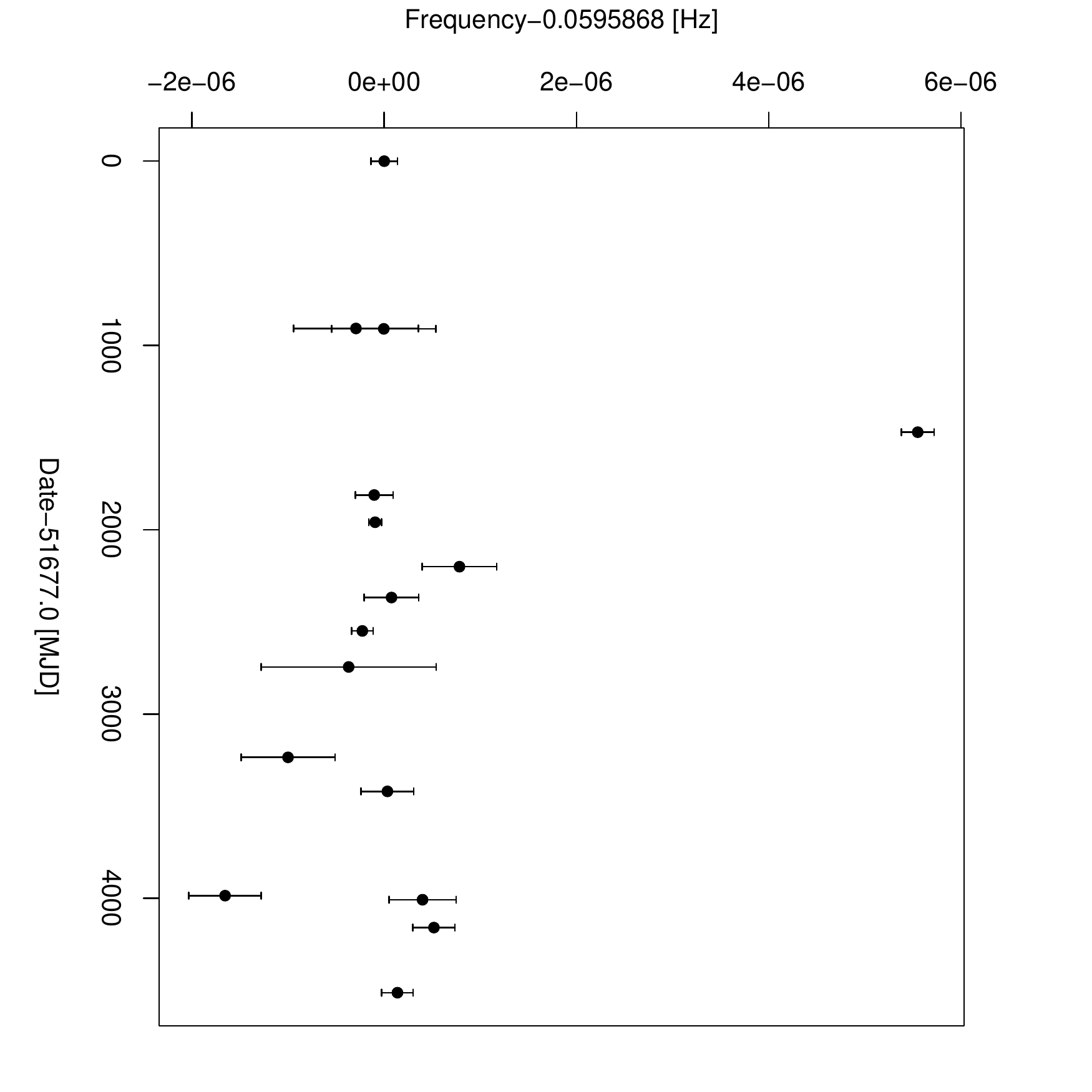}} 
\caption{Spin-frequency history of \xin\ (for details, see text and Table~\ref{ax2}) 
by application of the GL method for periodicity search in broad energy band for 
different \xmm\ EPIC-pn Full Frame mode observations. 
} 
\label{PvsTime} 
\end{figure}

In the GL method, periodic models are represented by a signal of folded photon
arrival times into trial frequency with a light curve shape as a stepwise
function with $m$ phase bins per period plus a noise contribution. With such a
model we are able to approximate a phase-folded light curve of any shape. 
Hypotheses for detecting periodic signals represent a class of stepwise, periodic models 
with the following parameters: trial period, phase, noise parameter, and number of bins ($m$). 
The most probable model parameters are then estimated by marginalization of the posterior 
probability over the prior specified range of each parameter. In Bayesian statistics, 
posterior probability contains a term that penalizes complex models (unless there is no significant
evidence to support that hypothesis), so we calculate the posterior
probability by marginalizing over a range of models, 
corresponding to a prior range of number of phase bins, $m$, from 2 to 20.
For all cases having strong support for the presence of a periodic signal,
using posterior probability density functions 
we have estimated the most probable values and uncertainties of all parameters, 
i.e. frequency\footnote{In the GL method, a posterior probability
density function is evaluated for any parameter of the model by application of Bayes theorem, 
e.g. for the trial frequency ($\nu$):
\be
p(\nu | D, M_m)=\frac{C}{\nu}\int_{0}^{2\pi} d\phi \frac{1}{W_m(\nu,\phi)}, 
\ee
where $C$
and 
$W_m(\nu,\phi)=\frac{N!}{n_1!n_2!\cdot\cdot\cdot n_m!}$ are the normalization constant and
number of ways the binned distribution could have arisen ``by chance''.
$n_j(\nu,\phi)$ is the number of events falling into the $j\,\rm{th}$
of $m$ phase bins given the frequency, $\nu$, and phase, $\phi$. 
$N$ is the total number of photons (for details, see Gregory \& Loredo, 1992, 1993).
For its uncertainty the highest posterior density (HPD) interval is determined as a probabilistic 
region around a posterior mode or moment, 
and is similar to a confidence interval in classical statistics.
}, phase and the most probable number of phase bins of a periodic signal. 
Results are presented in Table~\ref{ax2} and Figs.~\ref{PestimGL} and \ref{PvsTime}. 
Using these results, we created phase-folded light curves for all observations of \xin\ 
in three different energy bands (soft: 0.16$-$0.38~keV, hard: 
0.38$-$1.6~keV, broad: 0.16$-$1.6~keV, Figs.~\ref{resapp1},~\ref{resapp2},~\ref{resapp3},~\ref{resapp4}).

\begin{table*} 
\centering                          
\caption{Results of the application of the GL method in the frequency range of 0.05-0.07 Hz for the broad energy band of 0.16-1.60 keV.  
\label{ax2}}  
\begin{tabular}{cccccc} 
\hline\noalign{\smallskip} 
Obsid      & MJD \tablefootmark{*}    & Frequency & HPD$_{lo}$(68\%)  &  HPD$_{hi}$(68\%)   & Most probable \\ 
           & [mid-time]         & (Hz) &  [Hz]         &   [Hz]          &    number of phase bins  \\ 
\hline\noalign{\smallskip}  
0124100101  & 51677.485469  &  0.0595869 & 0.0595863 & 0.0595873  &  12  \\
0156960201  & 52584.936555  &  0.0595867 & 0.0595858 & 0.0595873  &  10  \\
0156960401  & 52587.011838  &  0.0595868 & 0.0595859 & 0.0595875  &  11  \\
0164560501  & 53147.743162  &  0.0595923 & 0.0595919 & 0.0595928  &  12  \\
0300520201  & 53488.686427  &  0.0595867 & 0.0595863 & 0.0595873  &  12  \\
0300520301  & 53636.310901  &  0.0595867 & 0.0595860 & 0.0595872  &  11  \\
0400140301  & 53877.355521  &  0.0595876 & 0.0595870 & 0.0595882  &  12  \\
0400140401  & 54044.616108  &  0.0595868 & 0.0595864 & 0.0595876  &  12  \\
0502710201  & 54225.859109  &  0.0595874 & 0.0595851 & 0.0595878  &  12  \\
0502710301  & 54421.380495  &  0.0595865 & 0.0595859 & 0.0595868  &  10  \\
0554510101  & 54911.796392  &  0.0595857 & 0.0595844 & 0.0595868  &  11  \\
0601170301  & 55096.379108  &  0.0595869 & 0.0595858 & 0.0595876  &  11  \\
0650920101  & 55662.143634  &  0.0595851 & 0.0595841 & 0.0595858  &  4   \\
0670700201  & 55684.233810  &  0.0595872 & 0.0595864 & 0.0595877  &  11  \\
0670700301  & 55835.339910  &  0.0595873 & 0.0595866 & 0.0595877  &  12  \\
0690070201  & 56188.514699  &  0.0595870 & 0.0595865 & 0.0595874  &  10  \\
\hline 
\end{tabular} 
\tablefoot{\tablefoottext{*}{The mid-time of the pointed observation.}}
\end{table*} 

\begin{figure*}
\centering
    \vspace{-0.2cm}
      \hbox{
        \includegraphics[width=8.0cm]{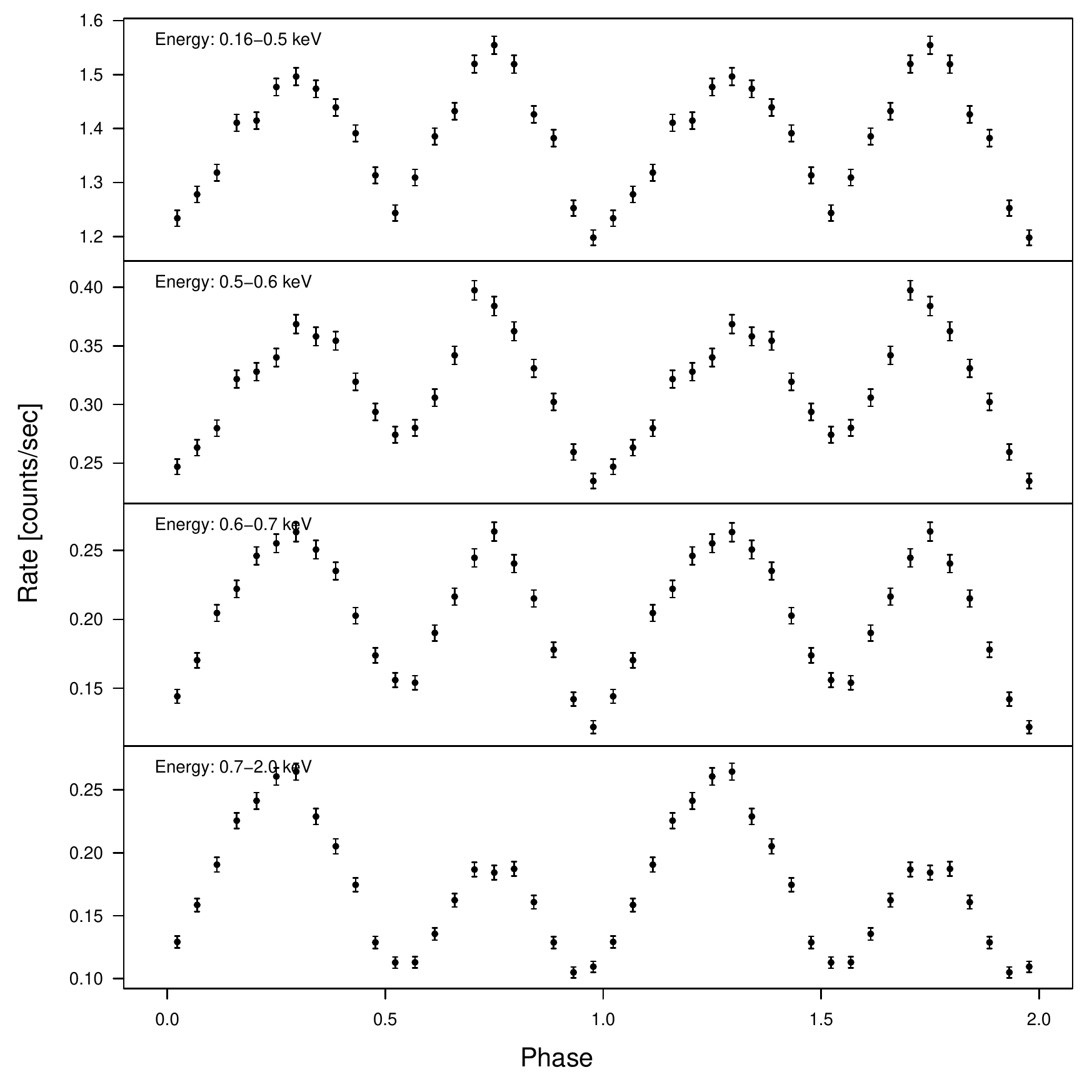}
        \includegraphics[width=8.0cm]{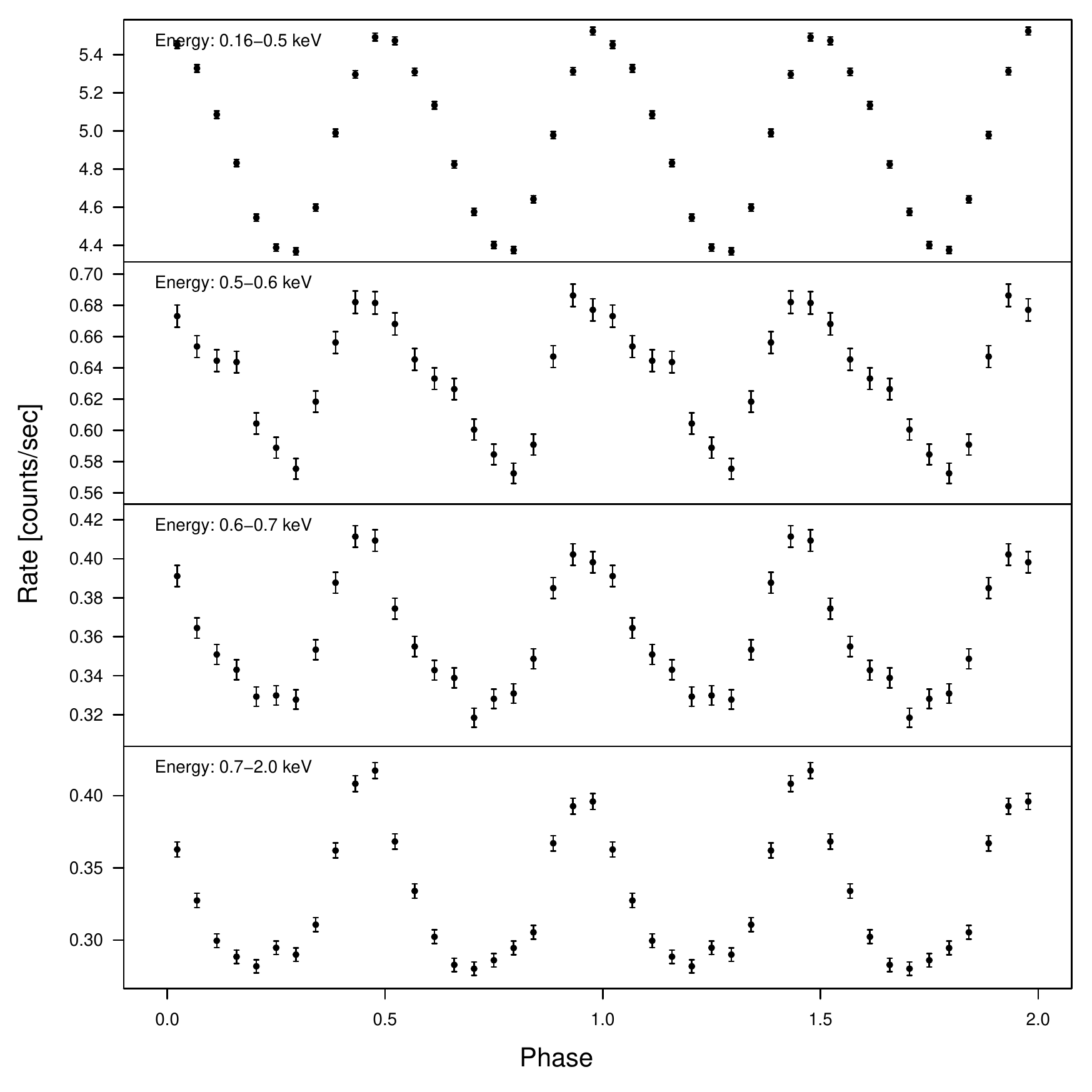}
      }
      \vspace{-0.2cm}
   \caption{Combined phase-folded light curves for RBS 1223 \citep[left
       panel][]{2011A&A...534A..74H} and \xin\ (right panel) in 
different energy bands (0.16$-$0.5\,keV, 0.5$-$0.6\,keV, 0.6$-$0.7\,keV, and 0.7$-$2.0\,keV)\label{fig:LCI_II}}
\end{figure*}

In addition, we constructed combined phase-folded light curves (Figs.~\ref{fig:LCI_II},~\ref{fig:groups}) 
for all observations and different data groups (see further Sect.~\ref{specphase}) which 
clearly show the dependence of the double-peak light curve profile on energy 
band \citep[cf, e.g][]{2011A&A...534A..74H}. 
Indeed, the relative strength of the modulation\footnote{We have employed the
  relative strength of modulation
\begin{equation}
A = \frac{\sum_i |{\rm CR}_i-{\rm CR}_{\rm mean}|}{\sum_i {\rm CR}_i},
\label{semi_amplitude}
\end{equation}
rather than the pulsed-fraction $PF=({\rm CR}_{\max}-{\rm CR}_{\min})/({\rm CR}_{\max}+{\rm CR}_{\min})$,
as an appropriate descriptor of the pulsed emission for complex shape light curves 
(the strength of variation are independent of the light curve shape, i.e. single or double pulse profile). 
Here ${\rm CR}$ is the count-rate.
Note that our defined relative strength of the modulation for a single-peaked
sinusoidal light curve differs by a factor of $2/\pi$ from the definition of
semi-amplitude or pulsed-fraction).}
for all the combined data groups are $0.07\pm0.02$, $0.05\pm0.02$,
$0.07\pm0.02$, and $0.12\pm0.04$ in the energy bands 
of 0.16$-$0.5~keV, 0.5$-$0.6~keV, 0.6$-$0.7~keV, 0.7$-$2.0~keV, respectively. 
Thus, the X-ray phase-folded light curves of \xin\  
\citep[cf, RBS1223 with the spin-period of 10.31~s,][]{2011A&A...534A..74H} have a markedly double-humped
shape with different ratios of maxima in different spectral bands.  
Moreover, they can be parameterized with a mixed von Mises distribution 
(see App.~\ref{meth}, Fig.~\ref{myrosenor1}, Table~\ref{fitmvM}), 
i.e. with double-peaked light curve profile showing a dependence of the estimated 
parameters (mean directions, concentrations, and proportion) upon the energy band, 
indicating that radiation emerges from at least two emitting areas.  These spin phase-folded 
light curves and the estimated parameters from the distribution of phases may serve 
for rough estimates of the viewing geometry and physical characteristics  
of the emitting areas (see below).

\subsection{Rotational phase-resolved X-ray spectral analysis}\label{specphase}

In order to create high S/N ratio spin phase-resolved spectra of \xin\ we coadded spectra for 
some sequential pointed observations where it shows approximately similar brightness in the broad 
energy band (0.16-1.6~keV). Namely, the total observational interval, spanned almost 12.5 years,
was divided into five groups (see, Fig.~\ref{obs_ctr} vertical dashed-lines) -- the second group is 
representing the brightest phase of \xin\ in the hard energy band.

\begin{figure}[t] 
\resizebox{\hsize}{!}{ 
        \includegraphics[clip=]{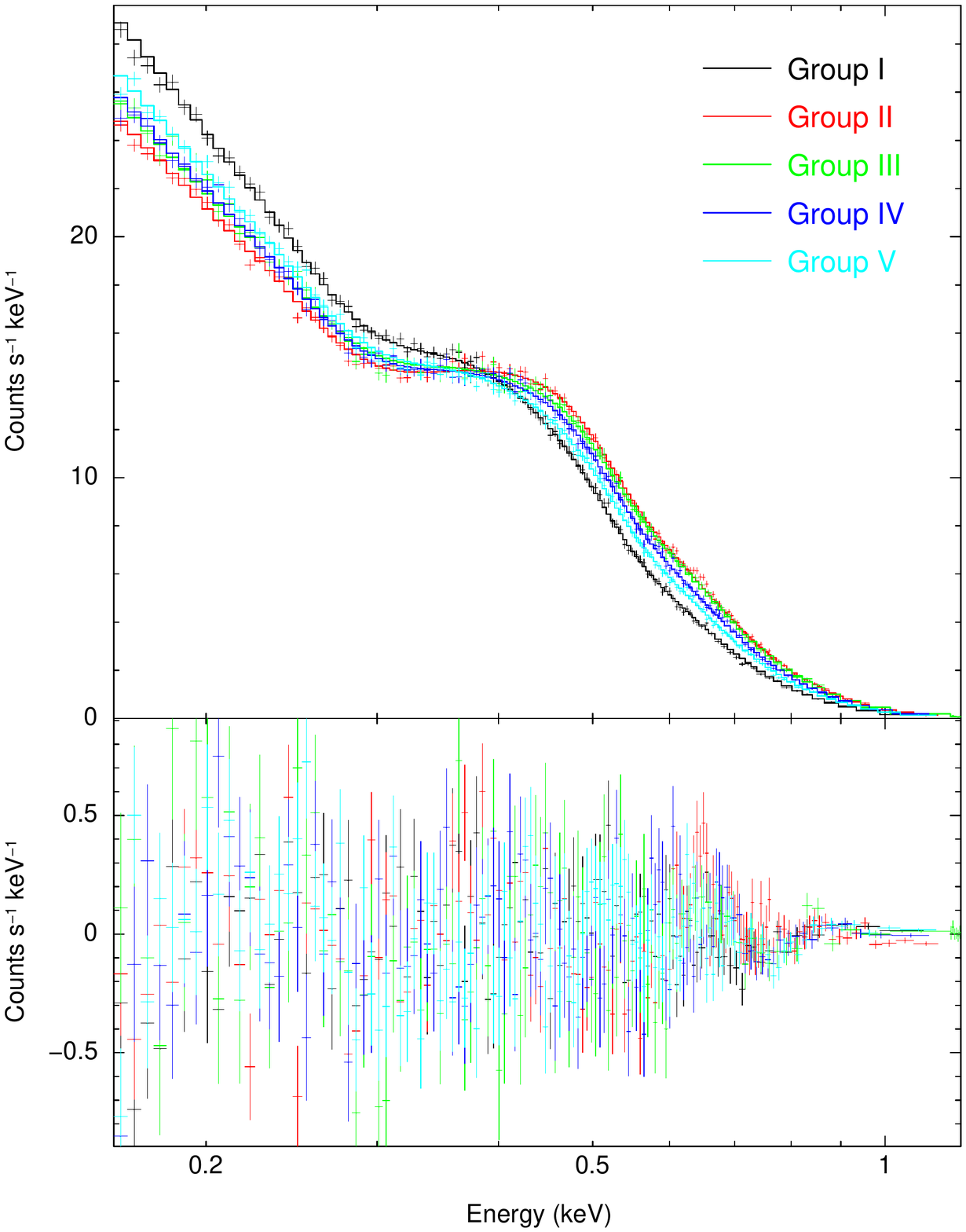}}
   \caption{Combined spin-phase averaged spectra and simultaneously fitted spectral model of an 
absorbed blackbody and a Gaussian absorption line for \xin. Note that the
  observed flux is almost the same before and after the 
variation at $E_{0} \approx \mathrm{ 0.38~keV}$. \label{fig:SPI_II_III_IV_V}}
\end{figure}

First, we have simultaneously fitted the phase-averaged,  high signal-to-noise 
ratio ($>$~30, each spectral bin containing at least 1000 counts) 
spectral data, collected from different observations (Table~\ref{obslog}),  
using a combination of an absorbed blackbody and a Gaussian absorption line  (a multiplicative component)  
in the model (in {\it XSPEC tbabs*bbodyrad*gabs}, 
see Fig.~\ref{fig:SPI_II_III_IV_V} and Table~\ref{fit1}).  
During the fitting the interstellar absorption 
and the Gaussian line energy was kept linked or free (see, Table~\ref{fit1}), 
while other parameters (blackbody temperature, Gaussian width, and normalizations) were left free 
for different data groups. 
Note, that if the linked  parameters also left free, the fitting results remained almost unchanged. 
It is worthwhile to note that these spectra cannot be fitted with the model 
of fixed parameters of the absorption line feature  at the pre-brightening phase.
 
\begin{table*} 
\caption[]{Simultaneous fitting results of combined, spin-phase averaged X-ray spectra of \xin \\ 
with spectral model of an absorbed blackbody with and without Gaussian absorption line (in {\it XSPEC tbabs*bbodyrad*gabs})} 
\label{fit1} 
\begin{tabular}{lcccccl} 
\hline\noalign{\smallskip} 
Fitted & \multicolumn{6}{c}{Groups\tablefootmark{*}} \\ 
\cline{2-7}\noalign{\smallskip}  
Parameter\tablefootmark{**} & I & II & III & IV & V & Remark \\ 
\cline{2-7}  
\hline\hline\noalign{\smallskip}  
Energy range & \multicolumn{6}{c}{0.16$-$1.2~keV} \\ 
\cline{2-7}\noalign{\smallskip}  
 $nH\times 10^{20}~\mathrm{ [ cm^{-2}]}$ & $\mathrm{ 1.06 \pm 0.06}$ & $\mathrm{ 1.09 \pm 0.11}$ & $\mathrm{ 0.87 \pm 0.10}$  &  $\mathrm{ 1.02 \pm 0.10}$ & $\mathrm{ 0.92 \pm 0.07}$ & free \\    
 $kT~\mathrm{ [eV]}$  & $\mathrm{ 85.2 \pm 0.4}$      & $\mathrm{91.3 \pm 0.4}$ &  $\mathrm{ 91.5 \pm 0.6}$  & $\mathrm{89.5 \pm 0.5}$ & $\mathrm{88.2 \pm 0.4}$ & free \\    
 $E_\mathrm{ line}~\mathrm{ [ eV]}$ & $\mathrm{ 329.0 \pm 3.3}$  & {\em ibid} & {\em ibid}  & {\em ibid} & {\em ibid} & linked\\    
 $\sigma_\mathrm{ line}~\mathrm{ [ev]}$  & $\mathrm{18.3 \pm 8.0}$  & $\mathrm{70.0 \pm 6.7}$  & $\mathrm{77.2 \pm 8.7}$  & $\mathrm{70.4 \pm 8.6}$  & $\mathrm{69.2 \pm 8.4}$  & free \\ 
 $D_\mathrm{ line}$\tablefootmark{**}  & $\mathrm{0.011 \pm 0.002}$   & $\mathrm{0.076 \pm 0.009}$  & $\mathrm{0.074 \pm 0.010}$  & $\mathrm{0.059 \pm 0.096}$  & $\mathrm{0.044 \pm 0.072}$  & free \\  
\hline\noalign{\smallskip} 
 Reduced $\chi^2$ & \multicolumn{6}{c}{$\chi^2/\mathrm{dof = 853.07/581 = 1.47}$} \\ 
\hline\noalign{\smallskip} 
 $nH\times 10^{20}~\mathrm{ [ cm^{-2}]}$ & $\mathrm{ 1.04 \pm 0.04}$ & {\em ibid} & {\em ibid}  & {\em ibid} & {\em ibid} & linked \\    
 $kT~\mathrm{ [eV]}$  & $\mathrm{ 85.2 \pm 0.4}$      & $\mathrm{90.6 \pm 0.3}$ &  $\mathrm{ 91.1 \pm 0.3}$  & $\mathrm{89.1 \pm 0.3}$ & $\mathrm{88.0 \pm 0.3}$ & free \\    
 $E_\mathrm{ line}~\mathrm{ [ eV]}$ & $\mathrm{ 328.0 \pm 6.2}$  & $\mathrm{ 323.0 \pm 3.1}$ & $\mathrm{ 340.7 \pm 3.7}$  & $\mathrm{ 329.9 \pm 4.0}$ & $\mathrm{ 342.9 \pm 4.3}$ & free\\    
 $\sigma_\mathrm{ line}~\mathrm{ [ev]}$  & $\mathrm{26.3 \pm 10.9}$  & $\mathrm{82.7 \pm 3.5}$  & $\mathrm{74.4 \pm 5.3}$  & $\mathrm{76.8 \pm 5.3}$  & $\mathrm{63.7 \pm 5.6}$  & free \\ 
 $D_\mathrm{ line}$\tablefootmark{**}  & $\mathrm{0.012 \pm 0.002}$   & $\mathrm{0.095 \pm 0.006}$  & $\mathrm{0.069 \pm 0.007}$  & $\mathrm{0.066 \pm 0.006}$  & $\mathrm{0.039 \pm 0.004}$  & free \\  
\hline\noalign{\smallskip} 
 Reduced $\chi^2$ & \multicolumn{6}{c}{$\chi^2/\mathrm{dof = 835.15/587 = 1.42}$} \\ 
\hline 
\end{tabular} 
\tablefoot{\tablefoottext{*}{See subsection~\ref{specphase}, Table~\ref{obslog} and Fig~\ref{obs_ctr} for definition of the data groups.}}
\tablefoot{\tablefoottext{**}{Parameter uncertainty corresponds to the 90\% of the confidence interval. $D_\mathrm{ line}$ presents the strength of a Gaussian line (for details see XSPEC manual).}}
\end{table*}

Next, we have extracted  high signal-to-noise ratio ($>$~10, each spectral bin
containing at least 100 counts) spectral data corresponding to the 8 rotational phase bins 
(the maxima, minima, and intermediate stages) of the double-peaked pulse profile and  
collected from different observations (see, Fig.~\ref{fig:LCI_II}~\ref{fig:groups}). 
These 40 spectra for the five data groups were subject of the simultaneous fitting with a number of 
highly magnetized INS surface/atmosphere models developed by us \citep{2010A&A...522A.111S} 
and implemented into the X-ray spectral fitting package {\it XSPEC} \citep{2011A&A...534A..74H}.  
They are based on various local surface models and compute rotational phase dependent 
integral emergent spectra of INS, using analytical approximations. The basic model includes 
temperature/magnetic field distributions over the INS surface,  
viewing geometry and gravitational redshift. Three local
radiating surface models are also considered, namely 
a naked condensed iron surface and partially ionized hydrogen model atmosphere,  
semi-infinite or finite on top of the condensed surface. 
To compute an integral spectrum, the model uses an analytical expression for the
local spectra, an emitting condensed iron surface and a diluted blackbody spectrum, transmitted 
through the thin magnetized atmosphere with an absorption feature. 
The center of the line depends on the local magnetic field strength, representing either 
the ion cyclotron line for completely ionized iron and/or the blend
of the proton cyclotron and bound-bound atomic transitions in neutral hydrogen. 
In the latter case, 
it is parameterized by half of a Gaussian line with parameters: $E_\mathrm{ line}$, 
optical depths $\tau = \tau^{0} \exp\left(-\frac{(E-E_\mathrm{line})^2}{2\sigma_\mathrm{ line}^2}\right)$  and 
the widths $\sigma_\mathrm{line}$ (for details and 
further references, see \cite{2010A&A...522A.111S,2011A&A...534A..74H}). 

Given the large number of free parameters in the model, we performed a preliminary analysis 
in order to get rough estimates  (or constraints) on some of them. From the observed  
double-peaked profile light curve shape in different energy bands 
(see, right panel Fig.~\ref{fig:LCI_II} and Fig.~\ref{fig:groups}),  
it has been already clear that the two emitting areas have slightly different spectral and geometrical  
characteristics (e.g., the relatively cooler one has a larger size). 
Secondly, some constraints (initial values for the fitting) on magnetic field strengths 
at the poles may be used on the base of  
period and its derivative values\footnote{A steady spin-down 
of \xin\ was derived \citep[][]{byur2016} based on the spin-frequency history
(Fig.~\ref{PvsTime}) by robust linear fitting, i.e. eliminating the outliers :
$\mathrm{P_{0}=16.78215 s, \dot{P}=1.86 \times 10^{-13} s s^{-1},
  MJD=51677.104103}$, and $\mathrm{\Delta \nu \sim 10^{-6}-10^{-5} Hz}$ at MJD=53147.442236}
 assuming magnetic dipole braking as a  main mechanism of the spin-down of
 \xin\ \citep[see, e.g.][]{2012MNRAS.423.1194H,byur2016}. 
Moreover, from the peak-to-peak separation and the ratio of the minima in this double-humped light  
curve, we may locate the cooler one at some offset angle with respect to the  
magnetic axis and azimuth \citep{2005A&A...441..597S}. 

Having these general constraints on temperatures and  
magnetic field strengths at the poles, we simulated a large number 
of photon spectra (absorbed blackbody with Gaussian absorption line) folded with the response of 
\xmm\  EPIC-pn camera with the same observing mode of \xin. 
We included also the interstellar absorption (for parameters see Table~\ref{fit1}), 
and used a characteristic magnetic field strength of B$\sim$~3.0$\times$10$^{13}$ Gauss.
The predicted phase-folded light curves (with a purely dipolar magnetic field configuration, 
depending on the relative orientation between the magnetic axis, the rotation axis, 
and the observer line of sight, as well gravitational redshift, see upper left
panel of Fig.~\ref{myrosenor1} in Appendix~\ref{meth})
in four spectral ranges (0.16-0.5~keV, 0.5-0.6~keV, 0.6-0.7~keV, 0.7-2.0~keV)
normalized to the maximum of the brightness, were cross-correlated with
observed ones (Fig.~\ref{fig:LCI_II}).
We infered some constraints on the parameters from the unimodal  
distributions of them, when cross-correlation coefficients were exceeding 0.9 in  
the mentioned four energy bands simultaneously and used them as  
initial input value and as lower and upper bounds for fitting purposes.  
For example, gravitational redshift cannot exceed 0.3 (it is not possible
to obtain the observed relative strength of modulation at larger $z$ due to strong light bending) 
or the antipodal shift angle must be less than $5^\circ$ 
(due to the observed phase separation between the two peaks in the light curve).  
It is also evident that the sum of the inclination angle 
of the line of sight and magnetic poles relative to the rotational axis are already  
constrained by the light curve class \citep[see,][class III]{2006MNRAS.373..836P}.

Having the above-mentioned crude constraints and input values of the free parameters, we performed fitting  
with the models implemented in the {\it XSPEC} package of all the combined spectra of \xin\ 
(see Table~\ref{obslog} and Fig.~\ref{obs_ctr}) simultaneously, i.e. each of
those phase resolved spectra considered as different data set with the fixed spin-phase, and
with the linked parameters to the other data sets.  
 
The fitted model (Chi$-$Squared fit statistic value of 6070 for 4910  
degrees of freedom) with dipolar field configuration (magnetic field
strengths at the poles 
of $\mathrm{B_{p1} \approx B_{p2}\approx(0.3\pm0.02)\times 10^{14}~G}$), 
as an orthogonal rotator (angle between the magnetic and the rotation axis $\theta= (89 \pm 5)^{\degr}$), 
 with the inclination angle (angle between rotation axis and the observer line
 of sight) of $i=(42 \pm 3)^{\degr}$, and the shift angle of $\kappa \approx
 0^{\degr}$  is presented in Table~\ref{fit2}  (see also, upper left panel 
of Fig.~\ref{myrosenor1} in Appendix~\ref{meth}).

\begin{figure*} 
\centering
    \vspace{-0.2cm}
    \vbox{
      \hbox{
        \includegraphics[width=8.6cm,angle=90]{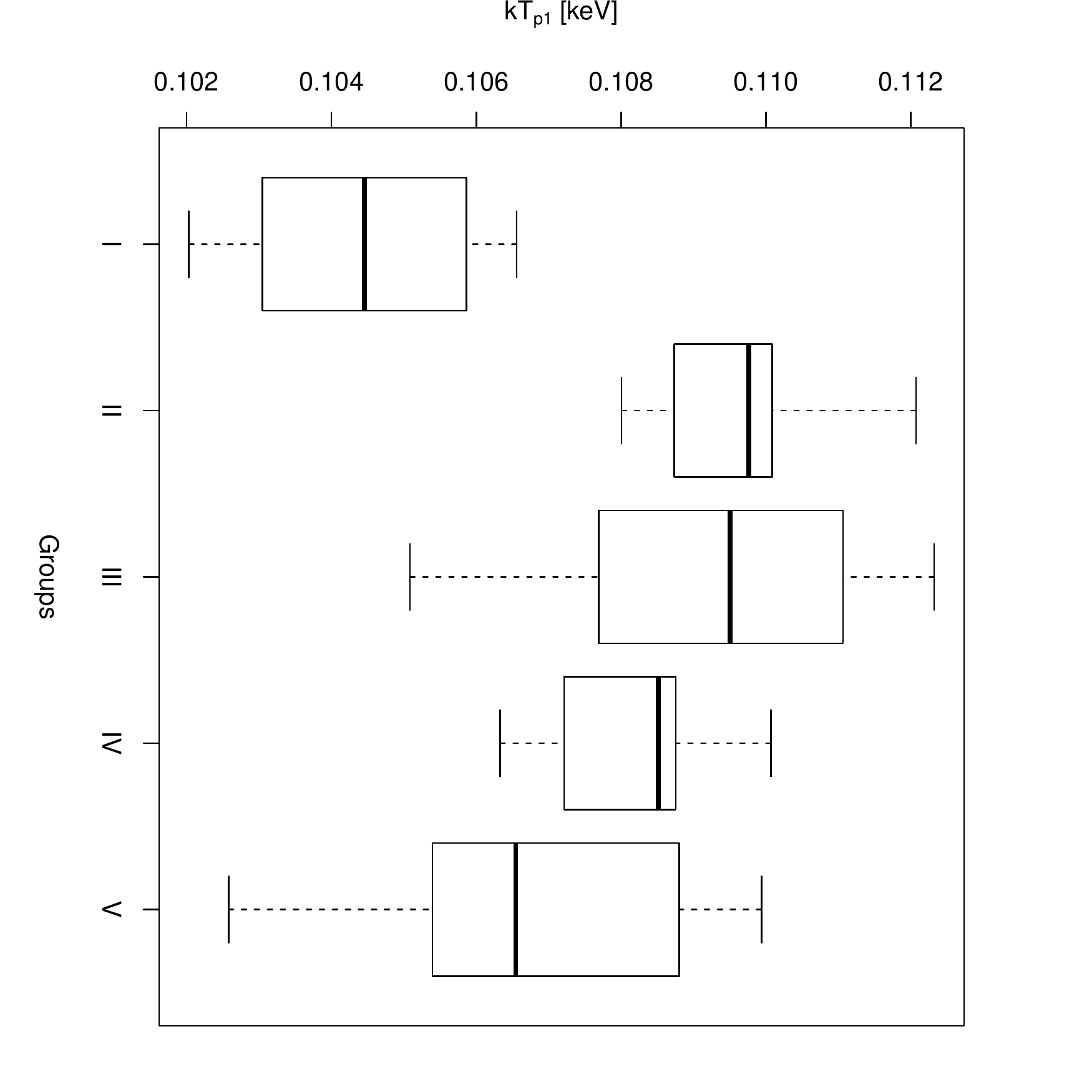}
        \includegraphics[width=8.6cm,angle=90]{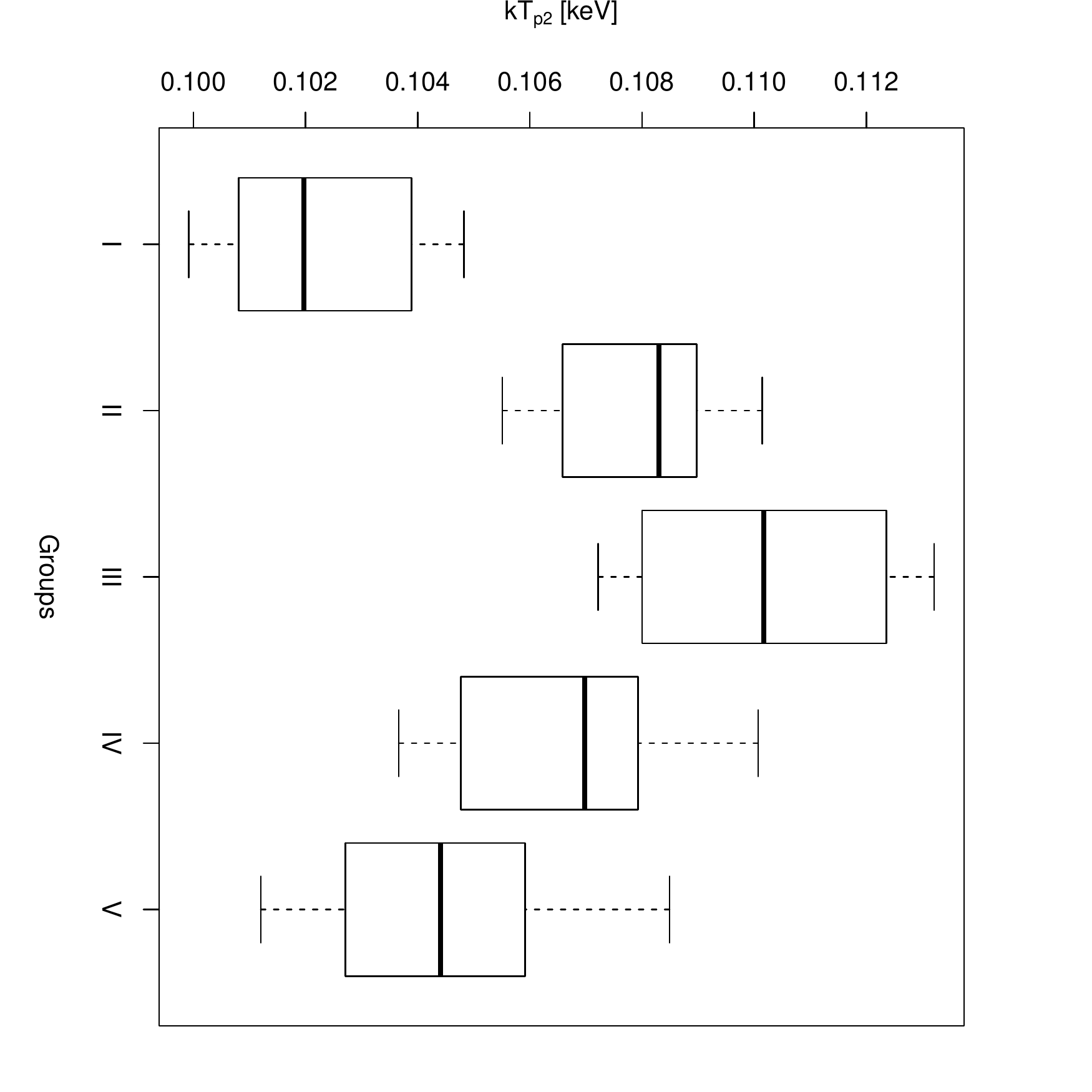}
      }
      \hbox{
        \includegraphics[width=8.6cm,angle=90]{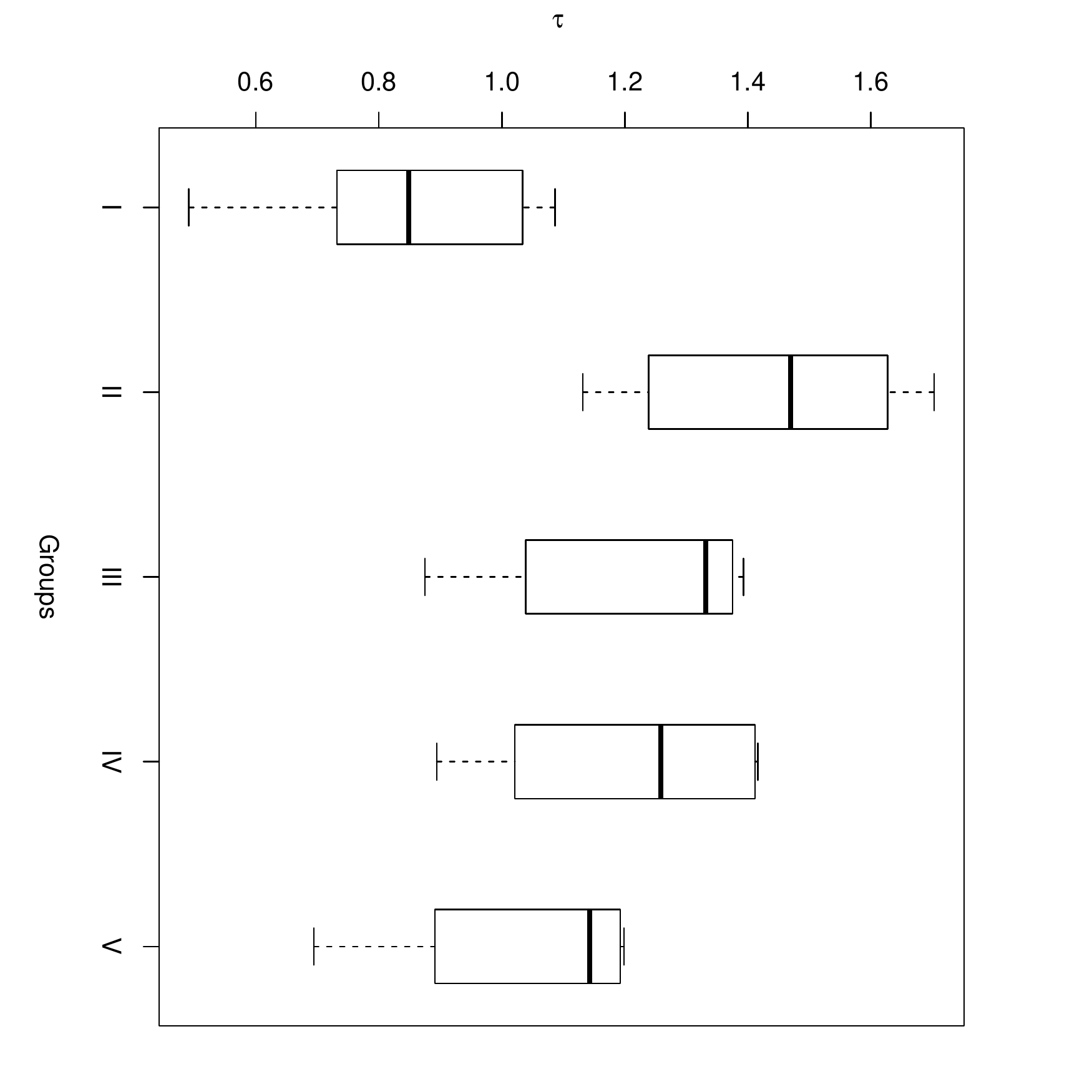}
        \includegraphics[width=8.6cm]{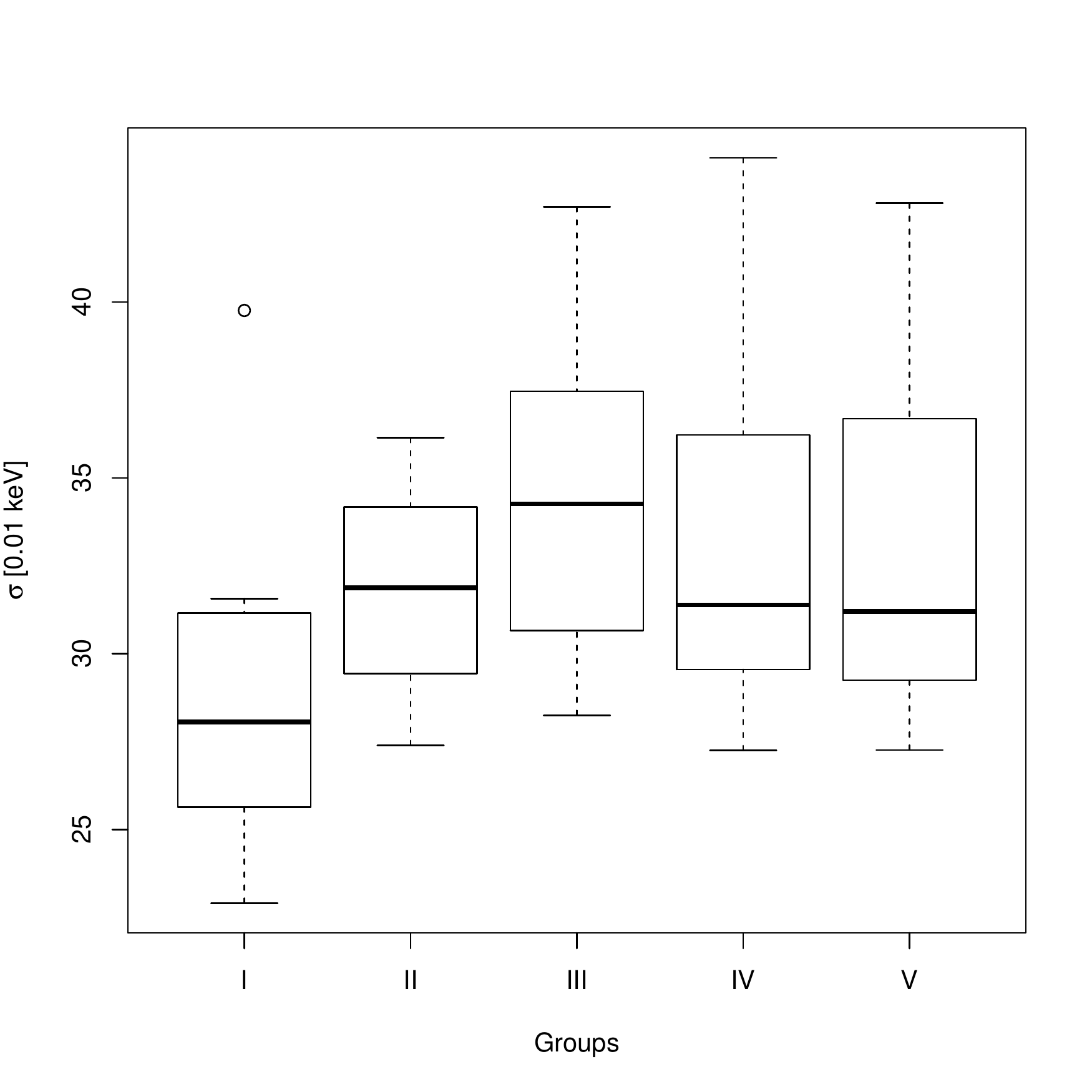}
      }
    }
    \vspace{-0.2cm}
\caption{Box-plots of the MCMC fitted parameters $kT_{p1},kT_{p2}$~(upper
  panel) and $\tau,\sigma$~(bottom panel) 
in the energy band of 0.16-2.0 keV of  \xin\ for different groups combined from multi-epoch pointed 
\xmm\  EPIC-pn observations with spectral model of condensed iron surface and partially ionized 
hydrogen atmosphere (for details see text and Table~\ref{fit2}).  
} 
\label{nktausigkT} 
\end{figure*}

In order to assess the degree of uniqueness and to estimate confidence intervals  
of the determined parameters, we have additionally performed a Markov Chain Monte Carlo (MCMC) fitting 
as implemented in {\it XSPEC}. 
The parameters of the fitted model are presented in Table~\ref{fit2} and Fig.~\ref{nktausigkT}.

\begin{figure}[t] 
\resizebox{\hsize}{!}{ 
\includegraphics[clip=]{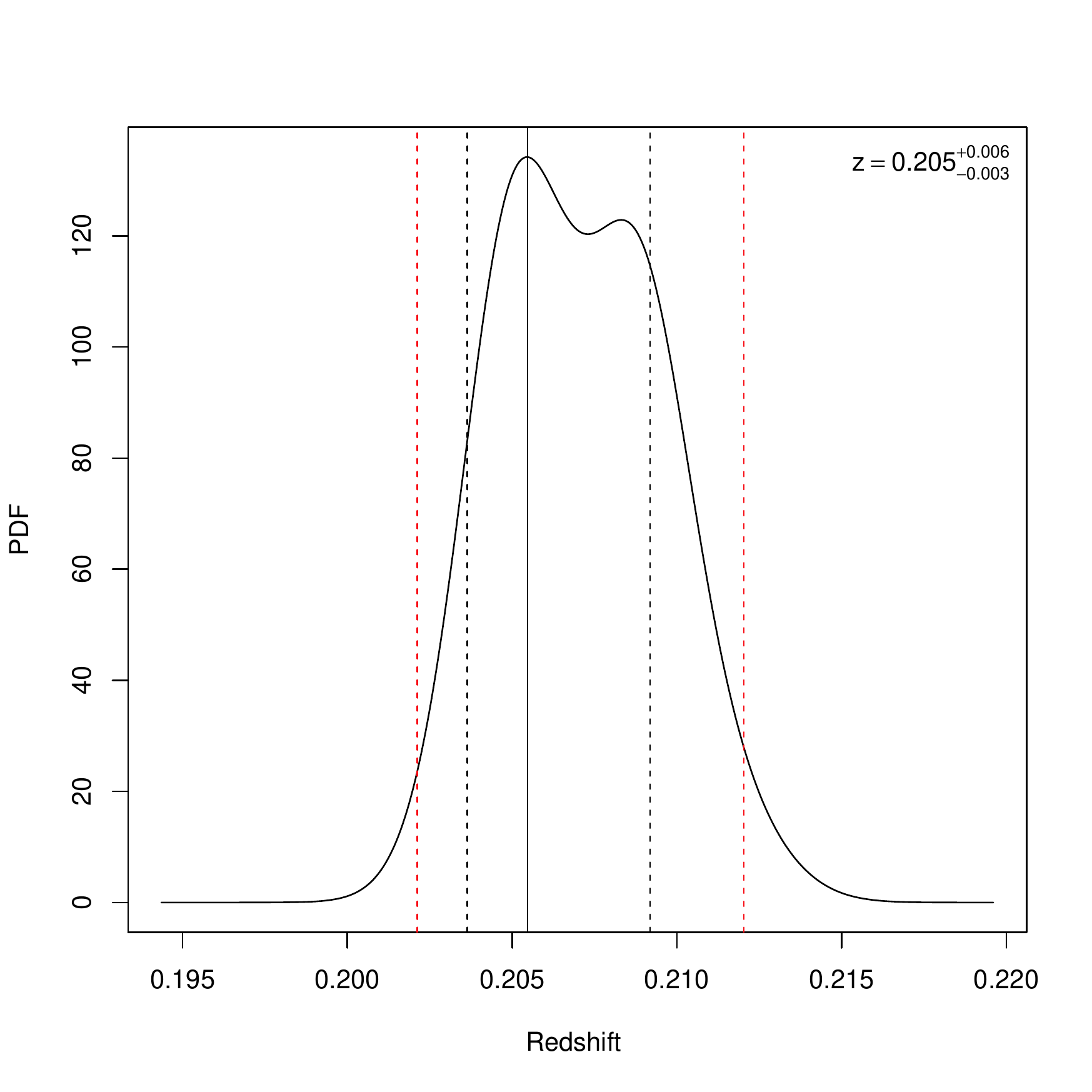}} 
\caption{Probability density distributions of 
gravitational redshift by MCMC fitting with the model of a neutron star with  
condensed surface and partially ionized hydrogen layer above it. The most
probable parameter value is indicated by the solid vertical line.  
Dashed vertical lines indicate the highest probability interval ($68\%
\mathrm{and} 95\%$, for details see text). 
} 
\label{mcmc_res} 
\end{figure}

In Fig.~\ref{mcmc_res} we present the probability density distributions of the gravitational redshift.  
Note, independent initial input values of parameters of the MCMC approach
converged to the same values, in different chains. 

\begin{table*} 
\caption[]{Simultaneous fitting results of combined, spin-phase resolved X-ray spectra of \xin \\
using a condensed iron surface and hydrogen atmosphere model.} 
\label{fit2} 
\begin{tabular}{lcccccl} 
\hline\noalign{\smallskip} 
Fitted & \multicolumn{6}{c}{Groups\tablefootmark{*}} \\ 
\cline{2-7}\noalign{\smallskip}  
Parameter\tablefootmark{**} & I & II & III & IV & V & Remark \\ 
\cline{2-7}  
\hline\hline\noalign{\smallskip}  
 $nH\times 10^{20}~\mathrm{ [ cm^{-2}]}$ & $\mathrm{ 1.20 \pm 0.06}$ & $\mathrm{ 1.17 \pm 0.06}$ & $\mathrm{ 1.17 \pm 0.06}$  &  $\mathrm{ 1.18 \pm 0.06}$ & $\mathrm{ 1.18 \pm 0.06}$ & free \\    
 $kT_{p1}~\mathrm{ [eV]}$  & $\mathrm{ 104.0 \pm 4.3}$      & $\mathrm{110.2 \pm 4.1}$ &  $\mathrm{108.3 \pm 4.5}$  & $\mathrm{107.5 \pm 4.0}$ & $\mathrm{106.6 \pm 3.6}$ & free \\    
 $kT_{p2}~\mathrm{ [eV]}$  & $\mathrm{102.5 \pm 4.2}$      & $\mathrm{107.7 \pm 3.9}$ &  $\mathrm{109.5 \pm 4.8}$  & $\mathrm{105.8 \pm 3.8}$ & $\mathrm{105.6 \pm 3.4}$ & free \\    
 $B_{p12}\times 10^{13}~\mathrm{ [G]}$  & $\mathrm{ 3.00 \pm 0.15}$ & {\em ibid} & {\em ibid}  & {\em ibid} & {\em ibid} & linked \\  
 $\theta~\mathrm{ [^{\degr}]}$  & $\mathrm{ 88.9 \pm 4.8}$      & {\em ibid} & {\em ibid}  & {\em ibid} & {\em ibid} & linked \\  
 $i~\mathrm{ [^{\degr}]}$  & $\mathrm{ 42.8 \pm 2.2}$      & {\em ibid} & {\em ibid}  & {\em ibid} & {\em ibid} & linked \\  
 $z$  & $\mathrm{ 0.205 \pm 0.001}$  & {\em ibid} & {\em ibid}  & {\em ibid} & {\em ibid} & linked \\  
 $\sigma_\mathrm{ line}~\mathrm{ [ev]}$\  & $\mathrm{268.3 \pm 50.9}$  & $\mathrm{313.4 \pm 38.6}$  & $\mathrm{329.0 \pm 56.9}$  & $\mathrm{302.7 \pm 60.3}$  & $\mathrm{310.6 \pm 60.4}$  & free \\ 
 $\tau_\mathrm{ line}$  & $\mathrm{0.80 \pm 0.24}$   & $\mathrm{1.56 \pm 0.26}$  & $\mathrm{1.38 \pm 0.26}$  & $\mathrm{1.37 \pm 0.26}$  & $\mathrm{1.18 \pm 0.22}$  & free \\  
\hline\noalign{\smallskip} 
 Reduced $\chi^2$ & \multicolumn{6}{c}{$\chi^2/\mathrm{dof = 6070/4910 = 1.24}$} \\ 
 C$-$statistic & \multicolumn{6}{c}{6050 using 5050 PHA bins and 4910 dof} \\ 
\hline 
\end{tabular} 
\tablefoot{\tablefoottext{*}{See subsection~\ref{specphase},
    Table~\ref{obslog}, Fig~\ref{obs_ctr}, and Fig.~\ref{myrosenor1} for
    definition of the data groups and parameters.}}
\tablefoot{\tablefoottext{**}{Parameter uncertainty corresponds to the 95\% of the HPD interval.}} 
\end{table*} 
 
\section{Discussion}\label{discus}

Our comprehensive X-ray timing analysis of \xin, based on the  
\xmm\ EPIC-pn long-term pointed observations, showed that the most probable, 
plausible spin frequency of this INS can be considered 0.0596~Hz 
(instead of 0.119~Hz reported in the literature). The phase-folded light curves of this period 
show a markedly double-peaked shape, which formally can be parameterized with a mixed 
von Mises distribution with significantly different components (more
  precisely, concentration and mixing proportion parameters, see Appendix~\ref{meth}). 

A similar picture we had seen with RBS 1223, another isolated neutron star. 
The pulsation discovery paper reported on a spin-period 5.16~sec 
\citep{2002A&A...381...98H}; however, a more detailed study showed that true period is 10.31~s 
\citep{2003A&A...403L..19H,2004AdSpR..33..638H,2005A&A...441..597S,2007Ap&SS.308..619S}.
Namely, this period with its double peaked profile was suggested by the presence of a second peak 
(with lower significance) in the power spectrum \citep{2004AdSpR..33..638H}. 
Further analysis of light curves in different energy bands and phase resolved spectroscopic study 
\citep{2007Ap&SS.308..619S,2011A&A...534A..74H}  provided additional confirmation on a plausibility
 of the model of RBS 1223 with emergent emission dominated by  two emitting
 areas with slightly different parameters. 
Yet another case, \cite{2011ApJ...743..183S} noted that without the radio period, 
they might have actually identified the pulse period of PSR J0726 as half of the true value.
\cite{2001A&A...365L.302C} performed spin pulse profile analysis of
\xin\ based on earlier \xmm\  observations, also folded photon arrival times
with twice of the detected period via Fourier periodogram which produced a
double-humped light curve with different count rates at the different peak; however, having no 
indication of subharmonics in the periodogram, they did not really consider 
twice of the period as a true one.
Another possible indication of the double-peaked light curve shape of \xin\ 
may be the non-sinusoidal light curve shape and the reported phase lag in different energy 
bands \citep{2006A&A...451L..17H,2009A&A...498..811H}. Indeed, slightly different
effective temperatures/sizes of two emitting areas at the magnetic poles, 
located not exactly podal and antipodal directions, may cause such kind of lag, if
photon arrival times were to fold into the phases with half of the value of
the genuine spin-period (see  Fig.~\ref{myrosenor1} in Appendix~\ref{meth}). 

The combined phase-resolved spectra of \xin\ can be simultaneously fitted by emergent radiation 
of a spectral model of an iron condensed surface with a partially ionized hydrogen 
atmosphere above it. 
Formally they can be fitted also by a blackbody spectrum with proton-cyclotron absorption  
Gaussian line and a peaked (typical for an electron scattering atmosphere) 
angular distribution of the emergent radiation. In both cases two emitting areas with 
slightly different characteristics are required (see Table~\ref{fit2}). 
 
It is worth to note, that the resulting fit parameters are very similar for different
spectral models, also confirmed by an MCMC approach with different input parameters. 
 
However, we believe that the emission properties due to the condensed surface
model with a partially ionized, optically thin hydrogen layer above it,
including vacuum polarization effects, is more physically motivated.
Moreover, semi-infinite atmospheres have rather fan-beamed emergent radiation 
\citep[see][and references therein]{2010A&A...522A.111S,2011A&A...534A..74H} and it seems 
impossible to combine a proton cyclotron line with a pencil-beamed emergent radiation.
 
Emission spectra based on realistic temperature and magnetic field  
distributions with strongly magnetized hydrogen atmospheres (or other light elements)  
are formally still an alternative\footnote{Our attempt to fit the combined,  
phase-averaged spectrum of \xin\ by  partially ionized, strongly  
magnetized hydrogen or mid-Z element plasma model  
\citep[{\it XSPEC nsmax},][]{2007MNRAS.377..905M,2008ApJS..178..102H},  
as well two spots or purely condensed iron surface models, failed. 
Noteworthy, an acceptable fit is obtained by {\it nsmax} model with  
additional, multiplicative Gaussian absorption line component (model {\it
  gabs}).},  but  it is unphysical because the absorption line is added by hand.
  
A pure proton-cyclotron absorption line scenario can be excluded owing to the 
width of the observed absorption spectral feature in the X-ray spectrum of
\xin. Magnetized semi-infinite atmospheres predict too low an equivalent width
of the proton cyclotron line in comparison with the observed one.
 
This result of spectral modelling (i.e the fitting with the condensed surface
model with partially ionized, optically thin hydrogen atmosphere above it,
including vacuum polarization effects)  suggests a true radius of \xin\ of
$13.3 \pm 0.5 \mathrm{km}$ for a standard neutron star of 1.4 solar mass,  
and indicates a stiff equation of state of \xin\ 
\citep[for similar results, see also][]{2007MNRAS.375..821H,2006ApJ...644.1090H,2006MNRAS.369.2036S,2011ApJ...742..122S}.

Our model fits provide the angles between our line of sight and both the rotation axis and 
the magnetic field axes, we obtain $42 \pm 3^{\circ}$ and $47 \pm 6^{\circ}$, respectively.
The direction of motion should be similar to the former angle in case of 
spin-orbit {\em almost} \citep[see, e.g.,][$\theta<15^{\circ}$]{2007ApJ...660.1357N} alignment.

In \cite{2011MNRAS.417..617T}, the direction of motion was determined to be
$62 \pm 9^{\circ}$, by tracing back the motion of \xin\ to its possible birth
association Trumpler$-$10. The direction of motion determined kinematically 
is thus roughly consistent with almost alignment of spin and orbit.

\section{Conclusions}\label{conc} 

We have carried out a complete X-ray timing re-analysis of the multi-epoch observational 
data sets of \xin\ for an identification of the {\em true} spin-period. 
The genuine spin-period of the isolated neutron star \xin\ is twice 
of the previously reported value in the literature (16.78~s instead of
8.39~s), with a markedly double-peaked light curve profile depending on time and energy.

The observed phase-resolved spectra of the INS \xin\   
are satisfactorily fitted with slightly different physical and geometrical  
characteristics of two emitting areas, for the model of a condensed iron surface, with partially ionized,  
optically thin hydrogen atmosphere above it, including vacuum polarization effects,  
as orthogonal rotator. The fit also suggests the absence of a strong toroidal  
magnetic field component. Moreover, the determined mass-radius ratio
($(M/M_{\sun})/(R/\mathrm{km})=0.105 \pm 0.002 $) suggests a very stiff equation of state of \xin. 
 
More work for detailed spectral model (also with quadrupole magnetic field configuration) 
computation will be certainly worth to do in the near future 
and its application to the phase-resolved spectra of other INSs and magnetars. In particular, 
including also high resolution spectra observed by \xmm\ and {\em Chandra} with possibly other absorption and emission features 
\citep{2009A&A...497L...9H,2010A&A...518A..24P,byur2016}. 
 
\begin{acknowledgements} 
      VH, VS, and RN acknowledge support by the German National Science Foundation
      ({\em Deut\-sche For\-schungs\-ge\-mein\-schaft, DFG\/}) through project 
      C7 of SFB/TR~7 ``Gravitationswellenastronomie''. 
\end{acknowledgements}

\begin{appendix} 
\section{True spin-period of \xin} \label{meth}

Most of the methods for the search and detection of a periodicity in X-ray astronomy 
are based on the analysis of the distribution of the phases ($\theta_i =
\nu\,(t_i-t_0) + \dot{\nu}\, (t_i-t_0)^2/2$), 
which is the result of folding photon arrival times ($t_i$ ($i=1,...,N$) 
with a trial frequency ($\nu$) and its derivative ($\dot{\nu}$) 
for the epoch of zero phase ($t_0$, e.g. \zm, H-test, the GL method, etc.).

This procedure transforms the time variable to the phases 
distributed on a circle. 

In order to disentangle between single- or multiple-peaked
profile of a light curve shape it has to be analyzed statistically. 
For this purpose, 
commonly used circular distributions are the von Mises distribution,
the wrapped normal distribution, the wrapped Cauchy distribution, etc.  
The von Mises distribution ($f(\theta;\mu,\kappa)=\frac{1}{I_{0}(\kappa)}e^{\kappa\cos(\theta-\mu)}$) 
has two parameters: the mean direction  and concentration  of the data distributed on a circle 
($I_{0}(\kappa)$ is the modified Bessel function of order 0).
 
\begin{figure*} 
\centering
    \vspace{-0.2cm}
    \vbox{
      \hbox{
        \includegraphics[width=8.6cm]{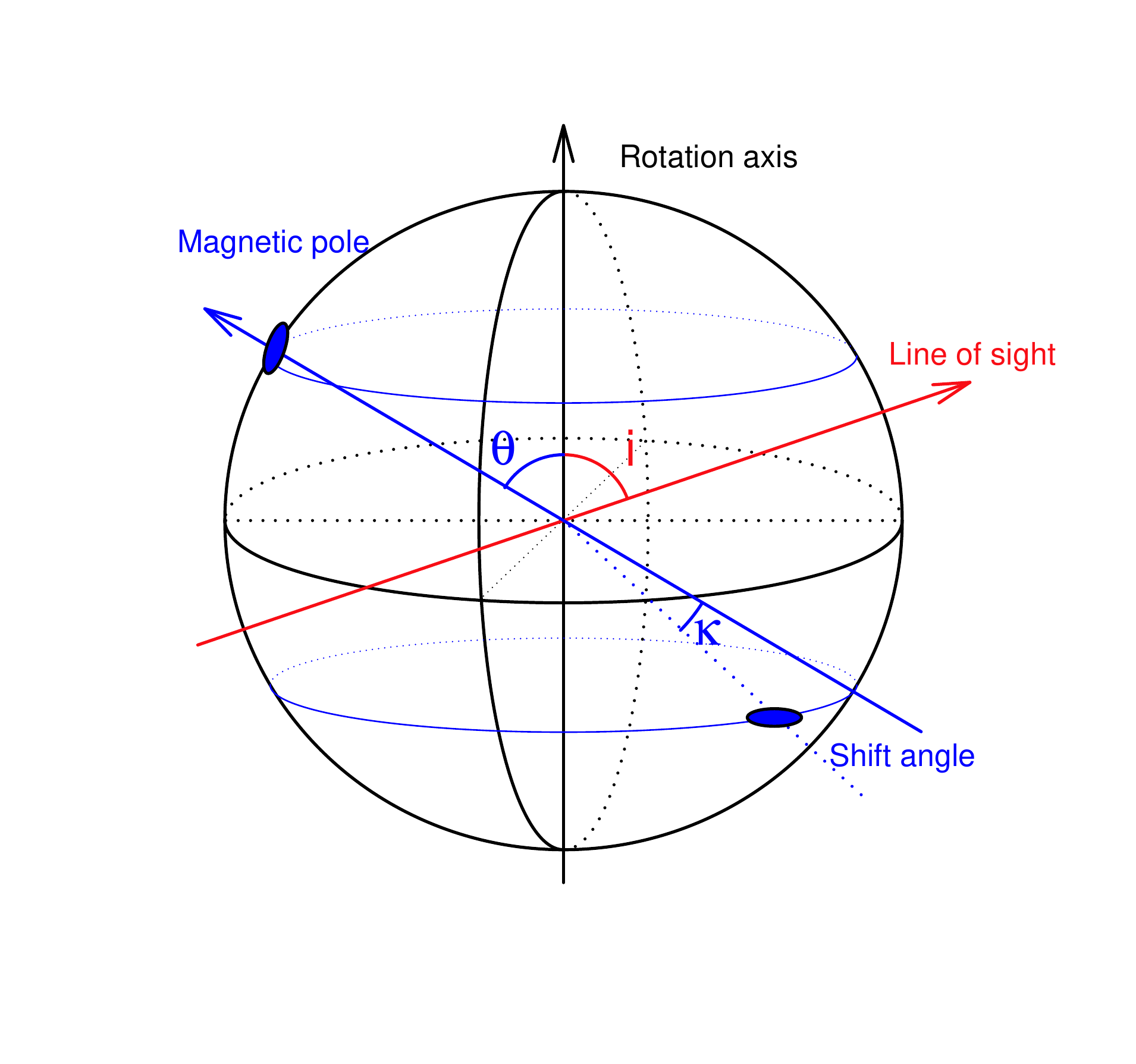}
        \includegraphics[width=8.6cm]{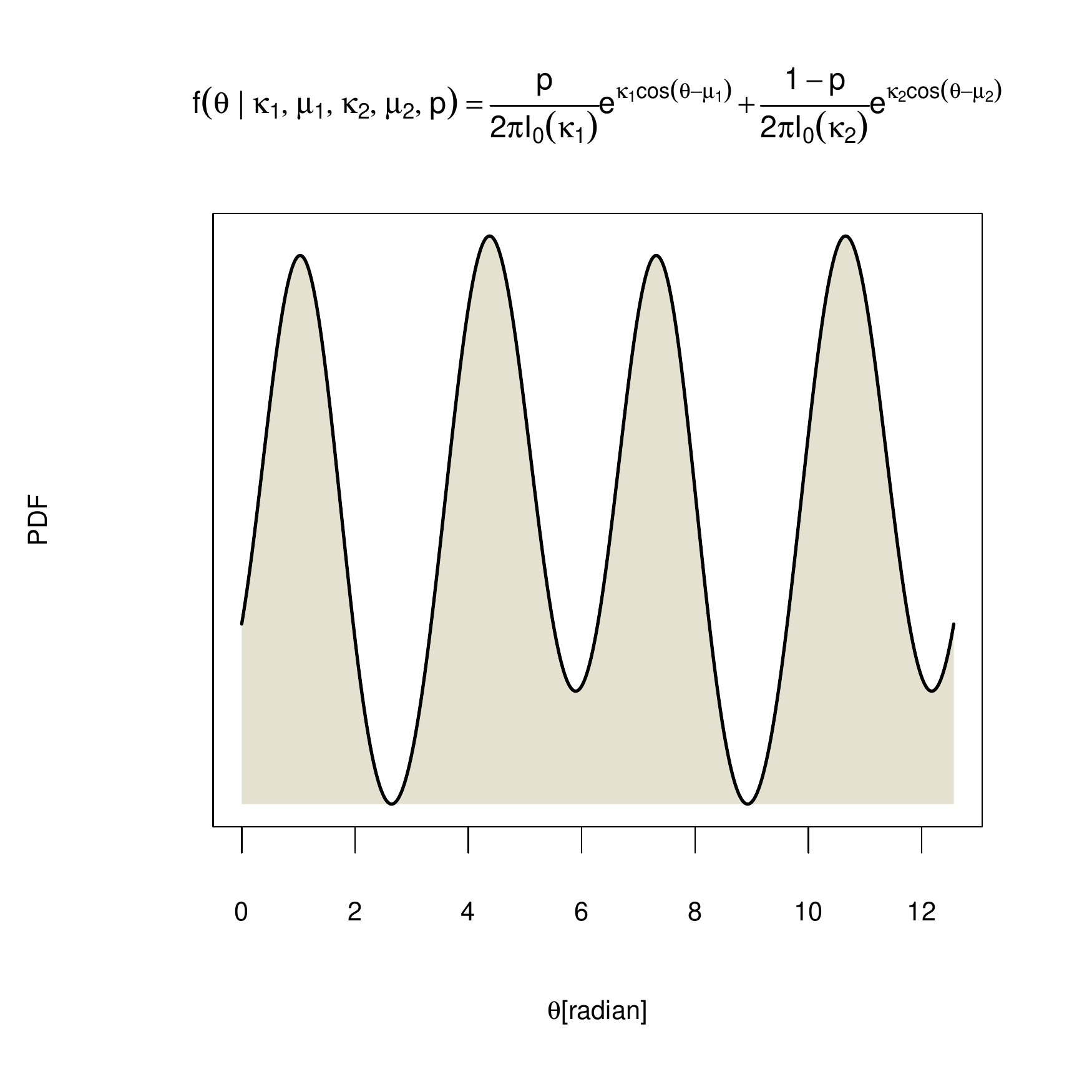}
      }
      \hbox{
        \includegraphics[width=8.6cm]{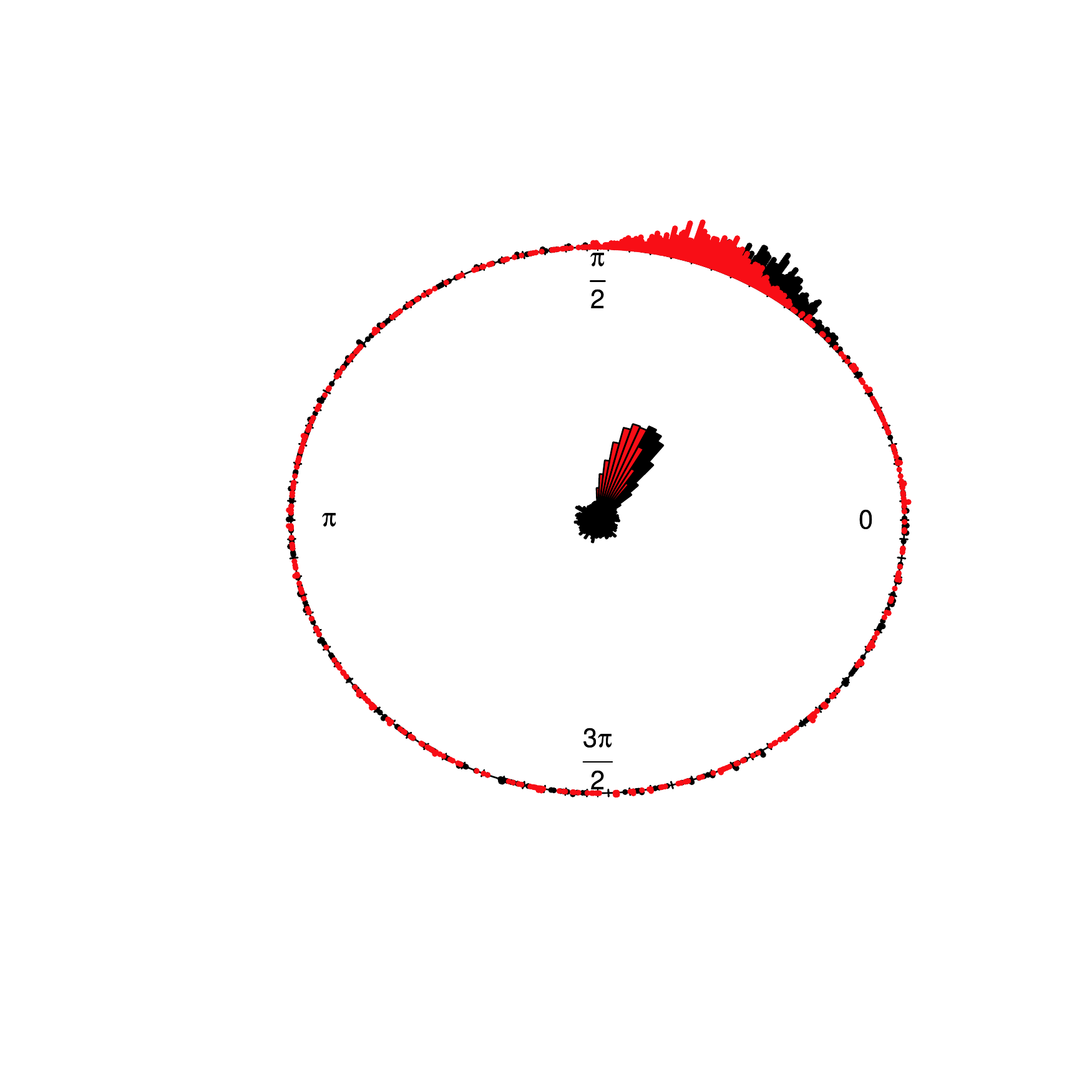}
        \includegraphics[width=8.6cm]{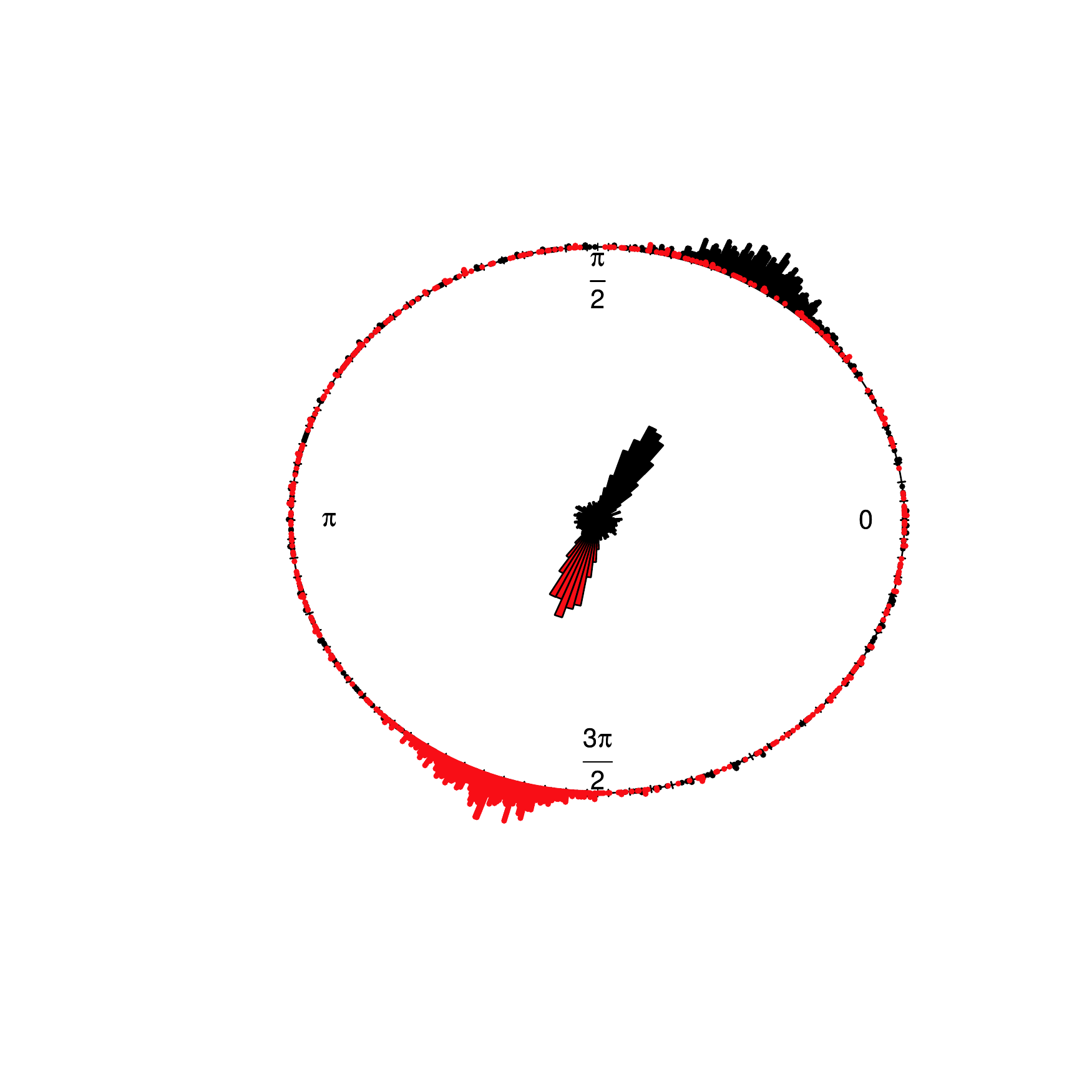}
      }
    }
    \vspace{-0.2cm}
\caption{{\em Upper panel:} An example of a simulated double-peaked light
  curve ($r(t)\propto e^{\sum_{j=1}^{2}[f_j\cos(\frac{2\pi t}{P}+\varphi_j)]} $ 
with pulsed fractions of each component $f_{1,2}$ of a periodic signal with
the period of $P$, right panel) and the geometry of the model (left panel).  
{\em Lower panel:} Distribution of phases (rose diagram-a circular histogram
plot of a mixed two von Mises distribution \citep{mardia2009directional,1997daa..conf..251C} 
$f(\theta;\mu_{1},\mu_{2}, \kappa_{1},\kappa_{2},p)=\frac{p}{I_{0}(\kappa_{1})}e^{\kappa_{1}\cos(\theta-\mu_{1})}+\frac{1-p}{I_{0}(\kappa_{2})}e^{\kappa_{2}\cos(\theta-\mu_{2})}$, 
 with the modes of $\mu_{1}=\pi/3,\mu_{2}=\pi+\pi/3+\pi/18$, 
concentrations $\kappa_1=1.2$, $\kappa_2=1.5$, and 
proportion $p=0.45$) for $\sim$ 2500 photon arrival times for an ideal
Poissonian process (right panel) for a periodic signal 
with a count rate similar to that observed from  \xin. 
The histogram plot displays directional data and the frequency of each class.
Distribution of the phases assuming that the period is equal to {\em half} of the true one (left panel). 
}
\label{myrosenor1} 
\end{figure*}

For demonstration purposes, in Figure~\ref{myrosenor1}, we show a sample 
for the light curve shape modeled by an exponentiated Fourier series, or generalized
von Mises distribution \citep{1997daa..conf..251C} for two component 
($r(t_{i})\propto e^{\sum_{j=1}^{2}{[-\kappa_j\cos(\frac{2\pi t_{i}}{P}+\varphi_j)]}}$, 
with phases $\varphi_{1,2}$ of photon arrival times ($t_{i}, i=1,N$) of a
periodic signal with the period of $P$) and  distribution of phases (rose
diagram-a circular histogram plot of a mixed two von Mises 
 distribution \citep{mardia2009directional,1997daa..conf..251C}, i.e. 
$f(\theta;\mu_{1},\mu_{2}, \kappa_{1},\kappa_{2},p)=\frac{p}{I_{0}(\kappa_{1})}e^{\kappa_{1}\cos(\theta-\mu_{1})}+\frac{1-p}{I_{0}(\kappa_{2})}e^{\kappa_{2}\cos(\theta-\mu_{2})}$, 
where $\mu_{1},\mu_{2}$ are the mean directions of the phases, $\kappa_{1},\kappa_{2}$ are concentrations, 
and $p$ is a mixing proportion. These parameters are determining the locations, peakedness, relative contribution of each component
of a double-profile light curve, i.e. are expressing the effective temperatures and sizes of the 
dominating areas resulting the emergent radiation.

\begin{table*}[ht]
\centering
\caption{Results of the fitting of spin-phases with a two-component mixed von Mises distribution.}  
\label{fitmvM} 
\begin{tabular}{lcccccc}
\hline\noalign{\smallskip}
 Groups & Energy range & N & $\mu_1$/$\mu_2$ & $\kappa_1$/$\kappa_2$ & p \\ 
        & [kev]        & [counts] &      &                 &  \\ 
  \hline\noalign{\smallskip}
GRI     & 0.16$-$0.5 & 294516 & 0.507$\pm$0.003 & 3.124$\pm$0.367 & 0.349 \\ 
"       & 0.50$-$0.6 & 30943  & 0.508$\pm$0.007 & 2.236$\pm$0.880 & 0.415 \\ 
"       & 0.60$-$0.7 & 16125  & 0.479$\pm$0.011 & 2.032$\pm$0.843 & 0.408 \\ 
"       & 0.70$-$1.2 & 14338  & 0.495$\pm$0.013 & 3.286$\pm$1.714 & 0.311 \\ 
\hline\noalign{\smallskip}
GRII    & 0.16$-$0.5 & 359528 & 0.494$\pm$0.002 & 3.101$\pm$0.282 & 0.352 \\ 
"       & 0.50$-$0.6 & 51604  & 0.496$\pm$0.005 & 2.622$\pm$0.732 & 0.392 \\ 
"       & 0.60$-$0.7 & 30956  & 0.514$\pm$0.008 & 1.333$\pm$0.108 & 0.511 \\ 
"       & 0.70$-$1.2 & 29568  & 0.502$\pm$0.006 & 2.159$\pm$0.483 & 0.402 \\ 
\hline\noalign{\smallskip}
GRIII   & 0.16$-$0.5 & 234977 & 0.502$\pm$0.003 & 3.138$\pm$0.271 & 0.347 \\ 
"       & 0.50$-$0.6 & 32629  & 0.500$\pm$0.011 & 2.397$\pm$1.151 & 0.381 \\ 
"       & 0.60$-$0.7 & 19045  & 0.520$\pm$0.009 & 1.363$\pm$0.294 & 0.512 \\ 
"       & 0.70$-$1.2 & 17789  & 0.524$\pm$0.012 & 1.454$\pm$0.200 & 0.514 \\ 
\hline\noalign{\smallskip}
GRIV    & 0.16$-$0.5 & 210538 & 0.508$\pm$0.003 & 2.796$\pm$0.330 & 0.360 \\ 
"       & 0.50$-$0.6 & 27712  & 0.498$\pm$0.008 & 1.911$\pm$0.728 & 0.505 \\ 
"       & 0.60$-$0.7 & 15814  & 0.508$\pm$0.009 & 1.467$\pm$0.476 & 0.507 \\ 
"       & 0.70$-$1.2 & 14840  & 0.508$\pm$0.007 & 1.346$\pm$0.368 & 0.506 \\ 
\hline\noalign{\smallskip}
GRV     & 0.16$-$0.5 & 352642 & 0.497$\pm$0.002 & 3.130$\pm$0.260 & 0.342 \\ 
"       & 0.50$-$0.6 & 43187  & 0.505$\pm$0.005 & 1.611$\pm$0.461 & 0.507 \\ 
"       & 0.60$-$0.7 & 23725  & 0.498$\pm$0.006 & 1.709$\pm$0.451 & 0.484 \\ 
"       & 0.70$-$1.2 & 21436  & 0.508$\pm$0.011 & 1.439$\pm$0.636 & 0.507 \\ 
   \hline\noalign{\smallskip} 
Student's t$-$test & & &H0: $\mu_1/\mu_2 \equiv 0.5$ & H0: $\kappa_1/\kappa_2 \equiv 1.0$ & H0: $p \equiv 0.5 $ \\
\cline{4-6}\noalign{\smallskip}
P$-$value & & & 0.14 & $\mathrm{9.8\times 10^{-9}}$ & 0.005 \\
\hline\noalign{\smallskip}
Wilcoxon test & & & & & \\
P$-$value & & & 0.08 & $\mathrm{1.9\times 10^{-6}}$ & 0.008 \\
\hline
\end{tabular}
\end{table*}

To fit the double-peaked light curve profile we coded an
Expectation-Maximization algorithm \citep[see, e.g.][an iterative approach to
  find the maximum of the likelihood, applied to a mixture models]{BIOM:BIOM682,CJS:CJS5550360110}
used to model rotational phases of all registered photons (column $N$ in the
Table~\ref{fitmvM}) as a mixture of two component von Mises distribution. 
The results of the fit are presented in the Table~\ref{fitmvM}, 
where for the estimation of uncertainties, an additional, bootstrap algorithm
was implemented and applied. 
We generate a bootstrap sample of 30  datasets for each data group, 
by randomly sampling from original spin-phases without replacement. Each of this bootstrap datasets 
fitted with a mixture of two component von Mises distribution. The standard
  deviations of the derived parameters accepted as an uncertainty measure.

 An application of so-called one sample t-test to the difference of the
  locations of the peaks (column $\mu_1/\mu_2$, Table~\ref{fitmvM}, or
  two-sample paired t-test to the $\mu1$ and $\mu2$) of the phase-folded light
  curve (double-peaked profile) is consistent to be about $\pi$, regardless of the energy band.  
Hence, the null hypothesis on the equality of the ratio of the mean directions
of 0.5 (i.e. podal and antipodal directions) is accepted. 
In contrary, for the parameters of concentration and proportion
($\kappa_1/\kappa_2$ and $p$) the null hypothesis is rejected ($\kappa_1$
and $\kappa_2$ are nonidentical populations and $p$ mean value is not equal to
0.5, see, Table~\ref{fitmvM}).
We arrive to the similar conclusions by an application of a non-parametric 
Wilcoxon test.

\section{Phase-folded light curves}
\label{Rapp}

\begin{figure*}
\centering
    \vspace{-0.2cm}
    \vbox{
      \hbox{
        \includegraphics[bb=5 108 527 649,width=5.4cm]{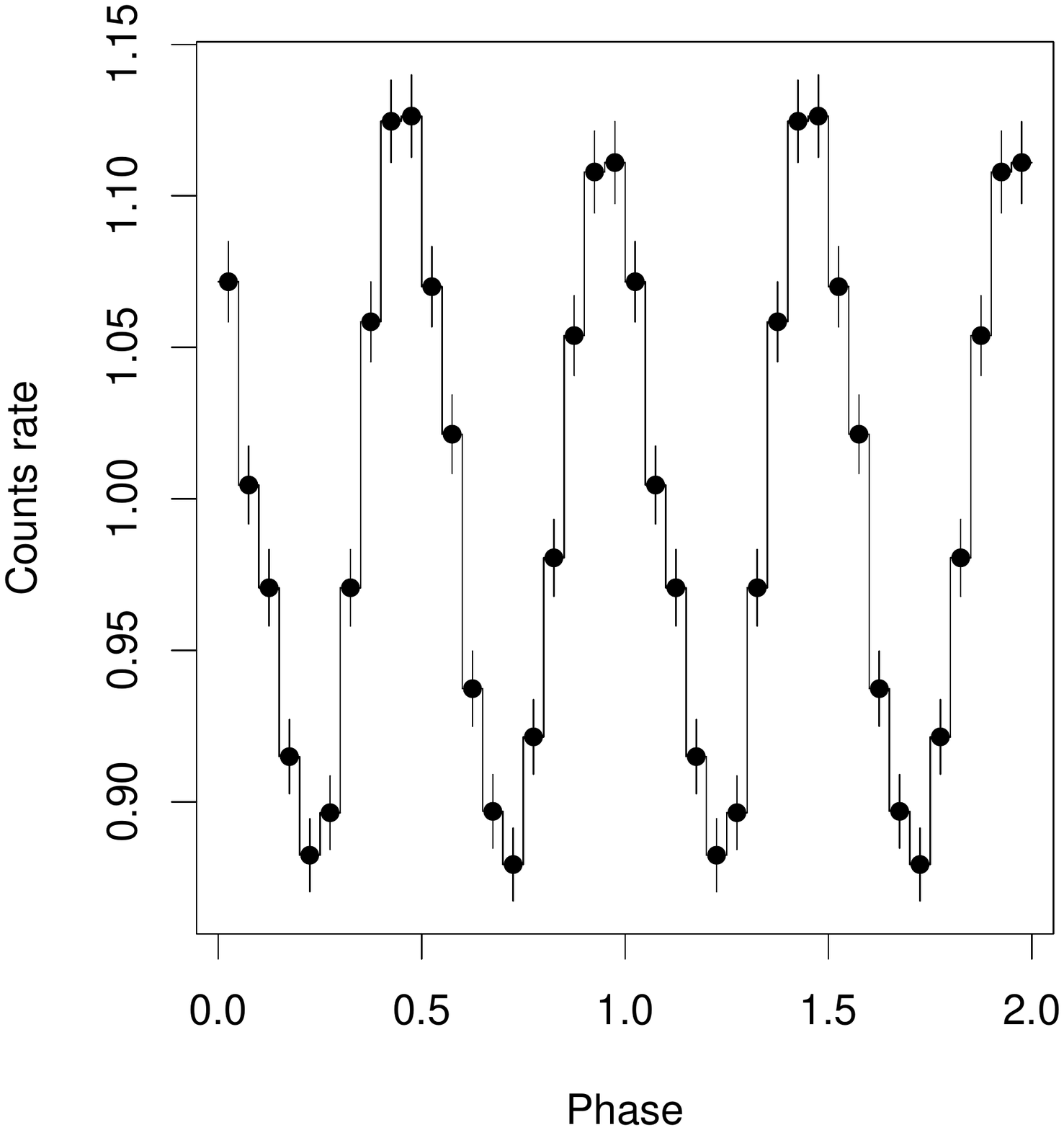}
        \includegraphics[bb=5 108 527 649,width=5.4cm]{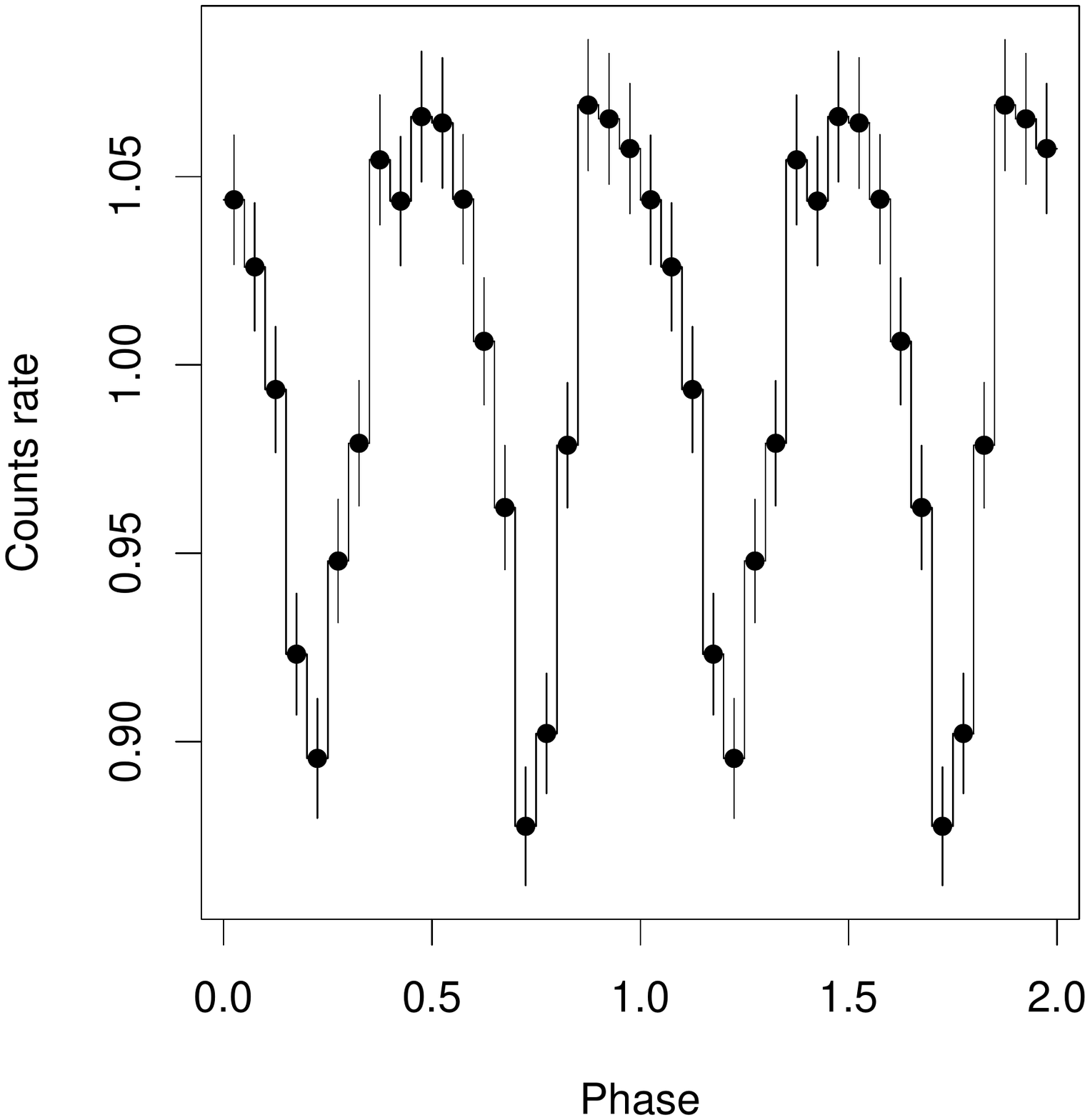}
        \includegraphics[bb=5 108 527 649,width=5.4cm]{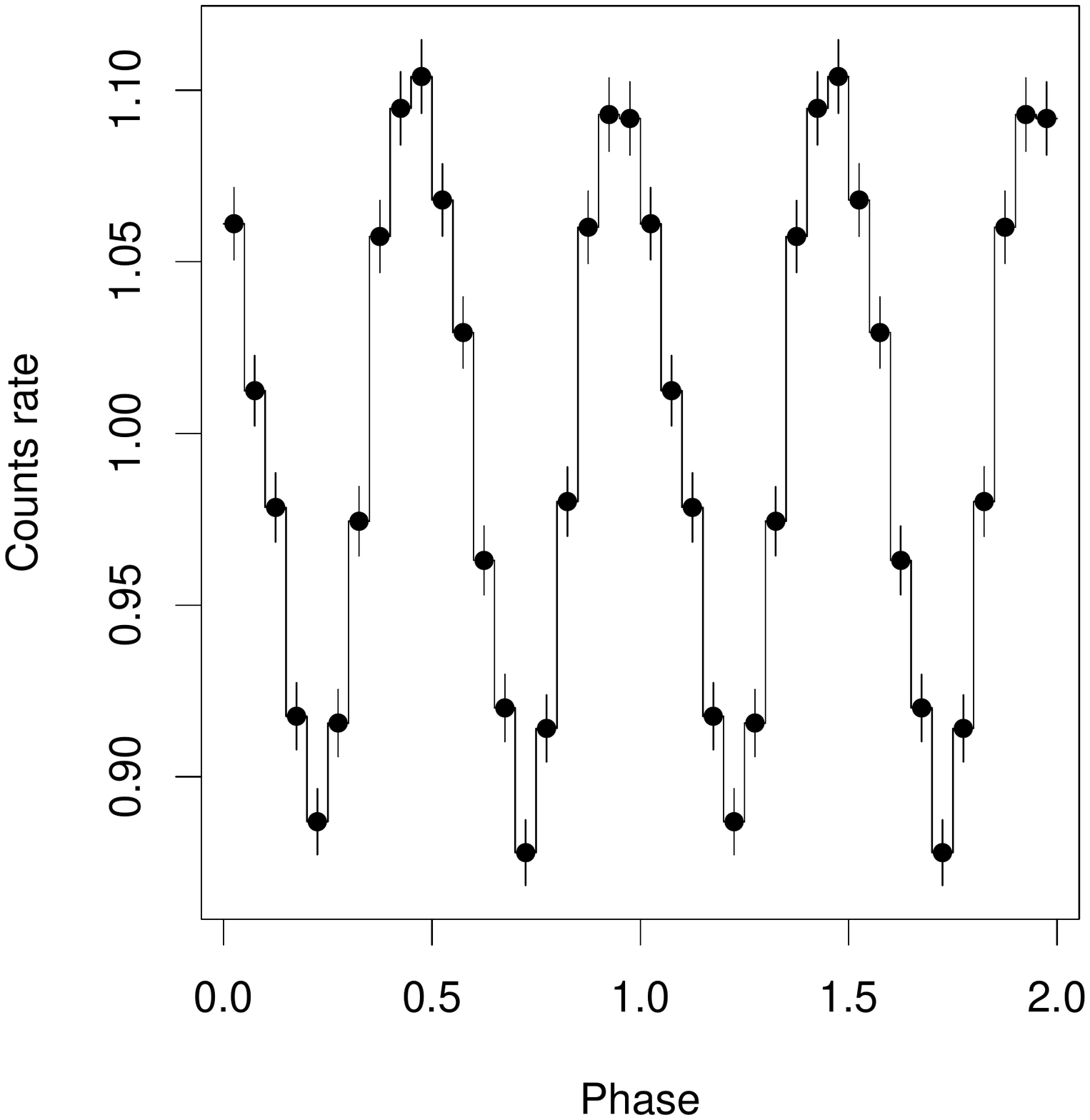}
      }
      \hbox{
        \includegraphics[bb=5 108 527 649,width=5.4cm]{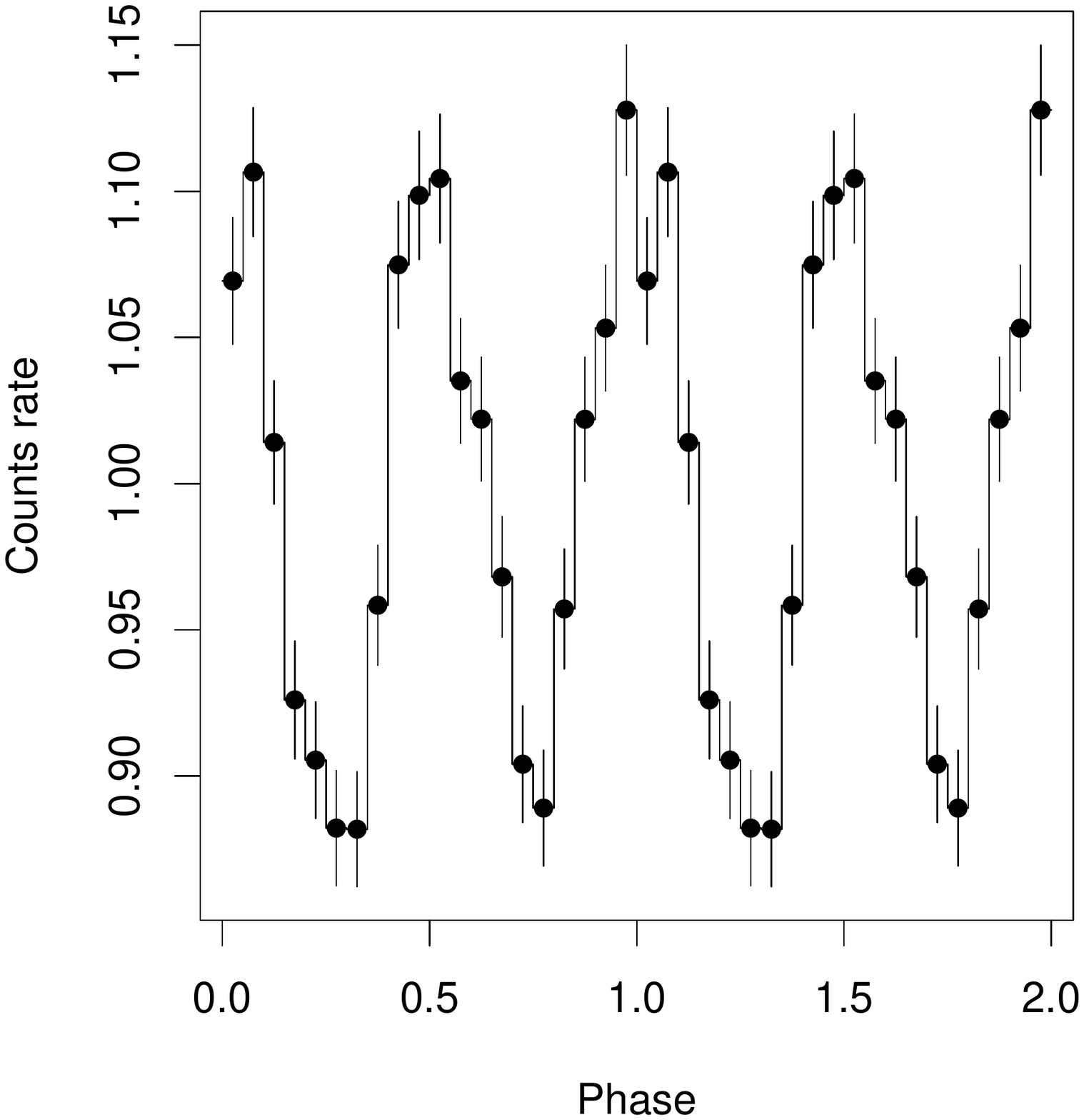}
        \includegraphics[bb=5 108 527 649,width=5.4cm]{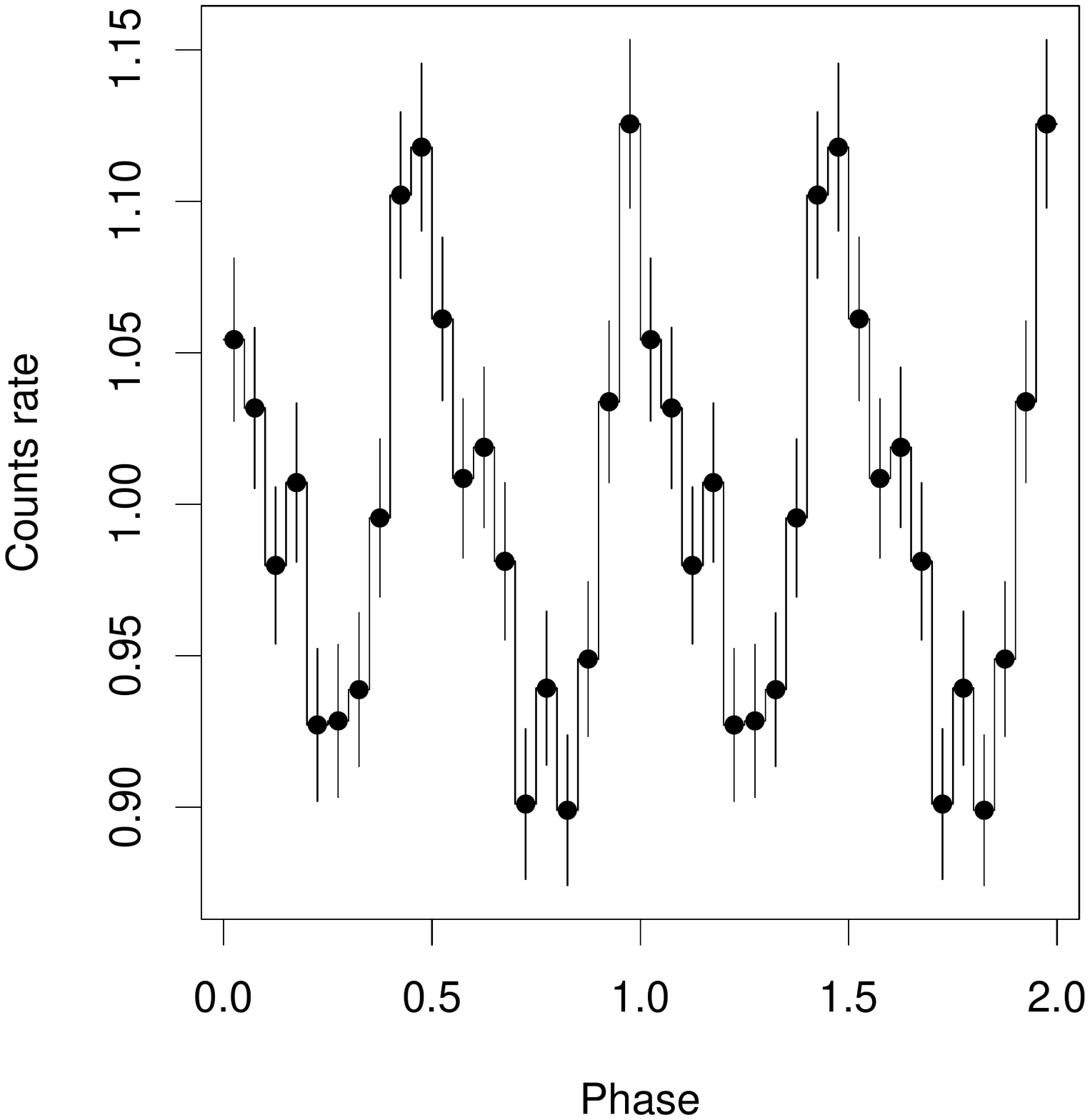}
        \includegraphics[bb=5 108 527 649,width=5.4cm]{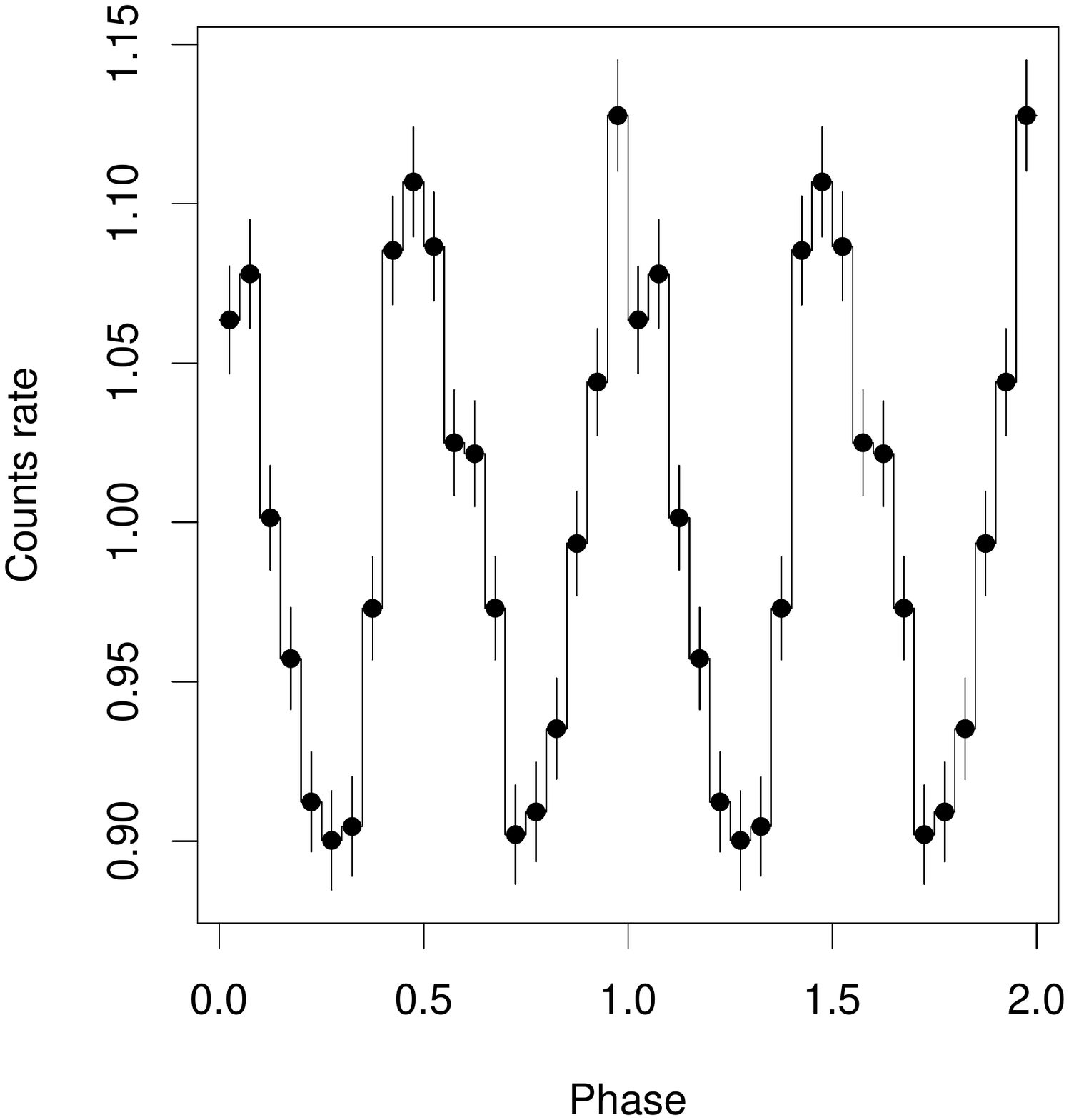}

      }
       \hbox{
        \includegraphics[bb=5 108 527 649,width=5.4cm]{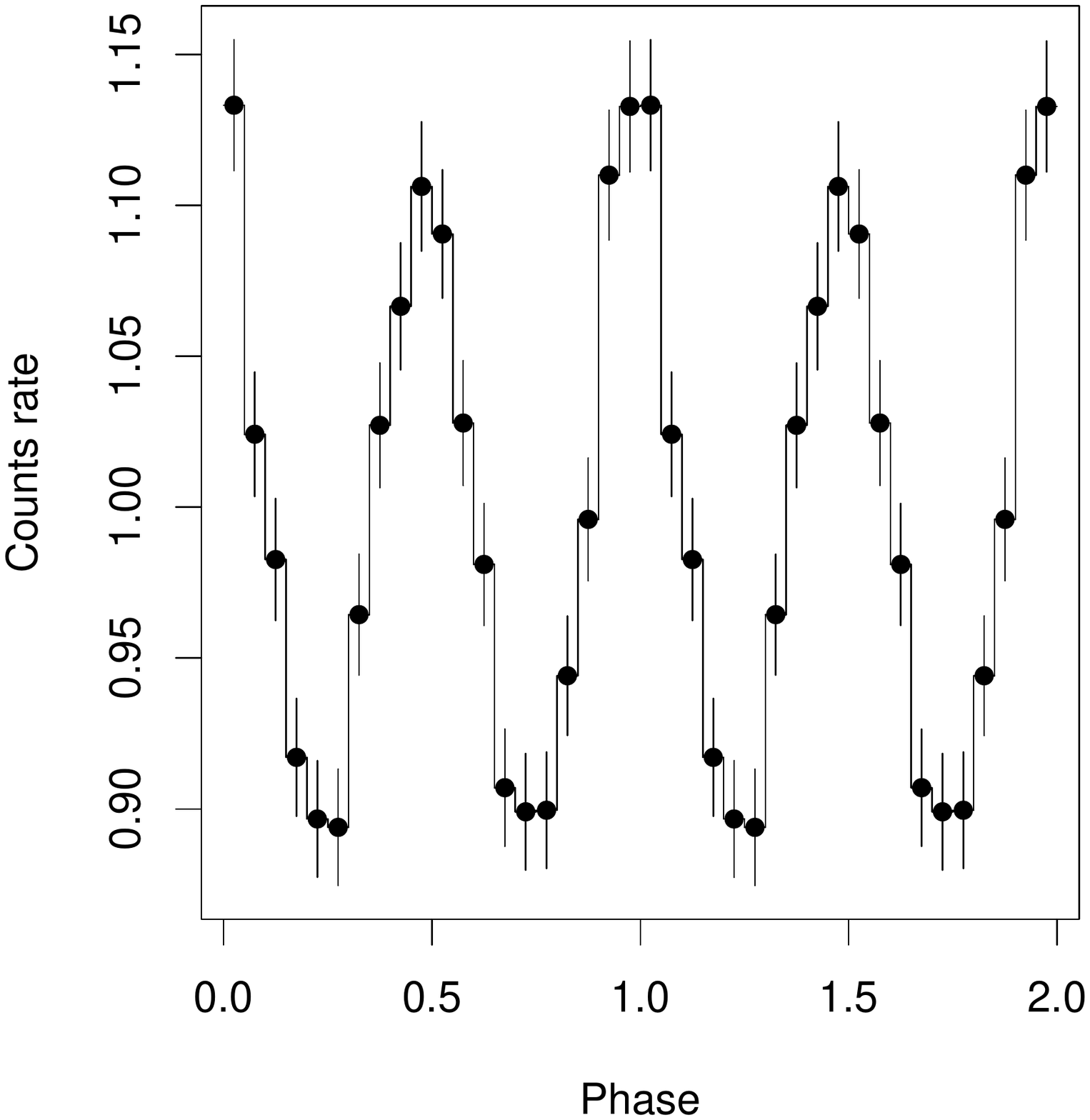}
        \includegraphics[bb=5 108 527 649,width=5.4cm]{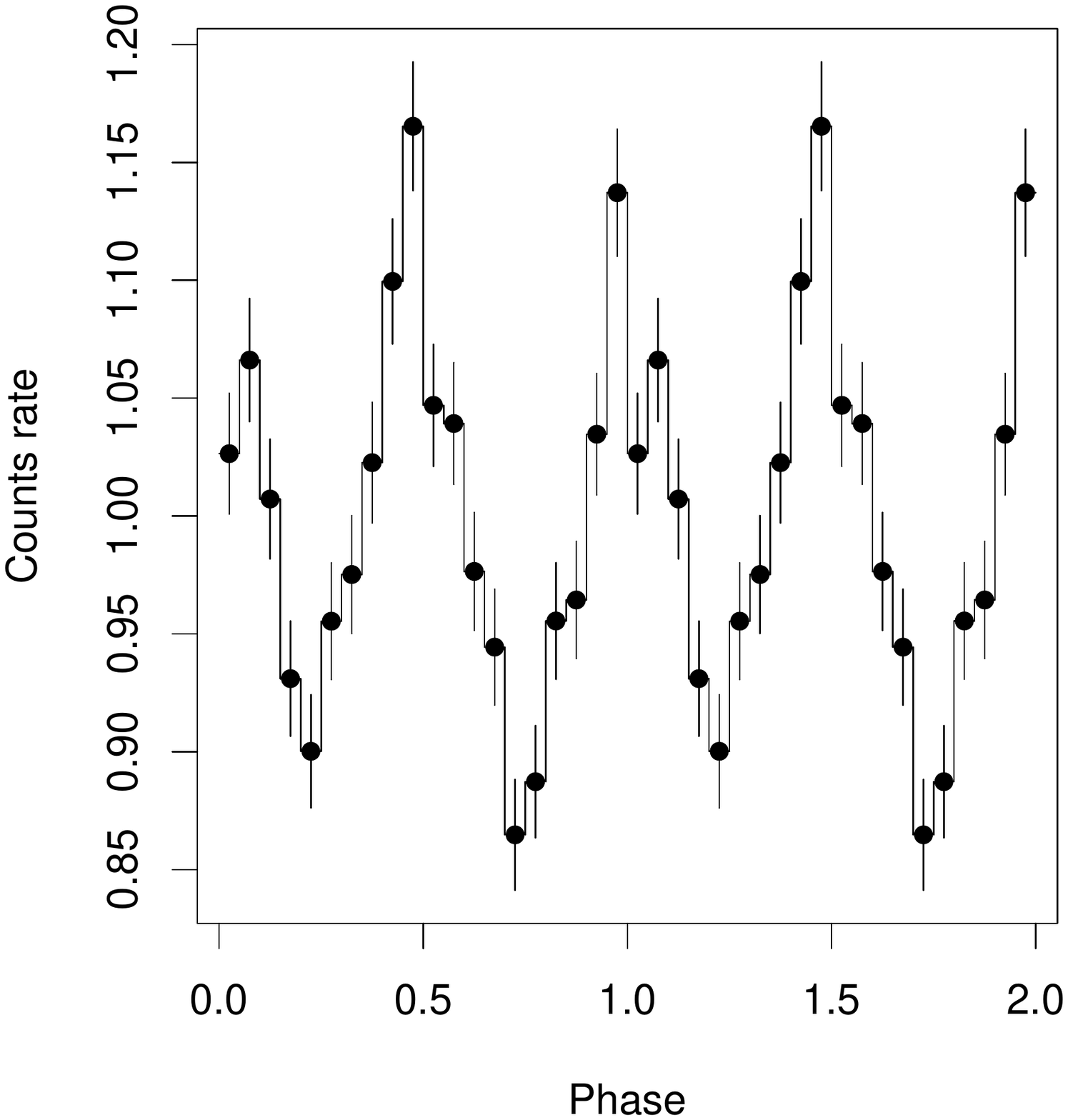}
        \includegraphics[bb=5 108 527 649,width=5.4cm]{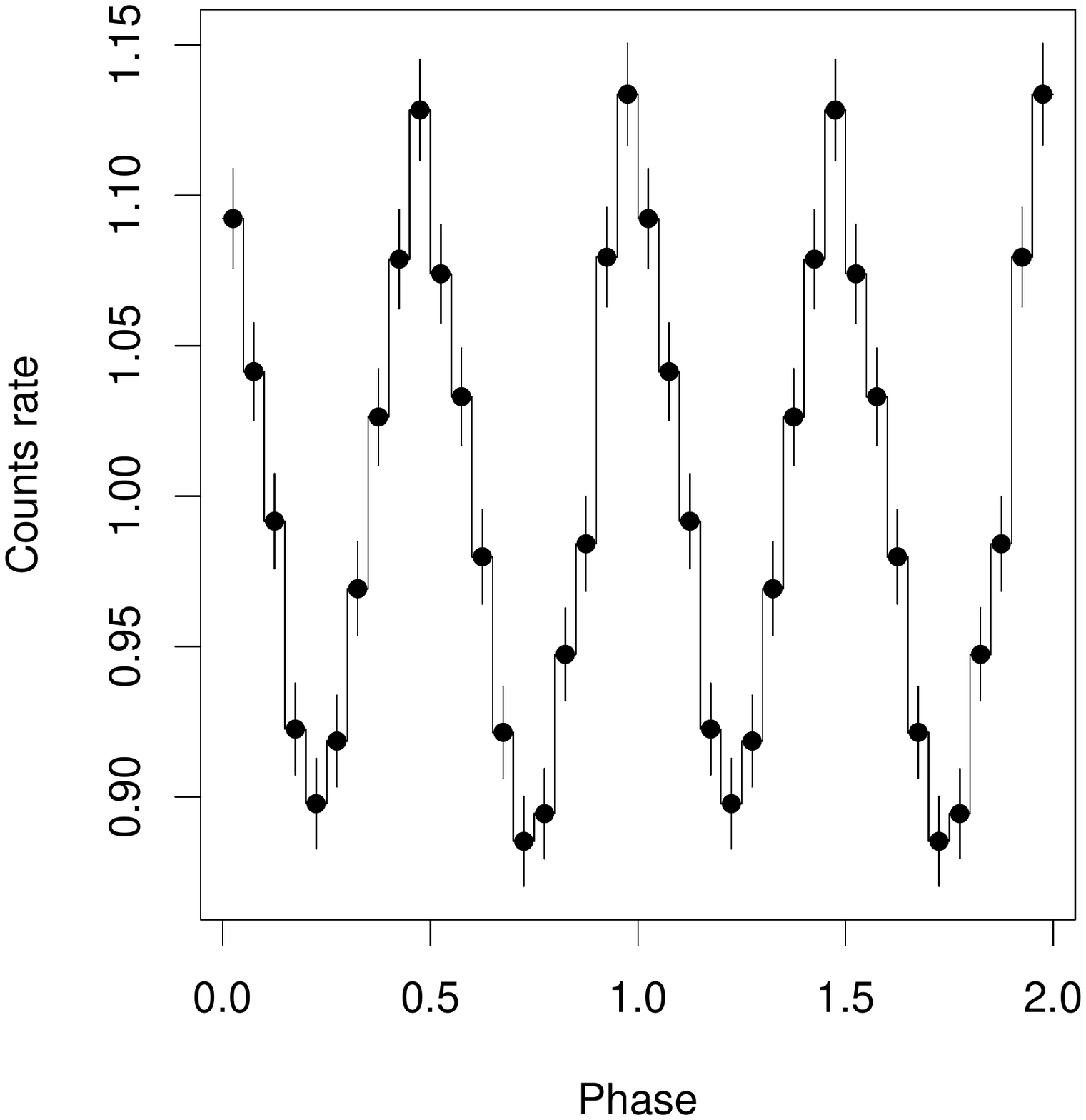}

      }
      \hbox{
        \includegraphics[bb=5 108 527 649,width=5.4cm]{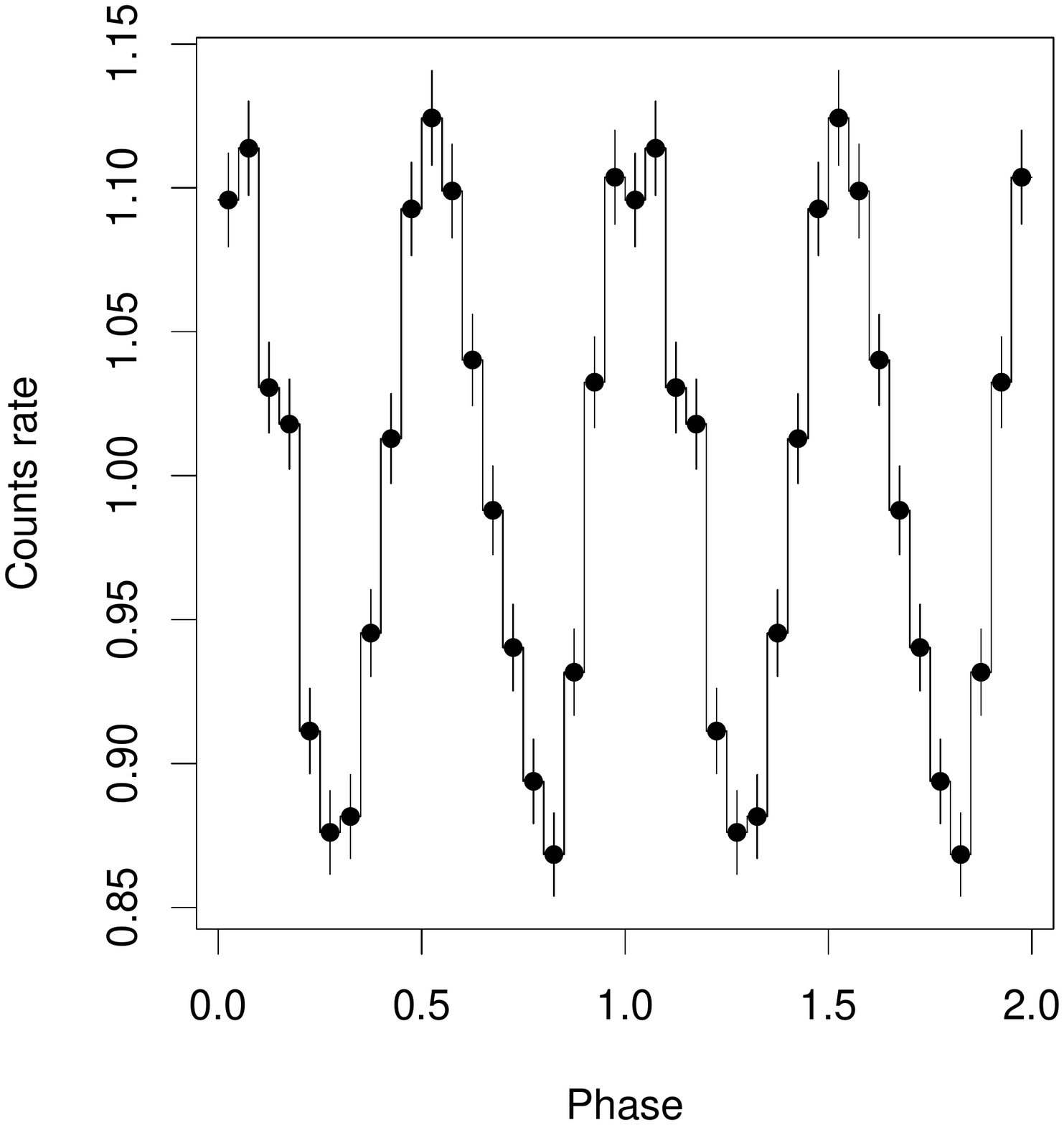}
        \includegraphics[bb=5 108 527 649,width=5.4cm]{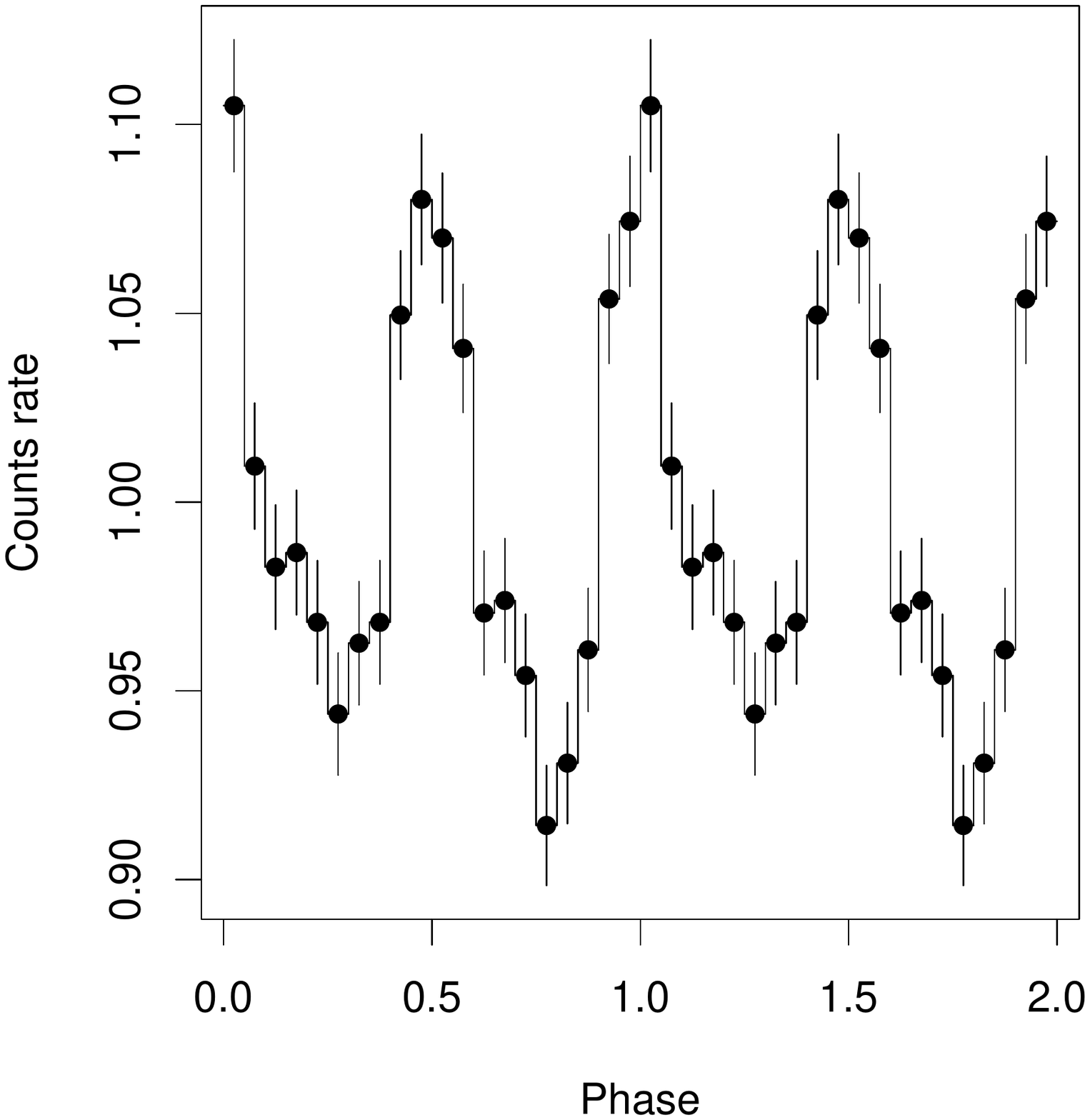}
        \includegraphics[bb=5 108 527 649,width=5.4cm]{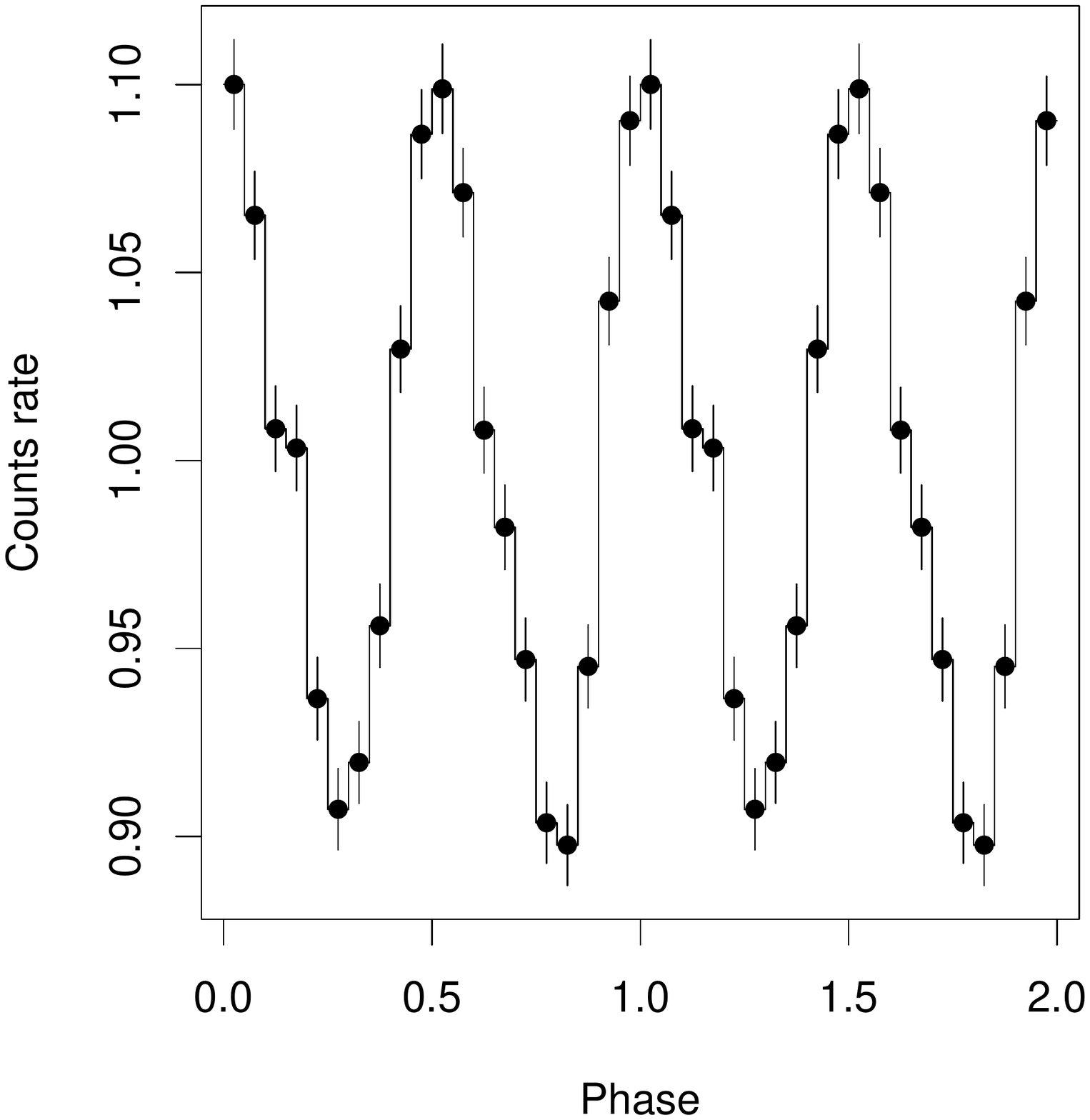}

      }
   }
    \vspace{-0.2cm}
    \caption{Phase-folded light curves in different energy bands 
(0.16-0.38\,keV, 0.38-1.6\,keV, and 0.16-1.6\,keV {\em left, middle, and right} panels, accordingly)
of \xin\ for pointed \xmm\  EPIC-pn observations 0124100101, 0156960201, 0156960401, and 0164560501.
All light curves normalized to the mean value of the corresponding energy band.
 Note the clear differences between the two peaks,indicating two emitting areas with different physical 
characteristics (for details, see text). \label{resapp1}}
\end{figure*}

Here (Fig.~\ref{resapp1}), we present rotational phase-folded light curves for all pointed observations
of \xin\ performed by \xmm\ EPIC-pn in different energy bands 
(0.16-0.38\,keV, 0.38-1.6\,keV, and 0.16-1.6\,keV) with spin-frequencies (Table~\ref{ax2}), 
 zero-phases (derived by an application of the GL method), and using 
  20 phase bins. All double-peaked profile light curves are  normalized to the mean count rate
of the pointed observations in the corresponding energy band.

\begin{figure*}
\centering
\vspace{-0.2cm}
\vbox{
  \hbox{
    \includegraphics[bb=5 108 527 649,width=5.4cm]{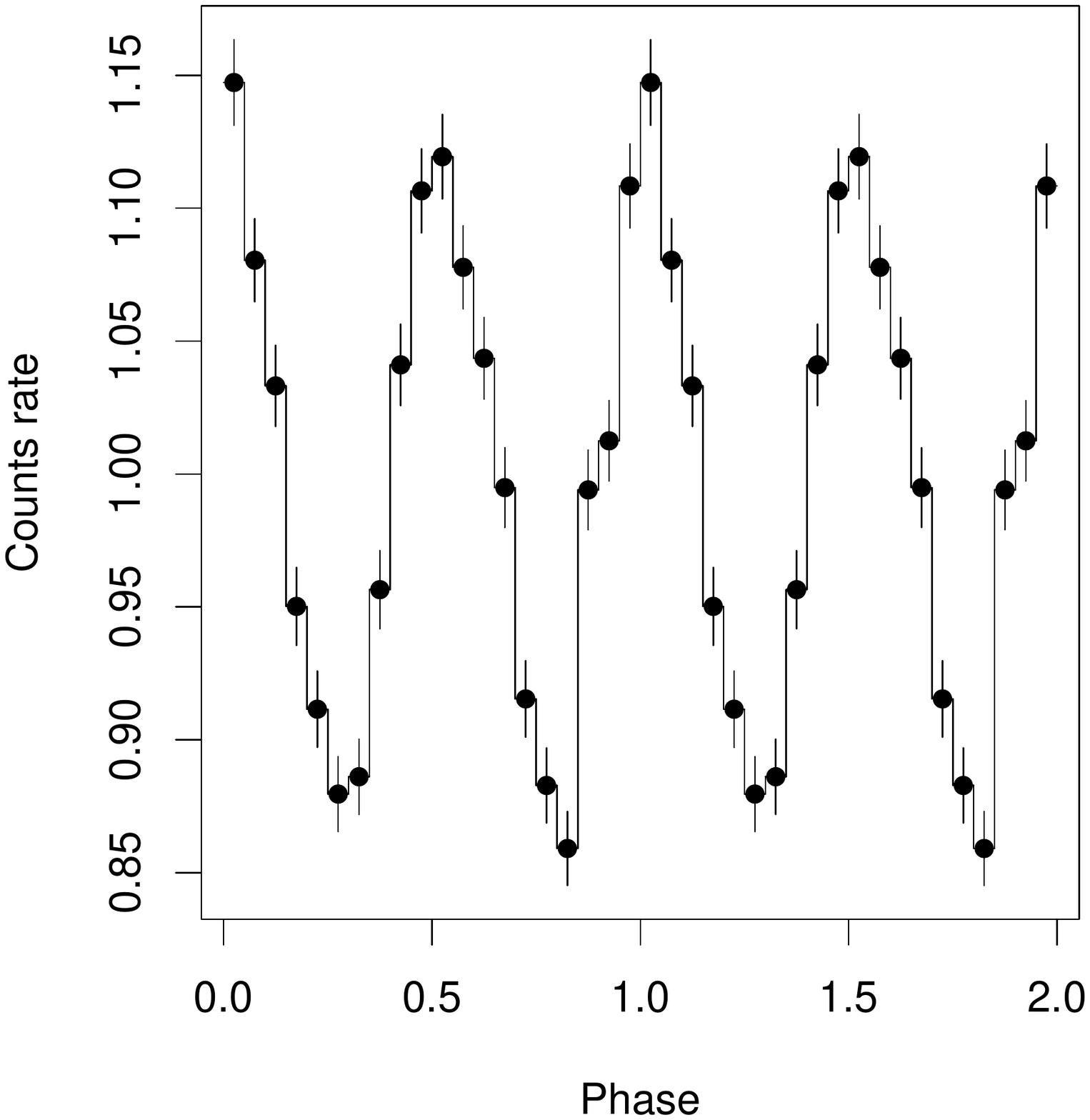}
    \includegraphics[bb=5 108 527 649,width=5.4cm]{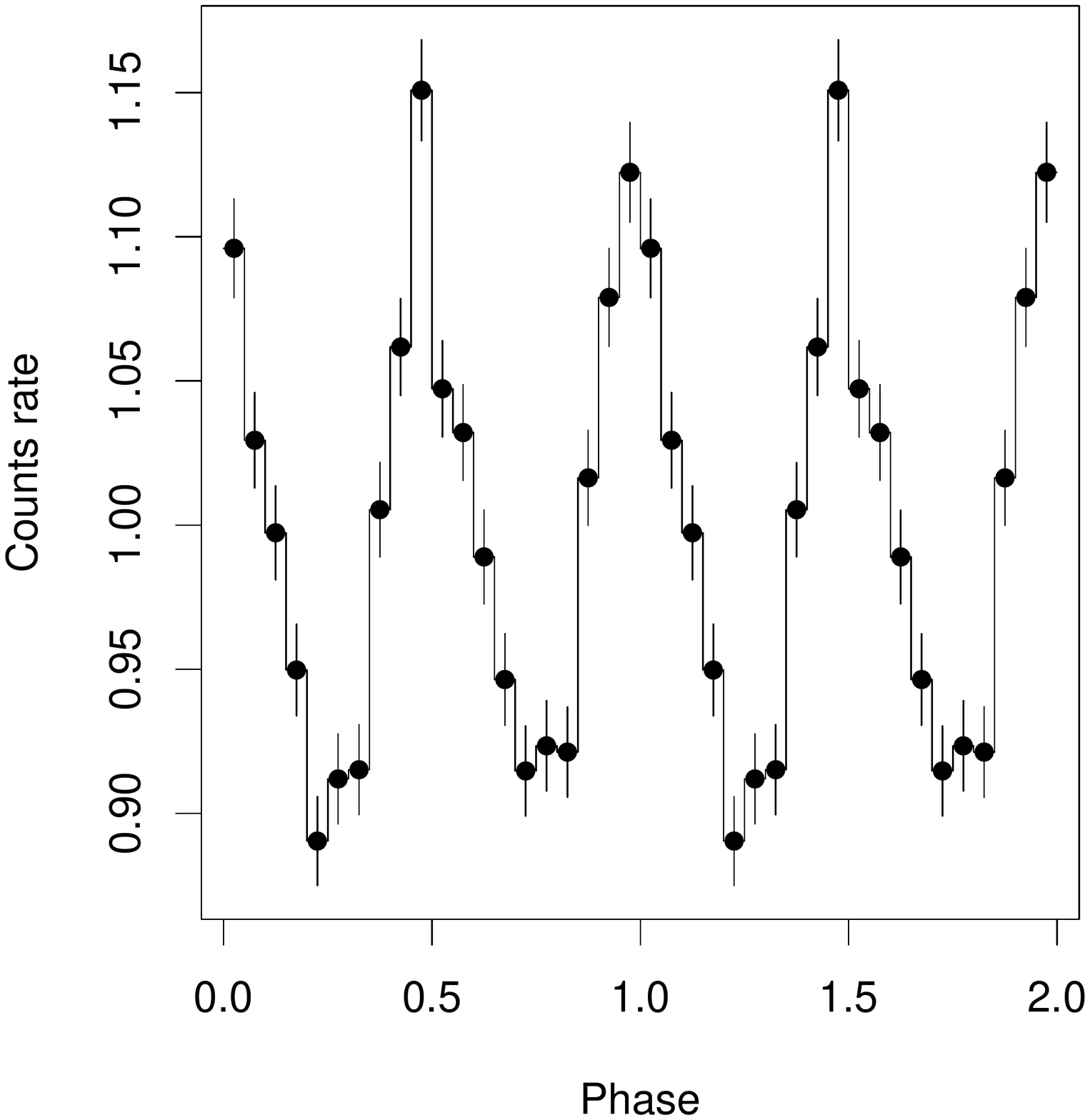}
    \includegraphics[bb=5 108 527 649,width=5.4cm]{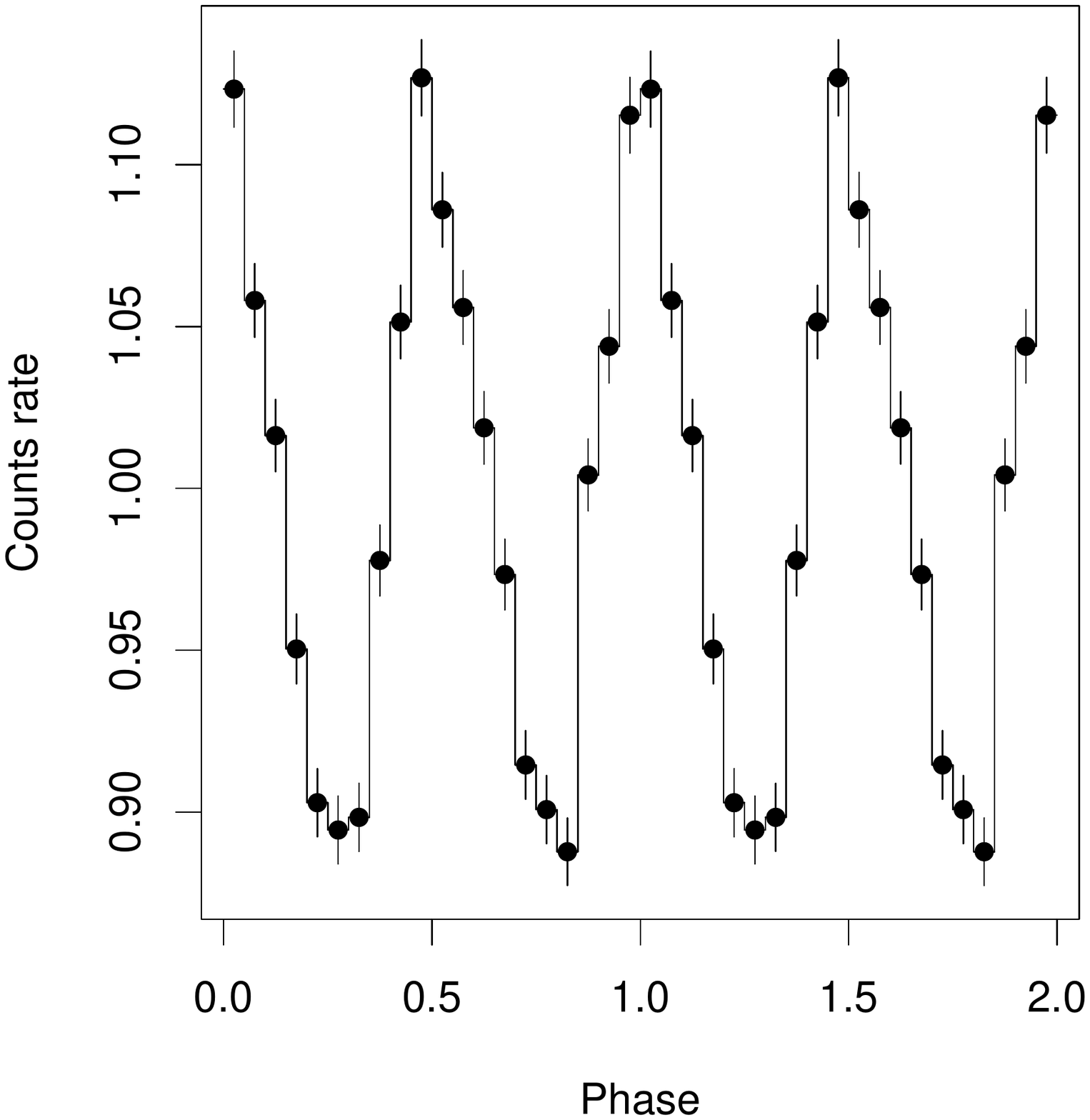}
  }
  \hbox{
    \includegraphics[bb=5 108 527 649,width=5.4cm]{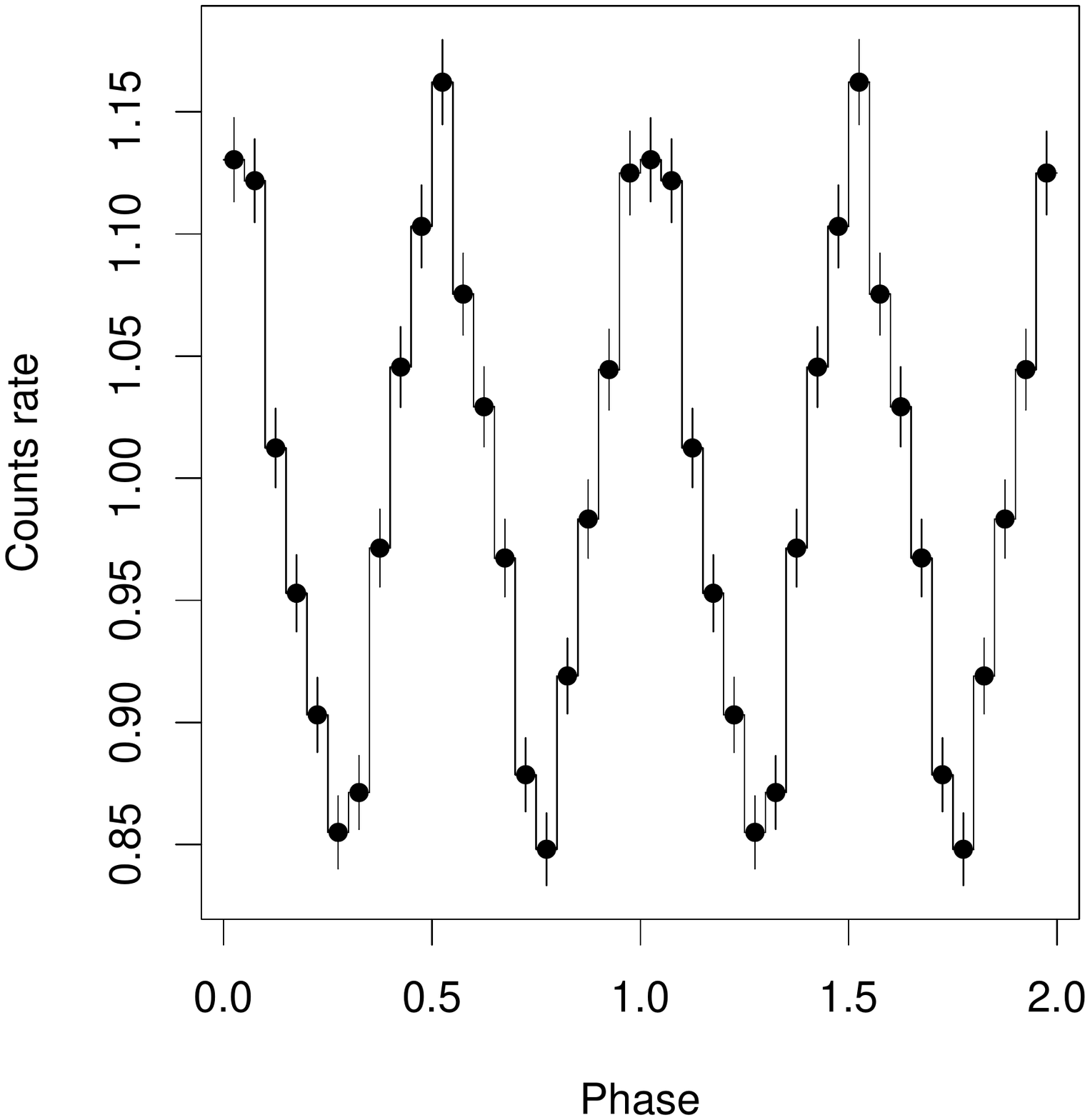}
    \includegraphics[bb=5 108 527 649,width=5.4cm]{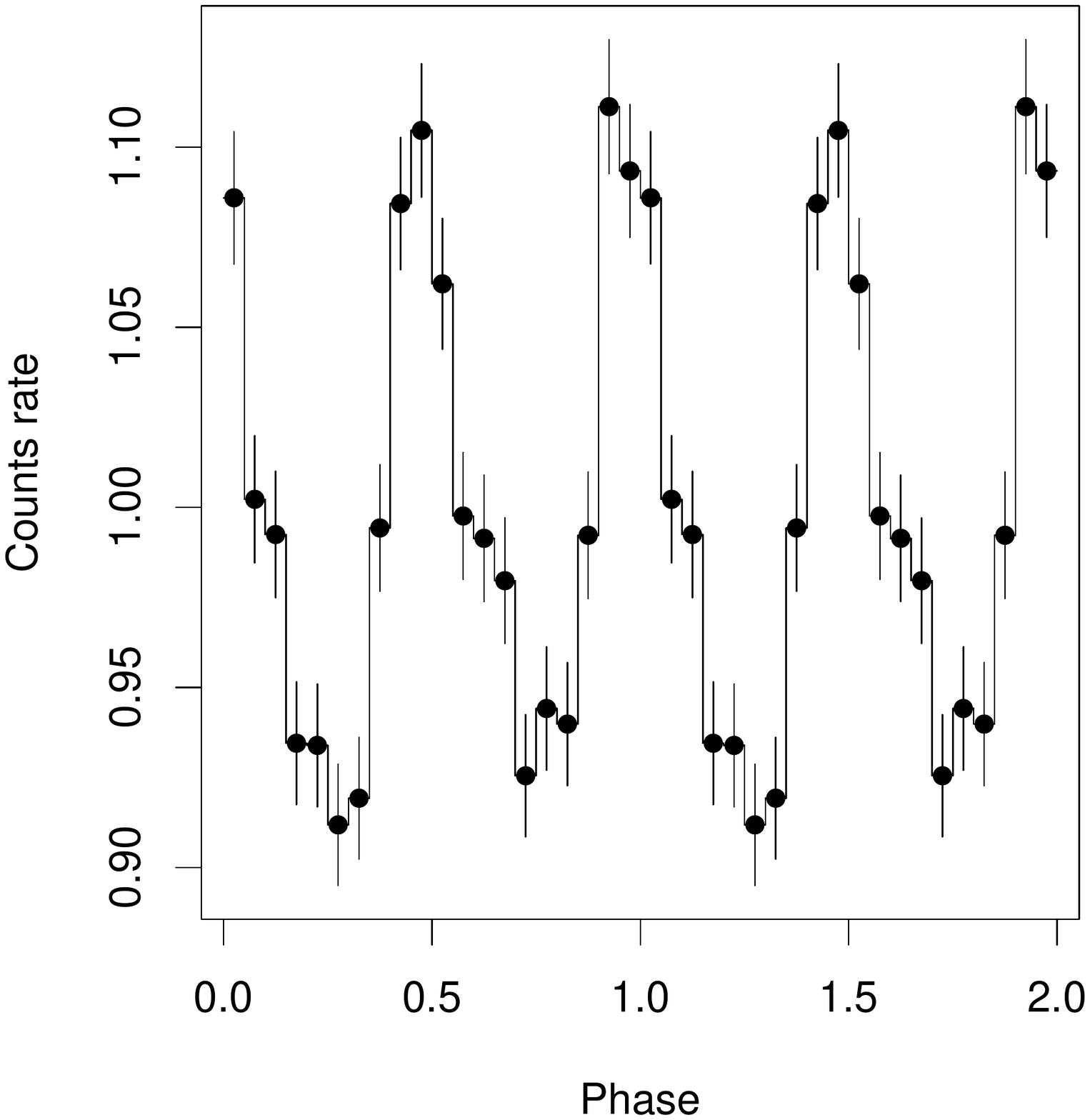}
    \includegraphics[bb=5 108 527 649,width=5.4cm]{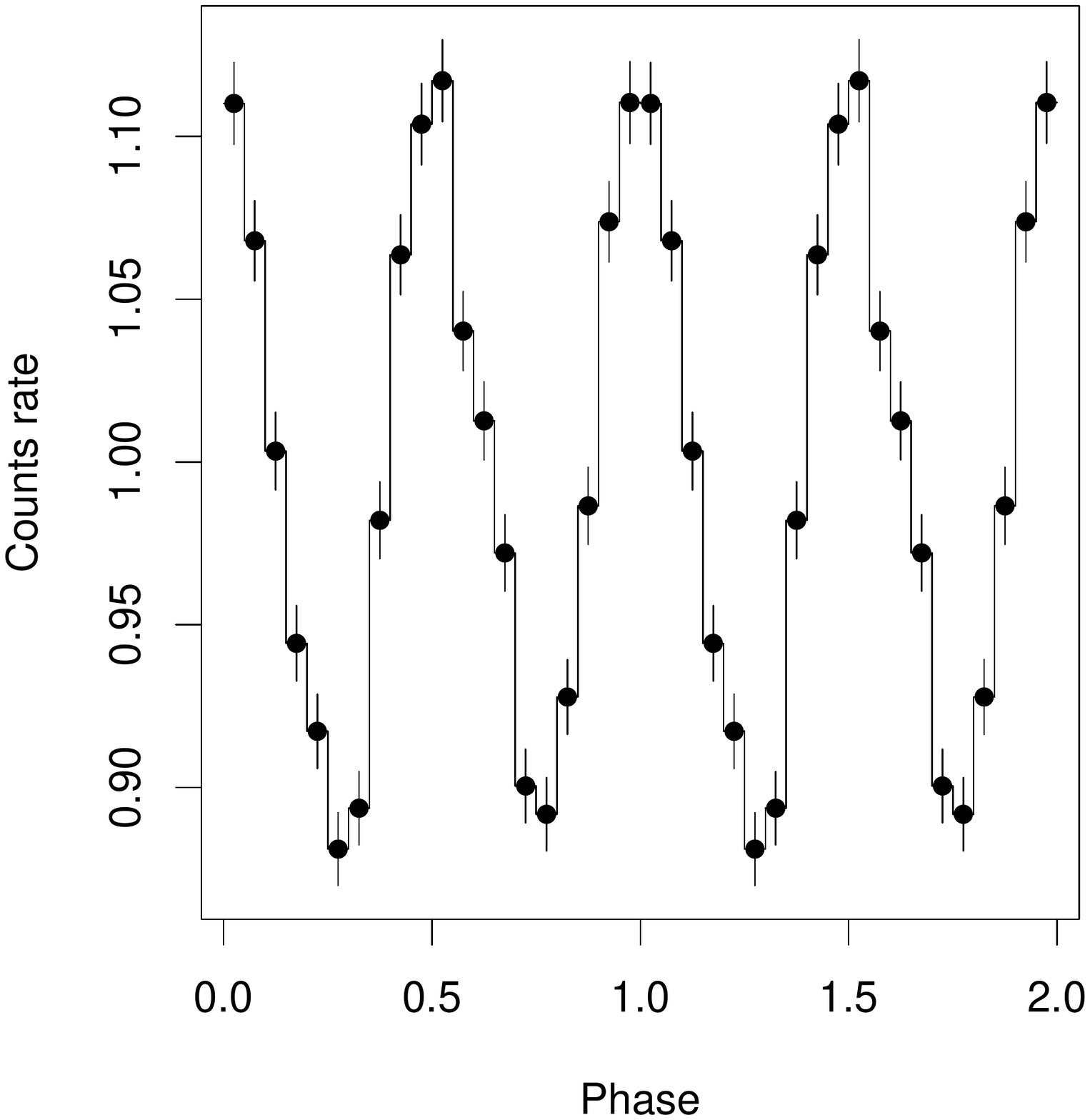}
  }
  \hbox{
    \includegraphics[bb=5 108 527 649,width=5.4cm]{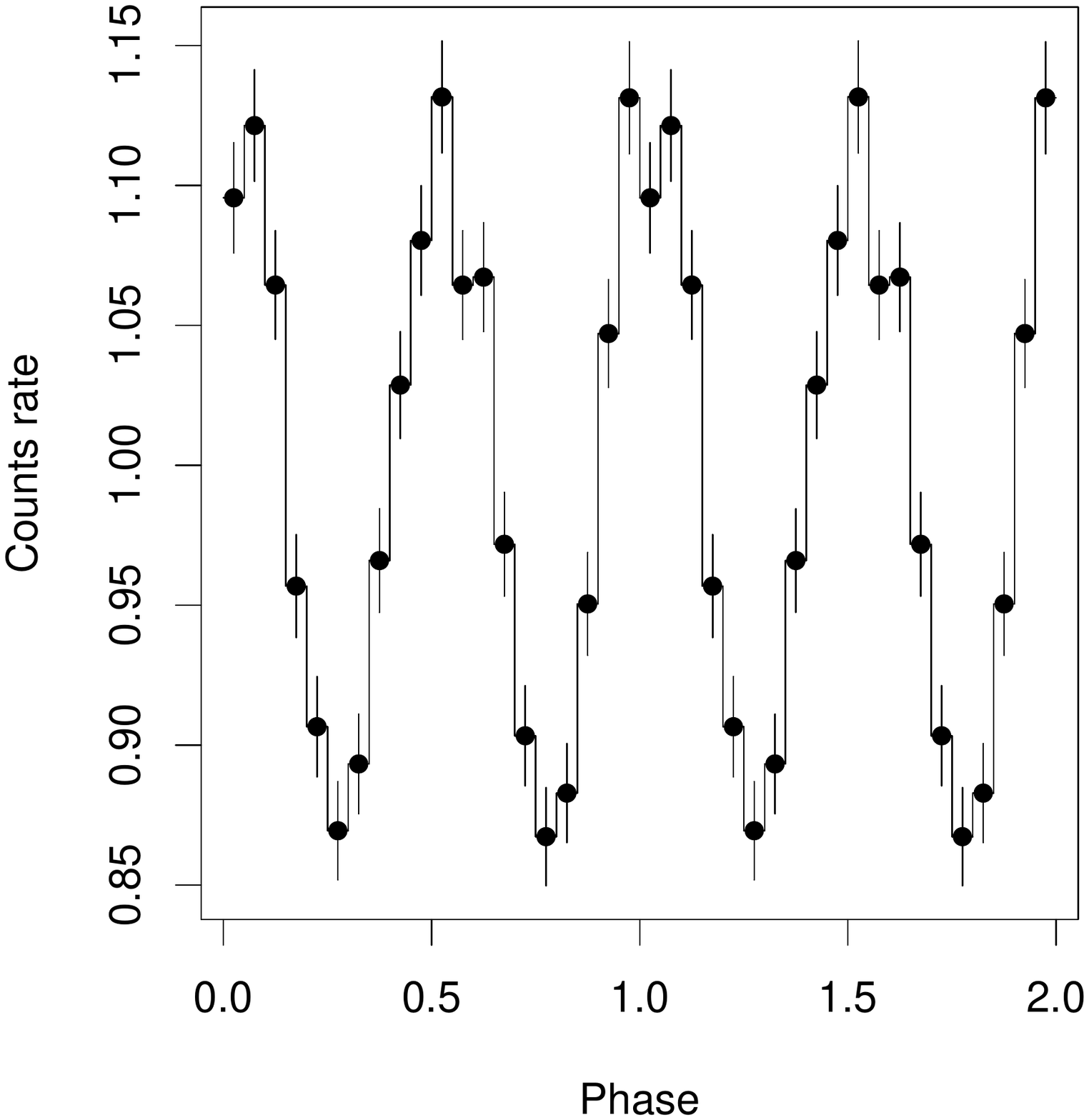}
    \includegraphics[bb=5 108 527 649,width=5.4cm]{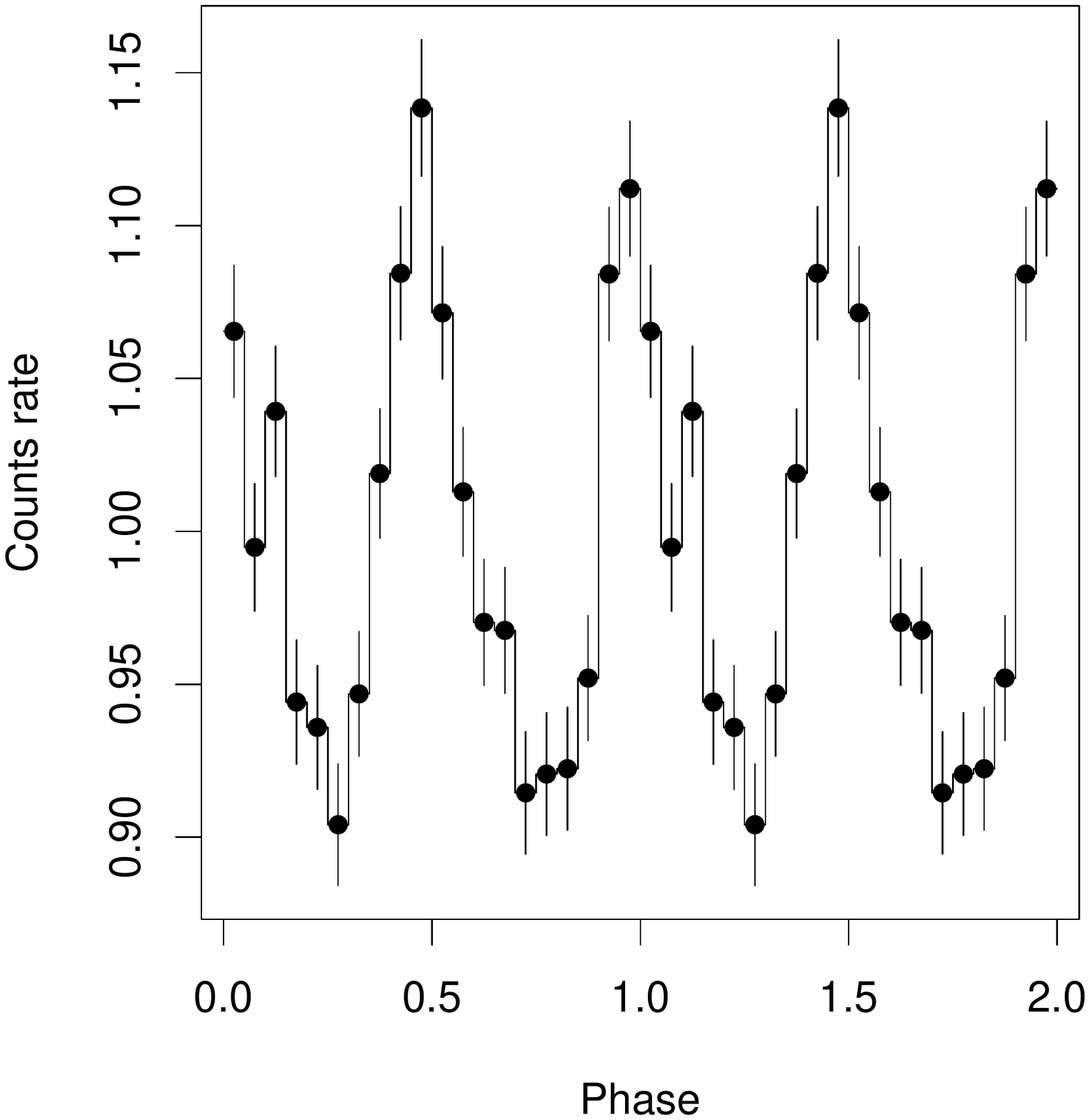}
    \includegraphics[bb=5 108 527 649,width=5.4cm]{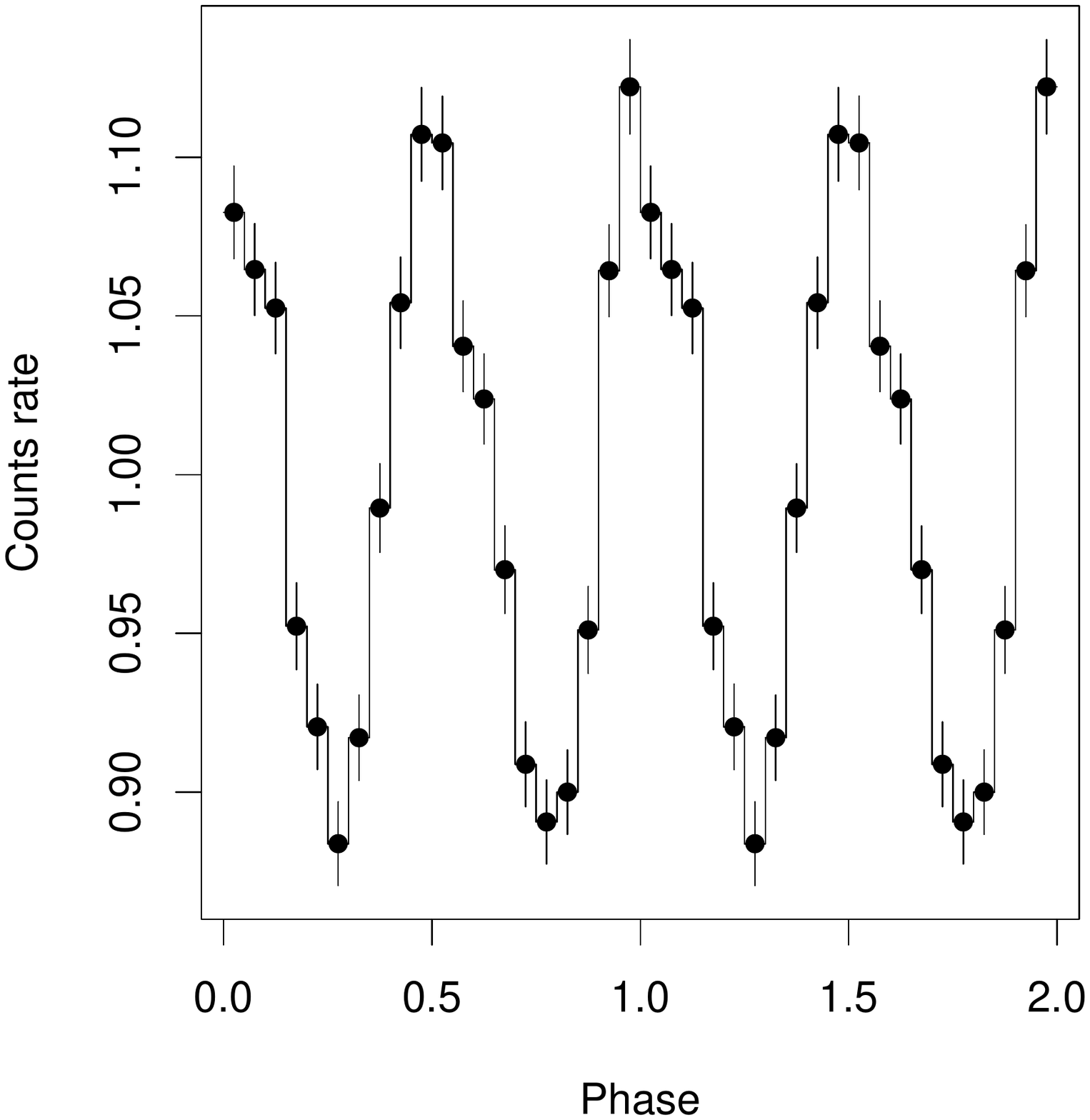}
  }
  \hbox{
    \includegraphics[bb=5 108 527 649,width=5.4cm]{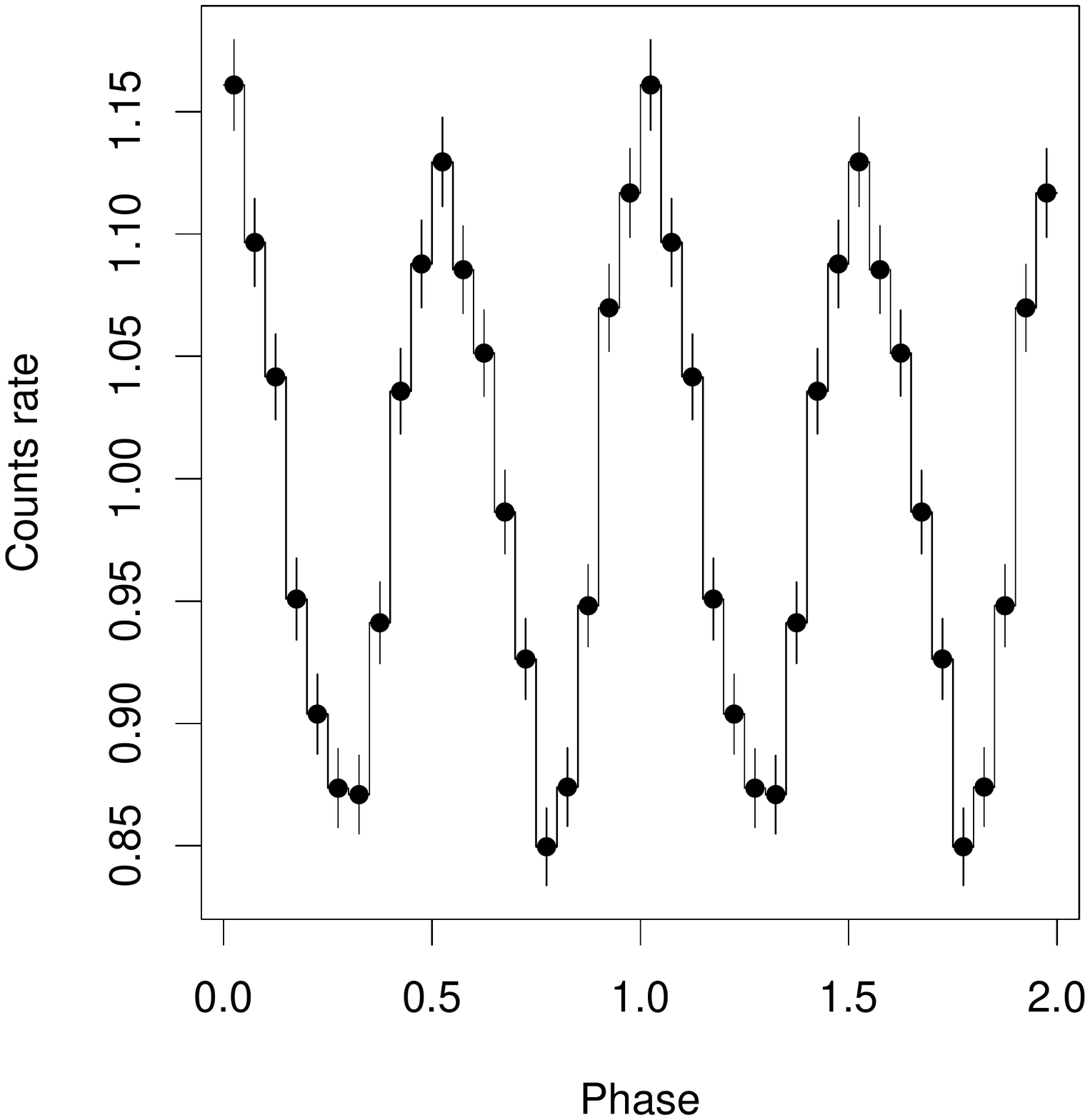}
    \includegraphics[bb=5 108 527 649,width=5.4cm]{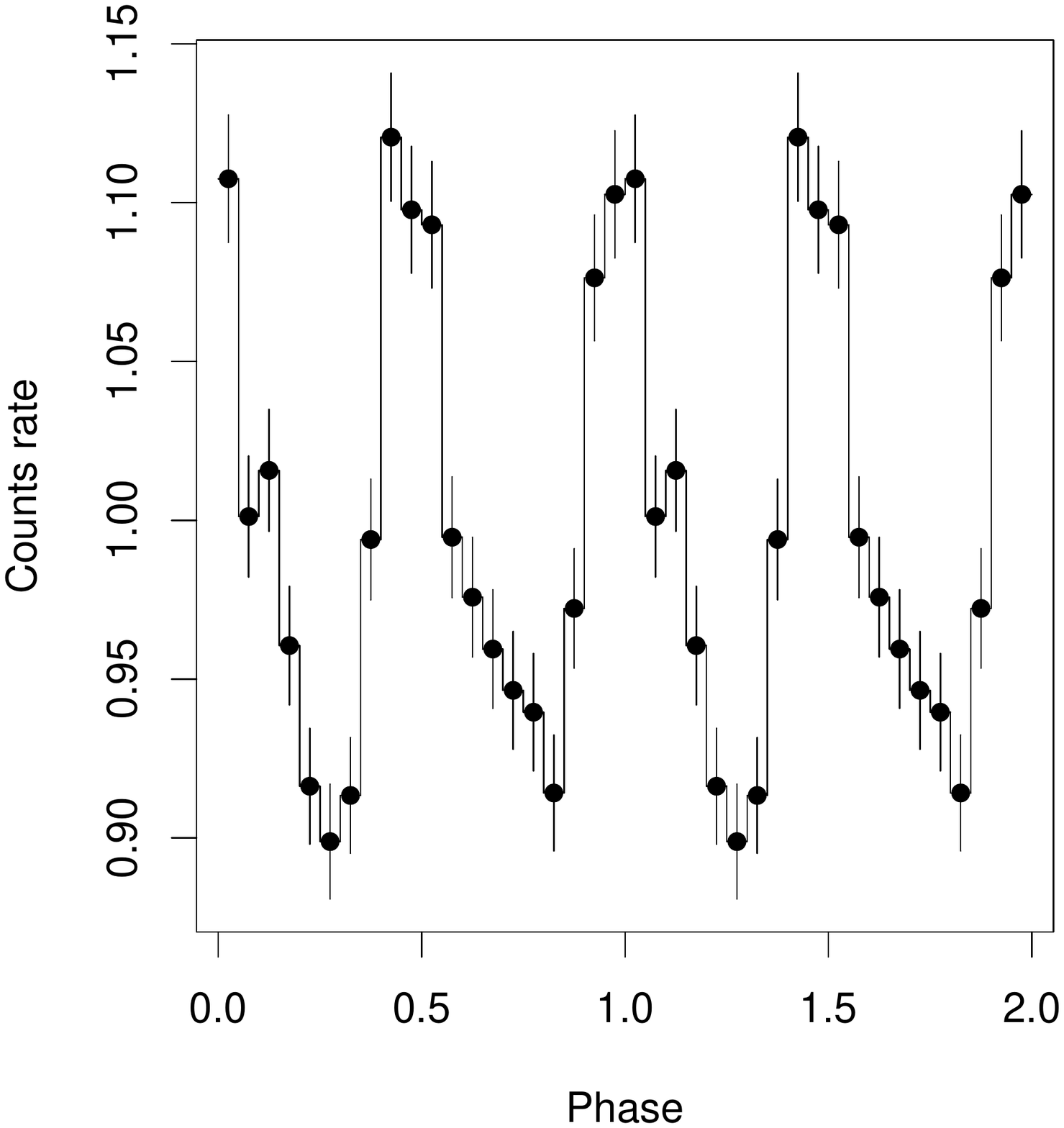}
    \includegraphics[bb=5 108 527 649,width=5.4cm]{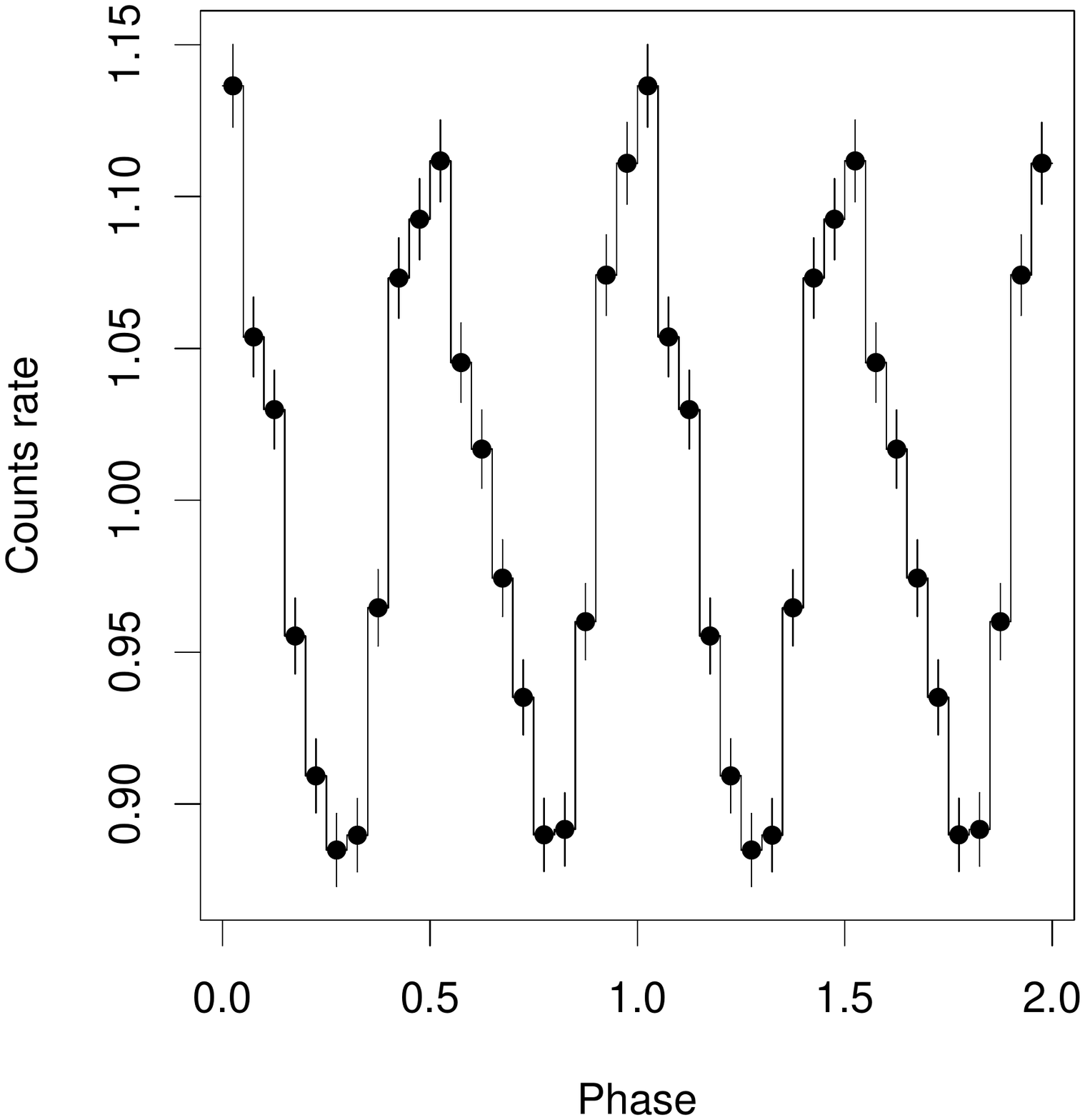}
  }
}
\vspace{-0.2cm}
\caption{For ObsIds 0300520201, 0300520301, 0400140301, and 0400140401 (see Fig.~\ref{resapp1}.)\label{resapp2}}
\end{figure*}

\begin{figure*}
\centering
\vspace{-0.2cm}
\vbox{
  \hbox{
    \includegraphics[bb=5 108 527 649,width=5.4cm]{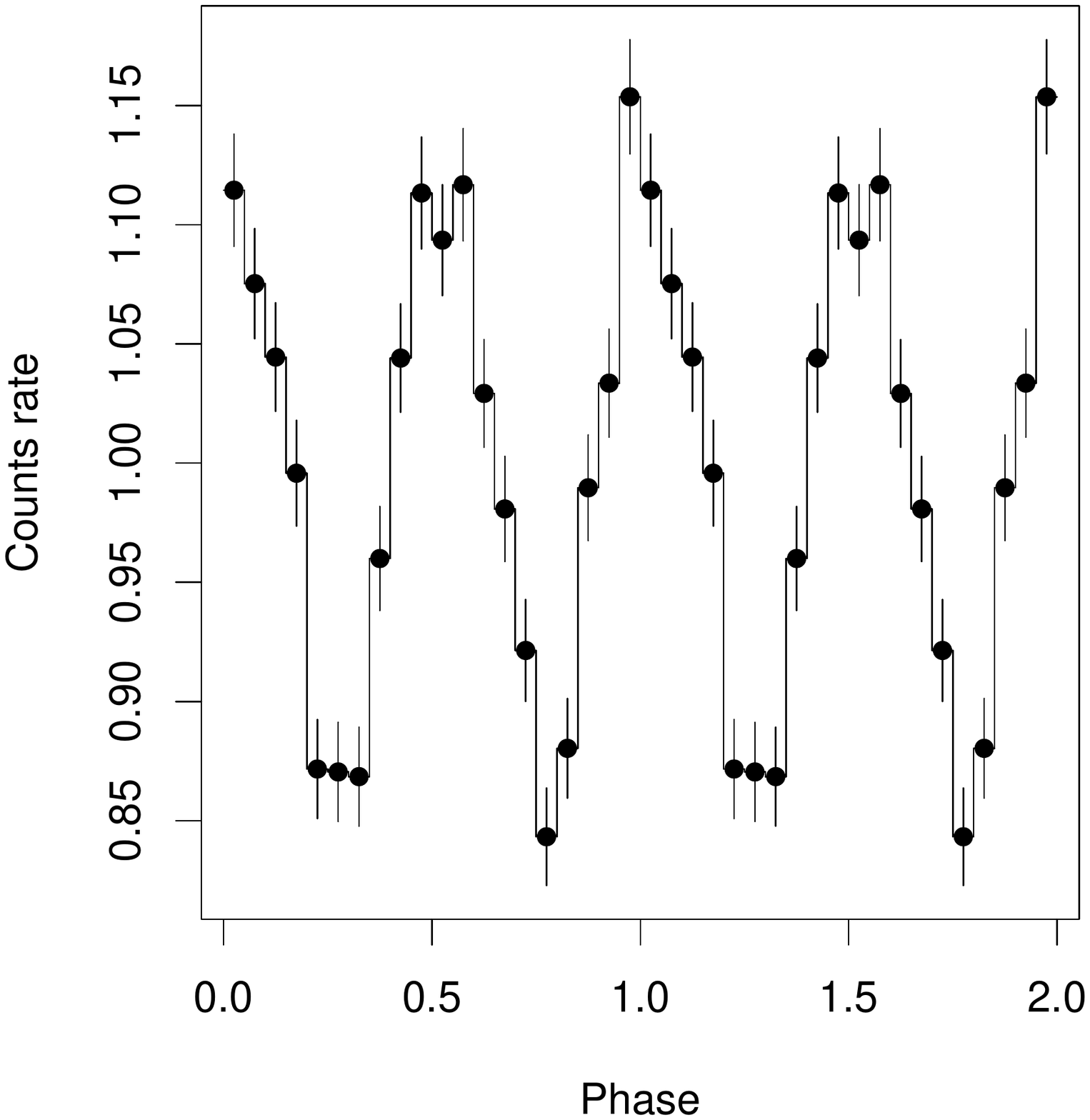}
    \includegraphics[bb=5 108 527 649,width=5.4cm]{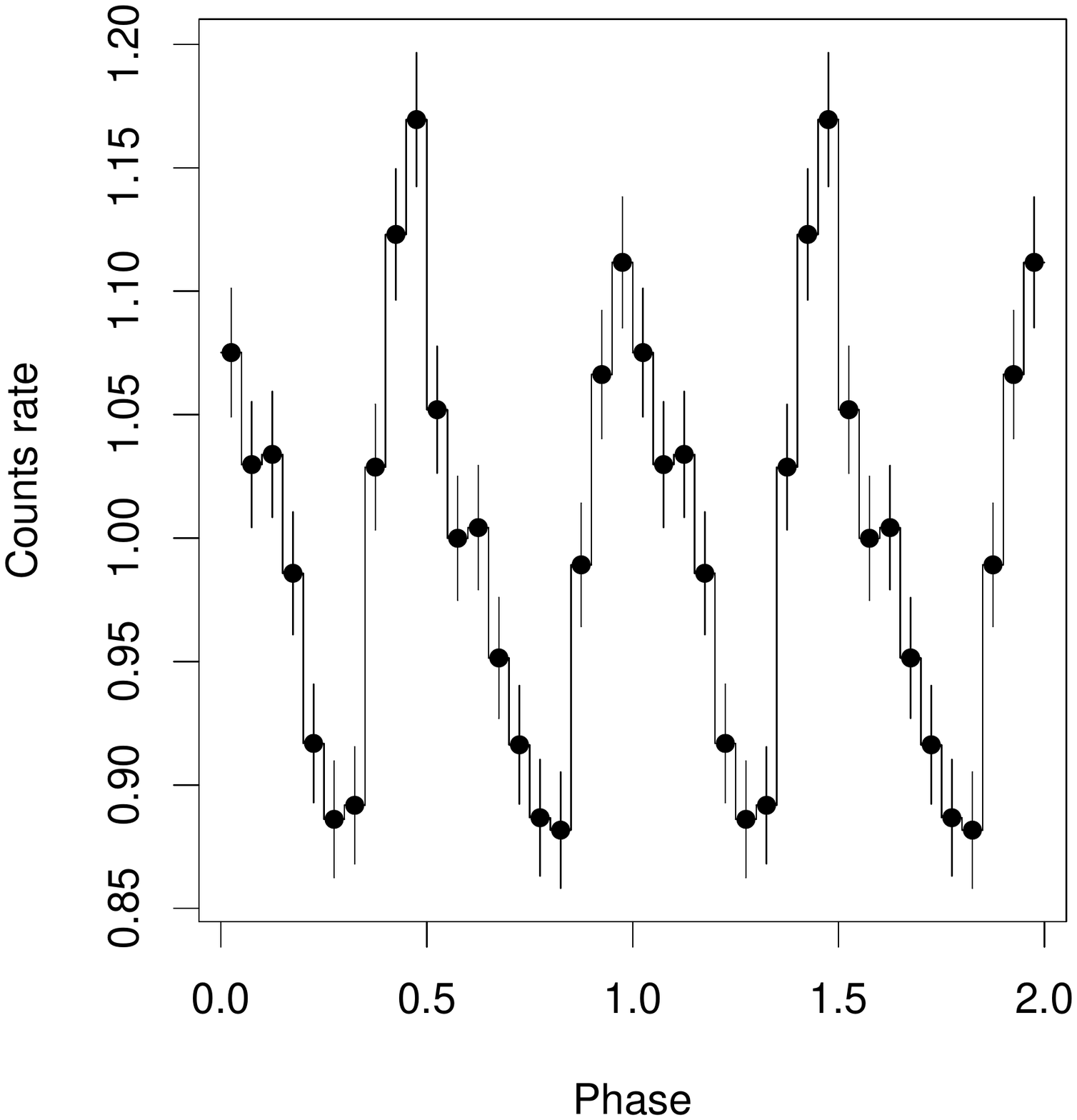}
    \includegraphics[bb=5 108 527 649,width=5.4cm]{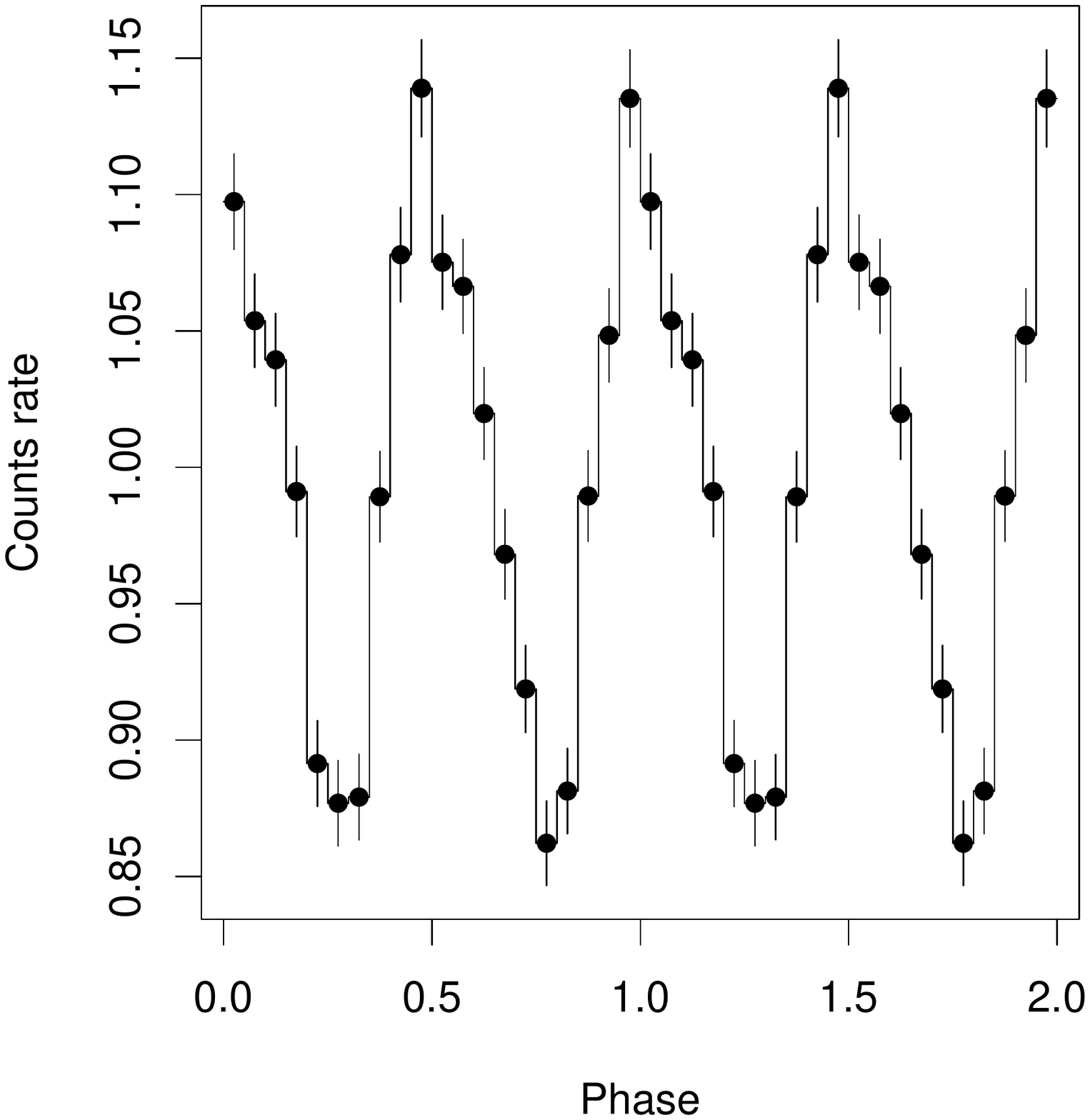}
  }
  \hbox{
    \includegraphics[bb=5 108 527 649,width=5.4cm]{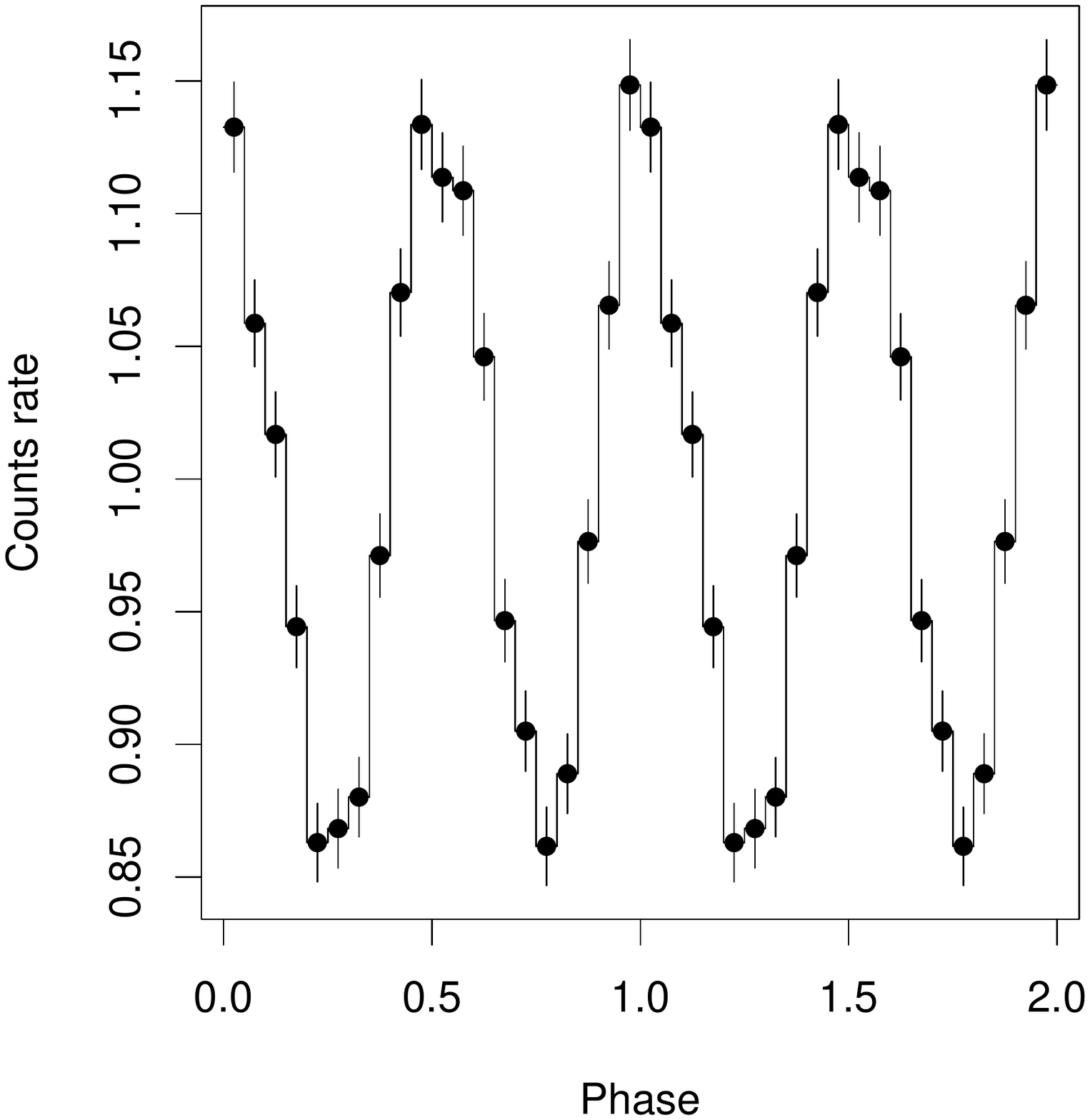}
    \includegraphics[bb=5 108 527 649,width=5.4cm]{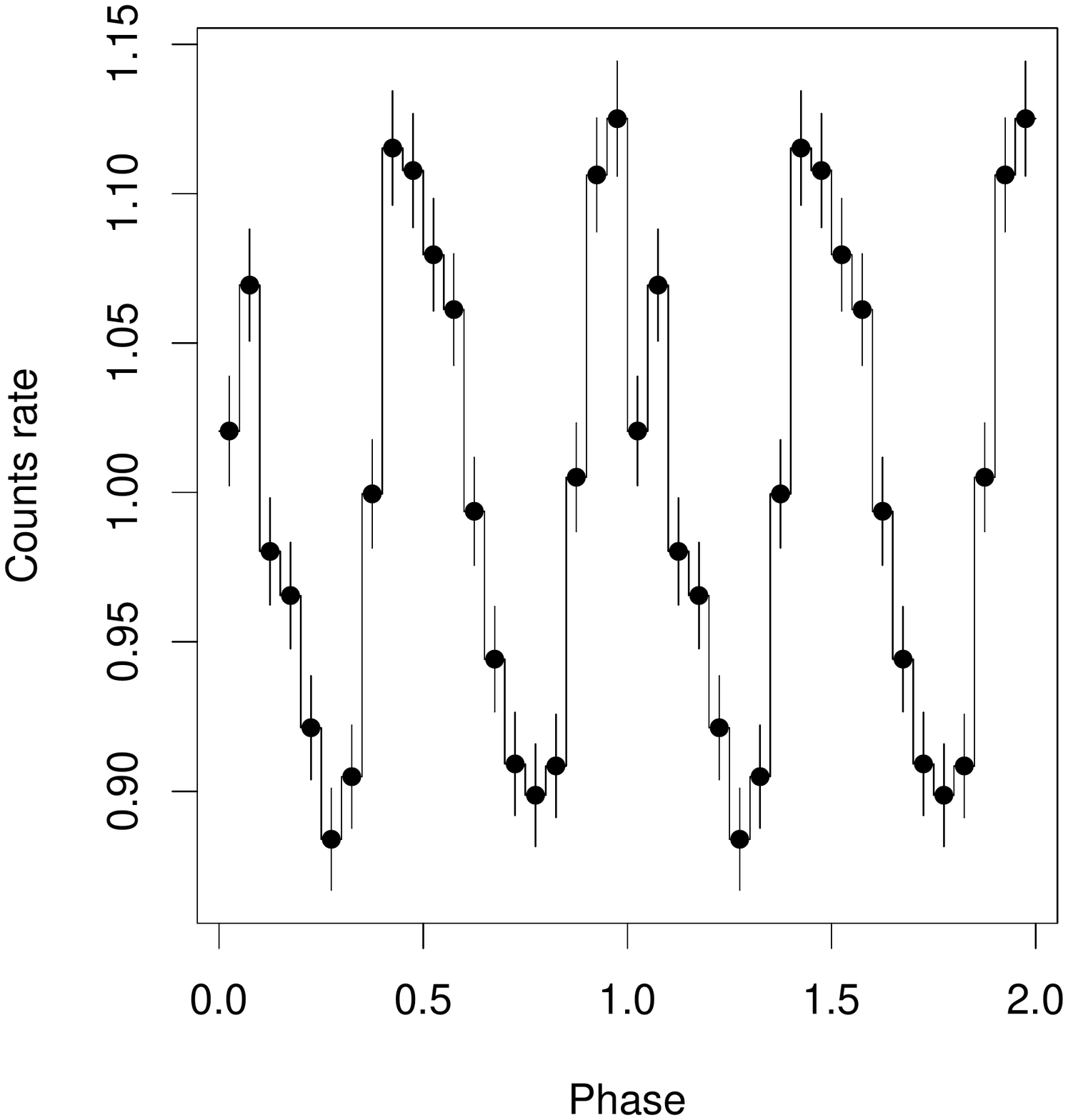}
    \includegraphics[bb=5 108 527 649,width=5.4cm]{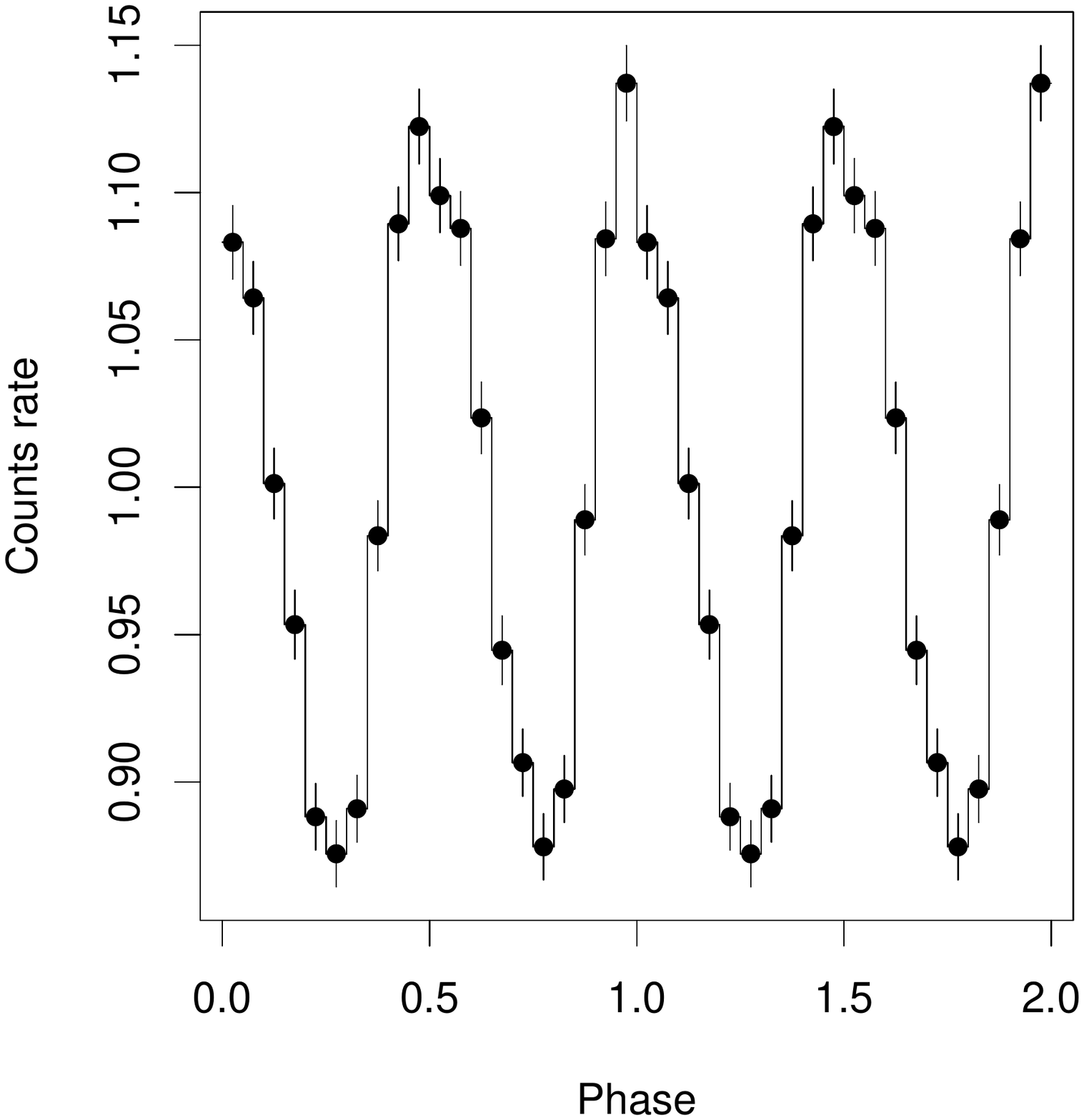}  
  }
  \hbox{
    \includegraphics[bb=5 108 527 649,width=5.4cm]{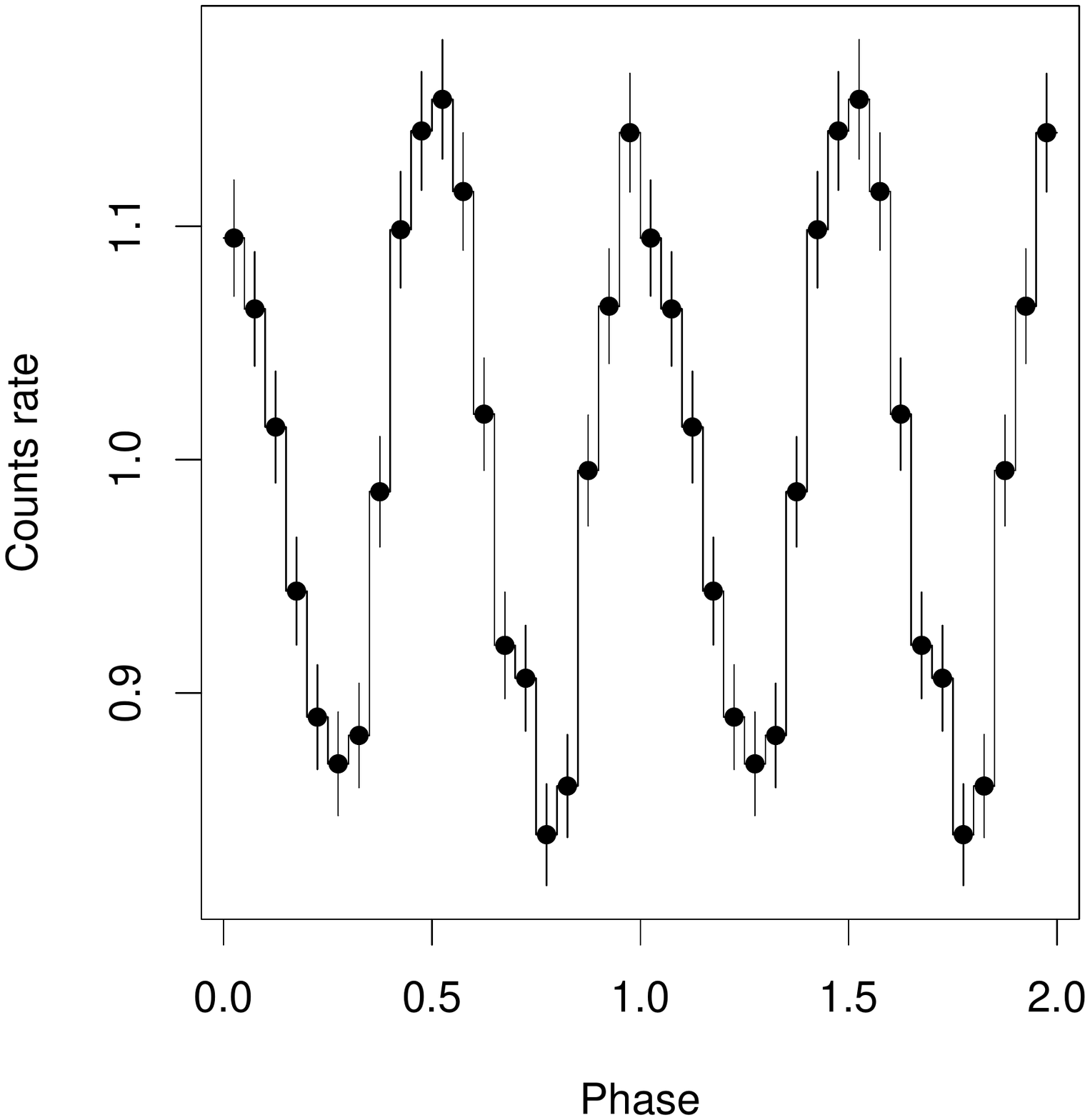}
    \includegraphics[bb=5 108 527 649,width=5.4cm]{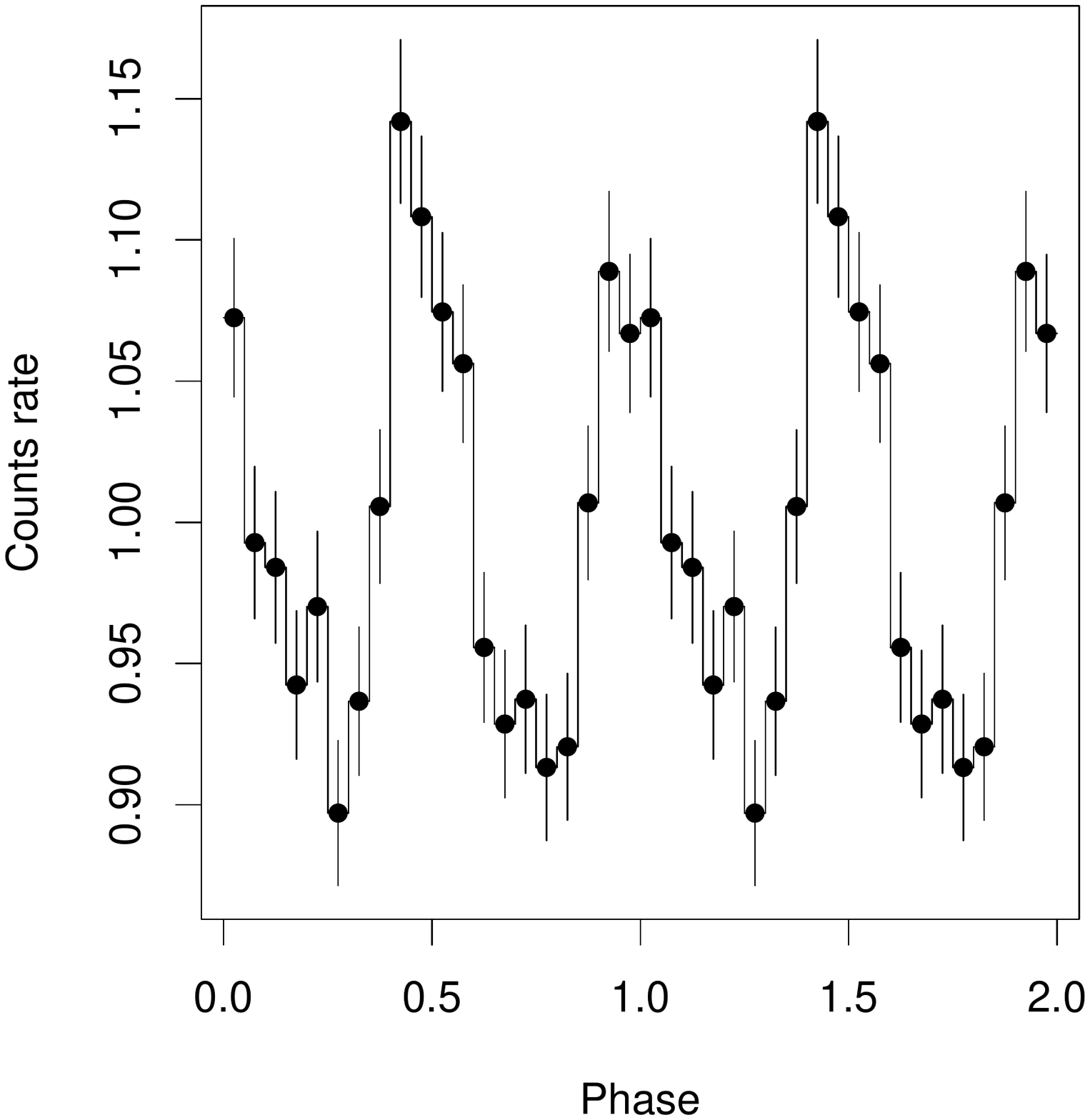}
    \includegraphics[bb=5 108 527 649,width=5.4cm]{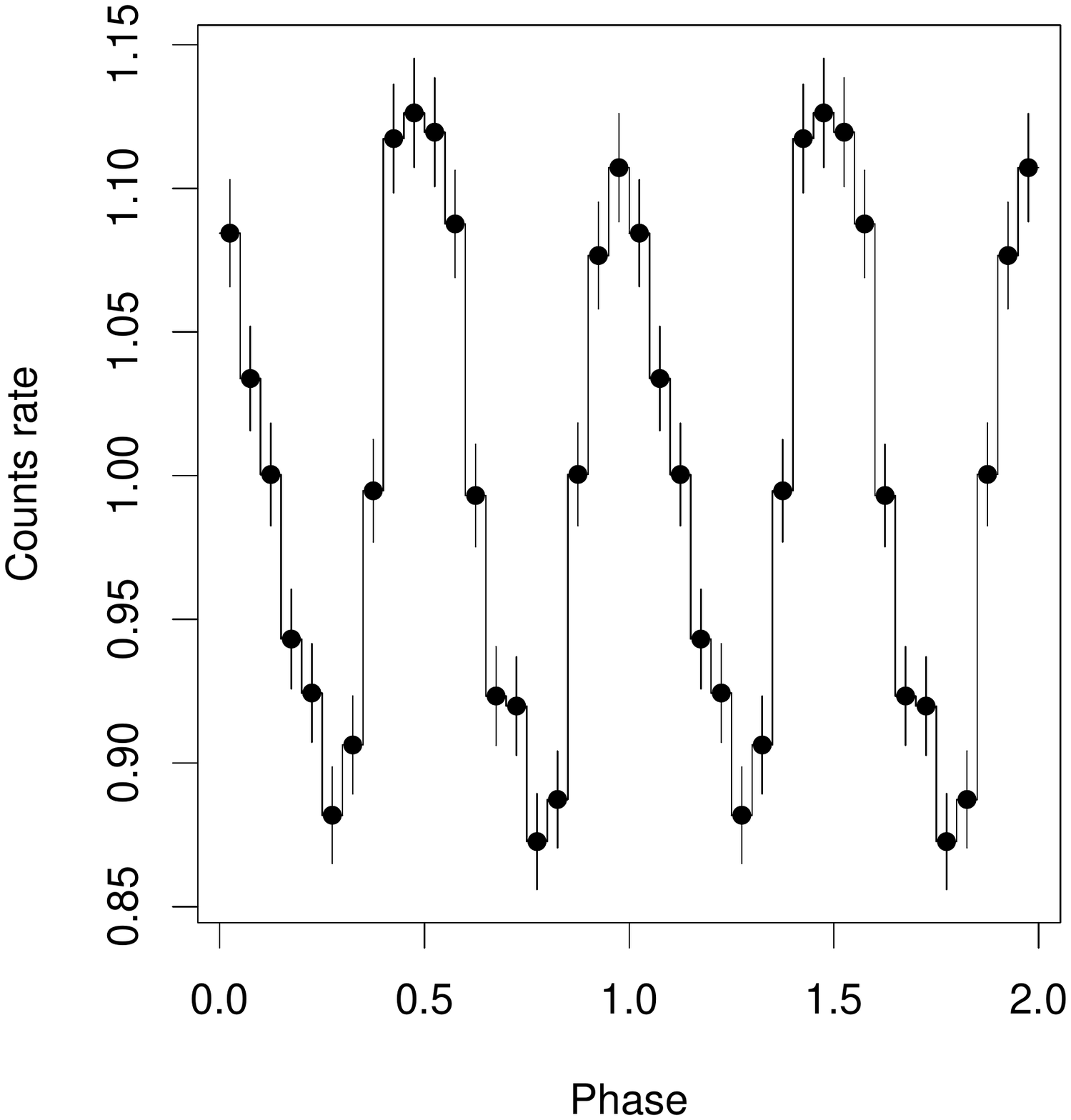}
  }
  \hbox{
    \includegraphics[bb=5 108 527 649,width=5.4cm]{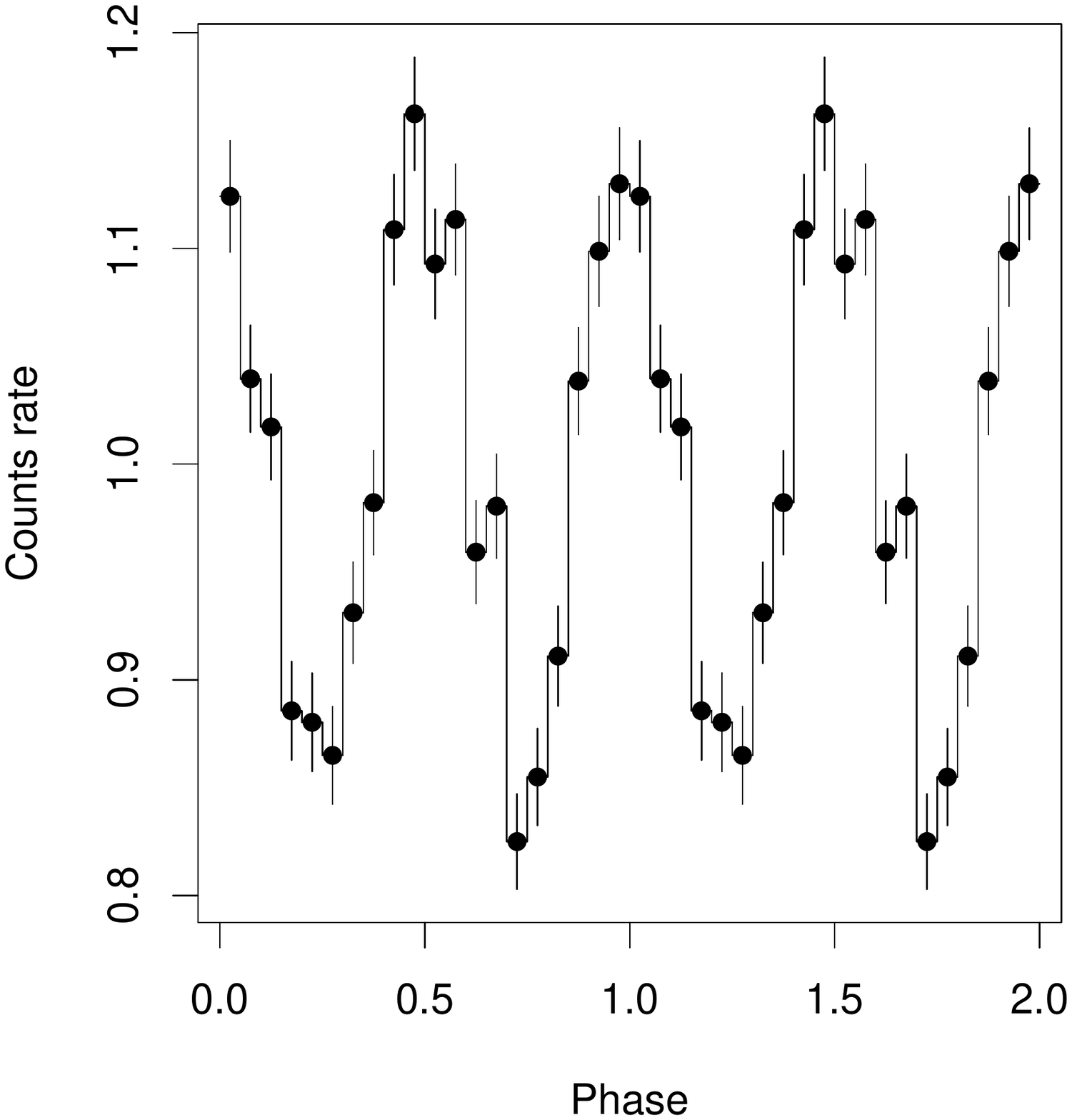}
    \includegraphics[bb=5 108 527 649,width=5.4cm]{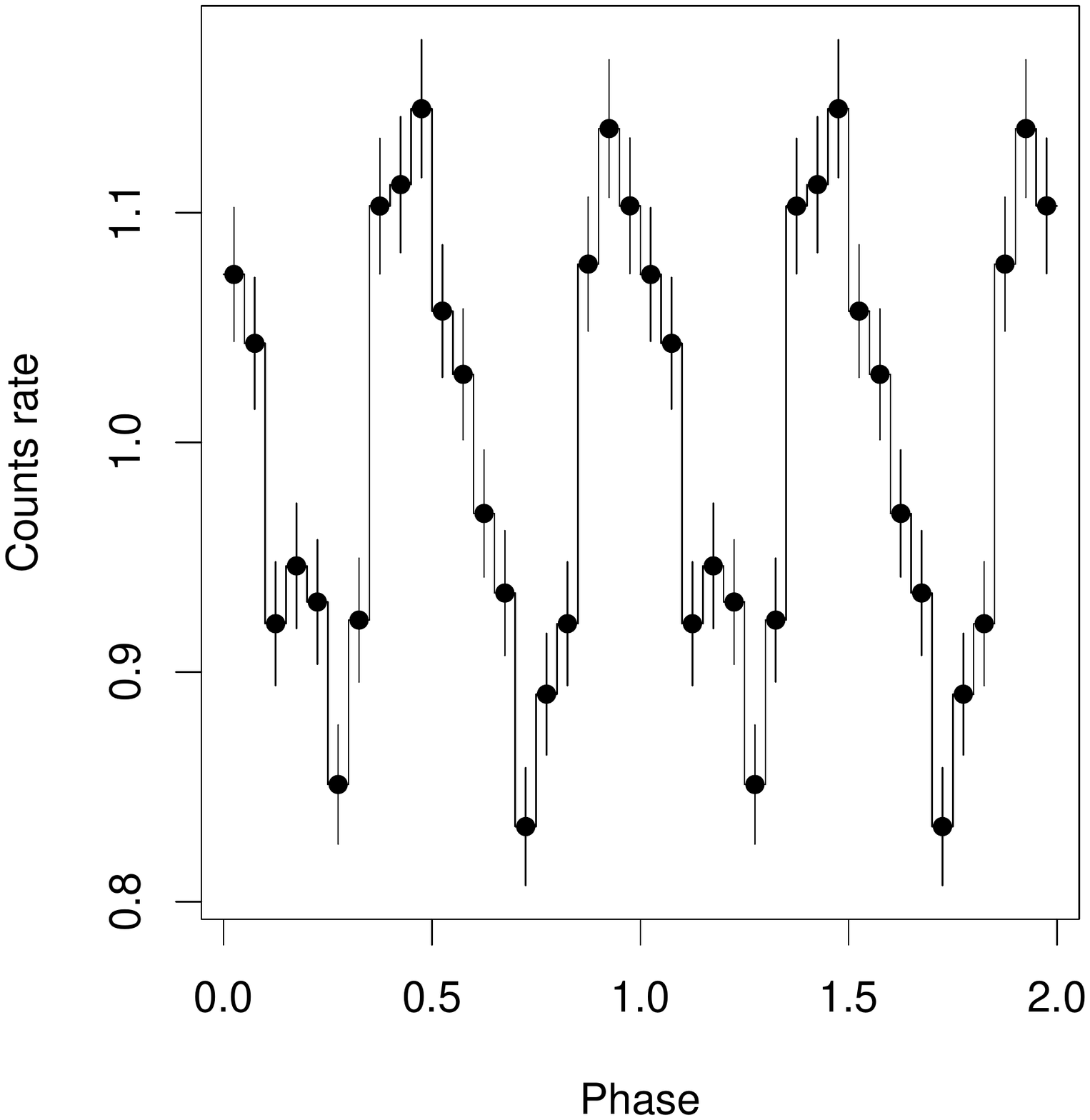}
    \includegraphics[bb=5 108 527 649,width=5.4cm]{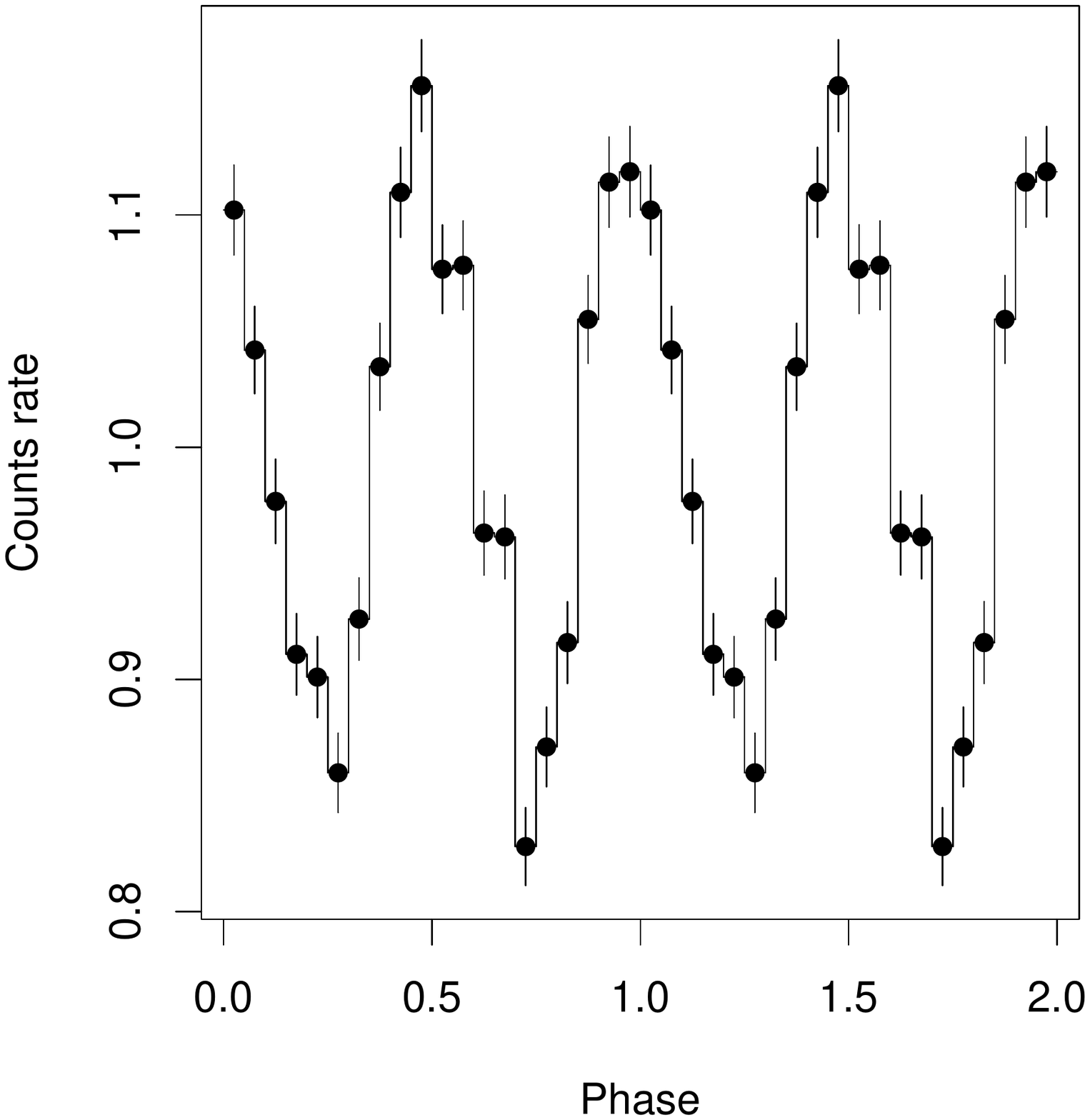}
  }
}
\vspace{-0.2cm}
\caption{For ObsIds 0502710201, 0502710201, 0554510101, and 0601170301 (see Fig.~\ref{resapp1}.)\label{resapp3}}
\end{figure*}

\begin{figure*}
\centering
\vspace{-0.2cm}
\vbox{
  \hbox{
    \includegraphics[bb=5 108 527 649,width=5.4cm]{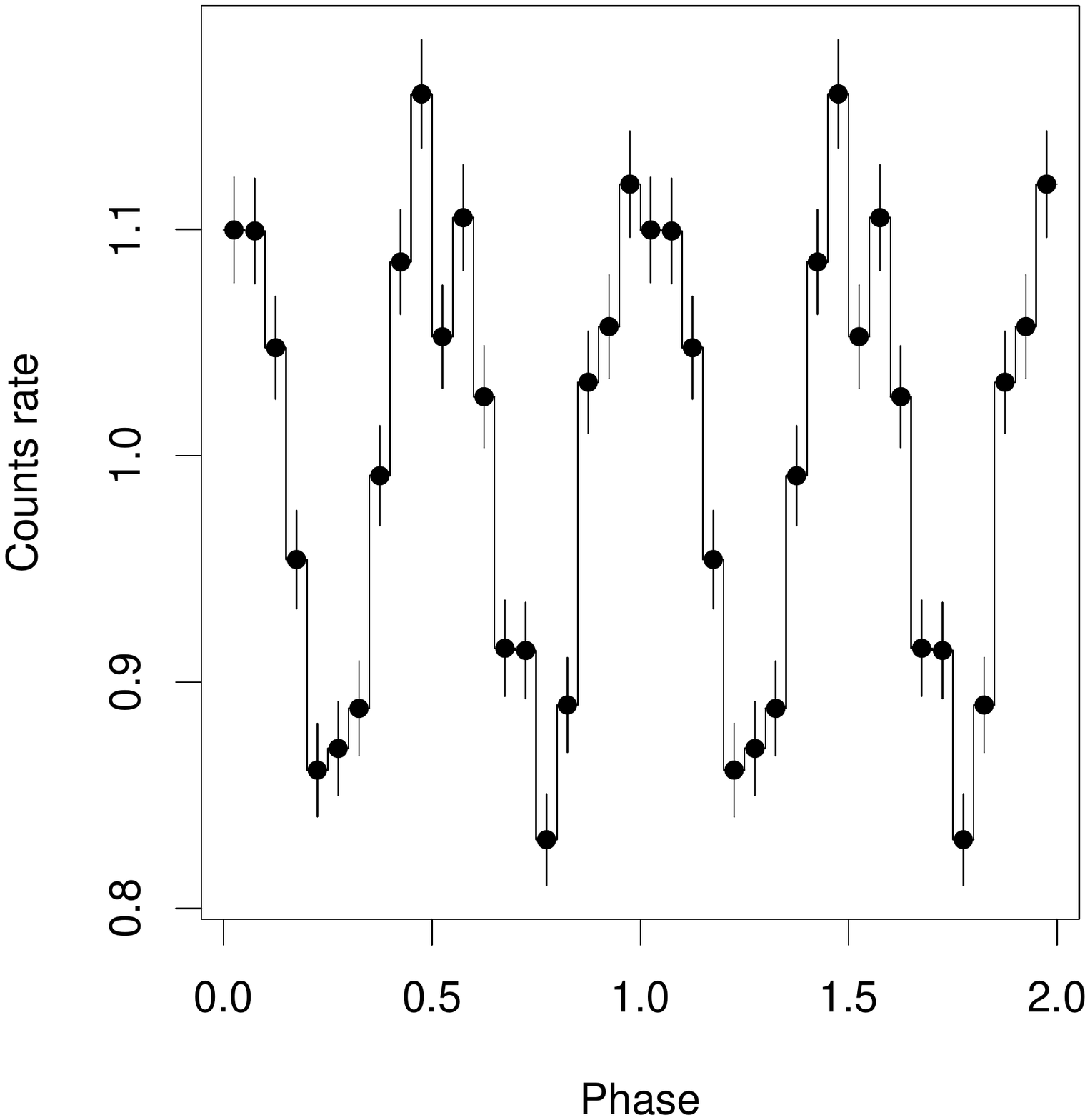}
    \includegraphics[bb=5 108 527 649,width=5.4cm]{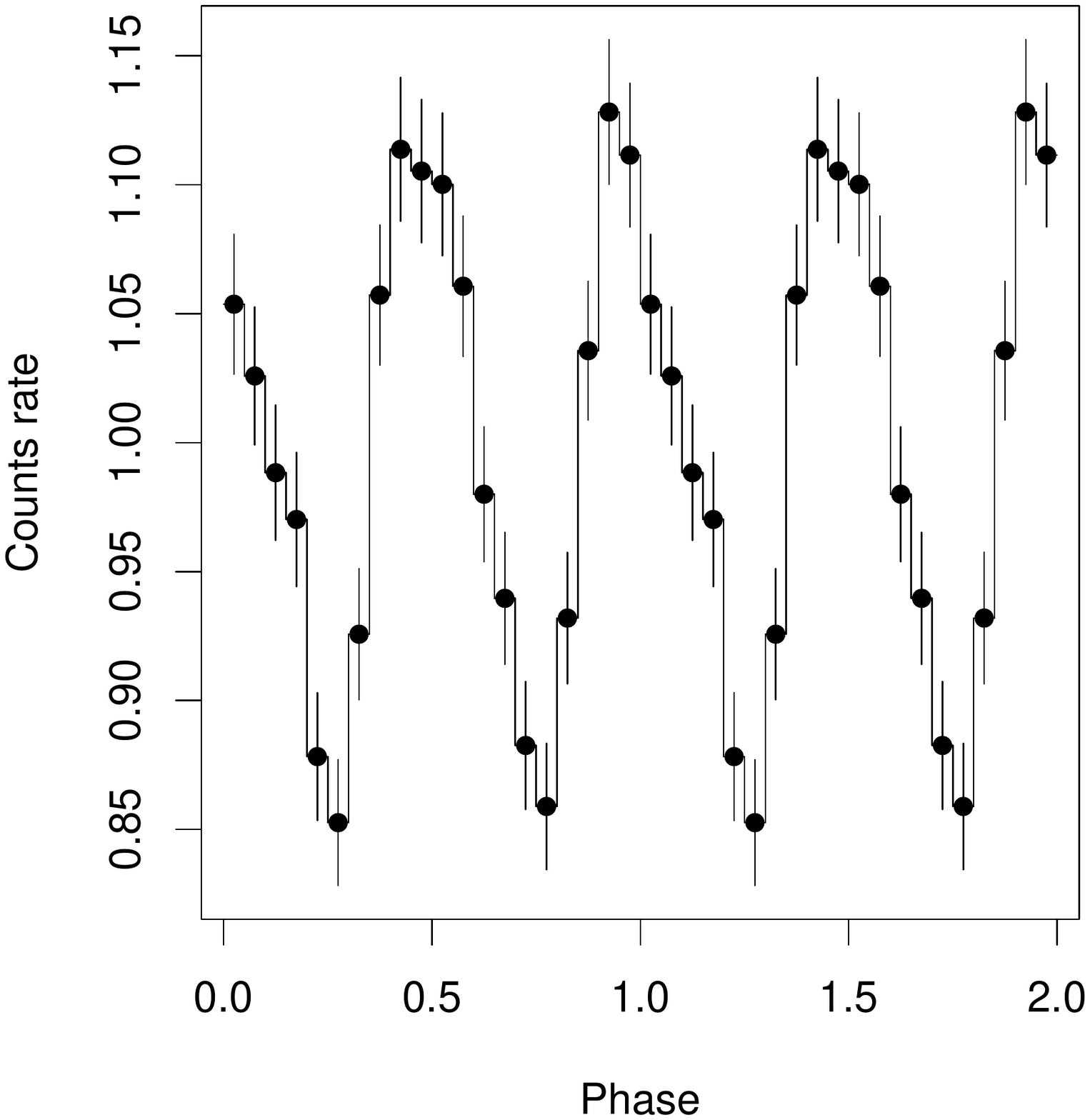}
    \includegraphics[bb=5 108 527 649,width=5.4cm]{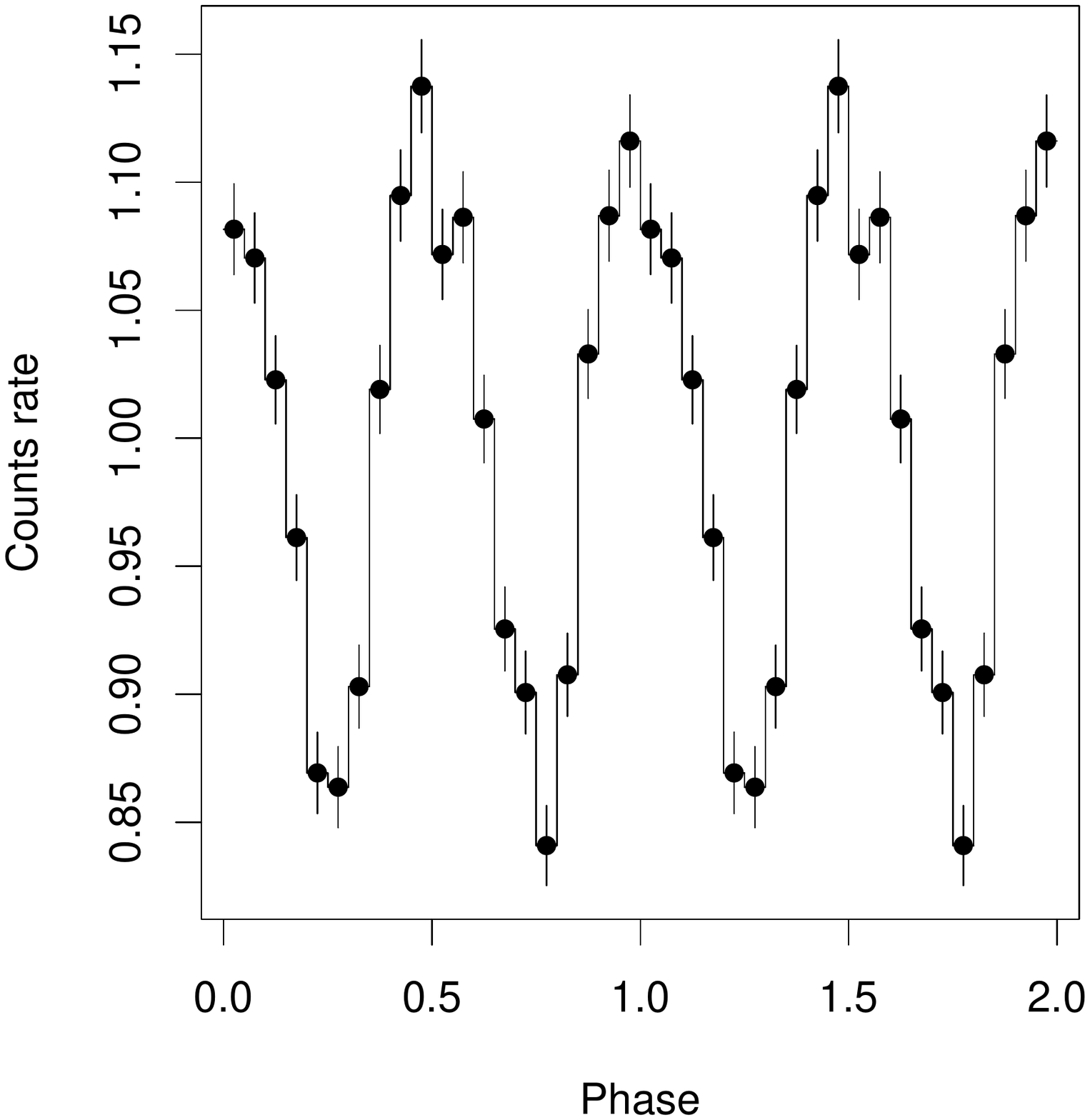}
  }
  \hbox{
    \includegraphics[bb=5 108 527 649,width=5.4cm]{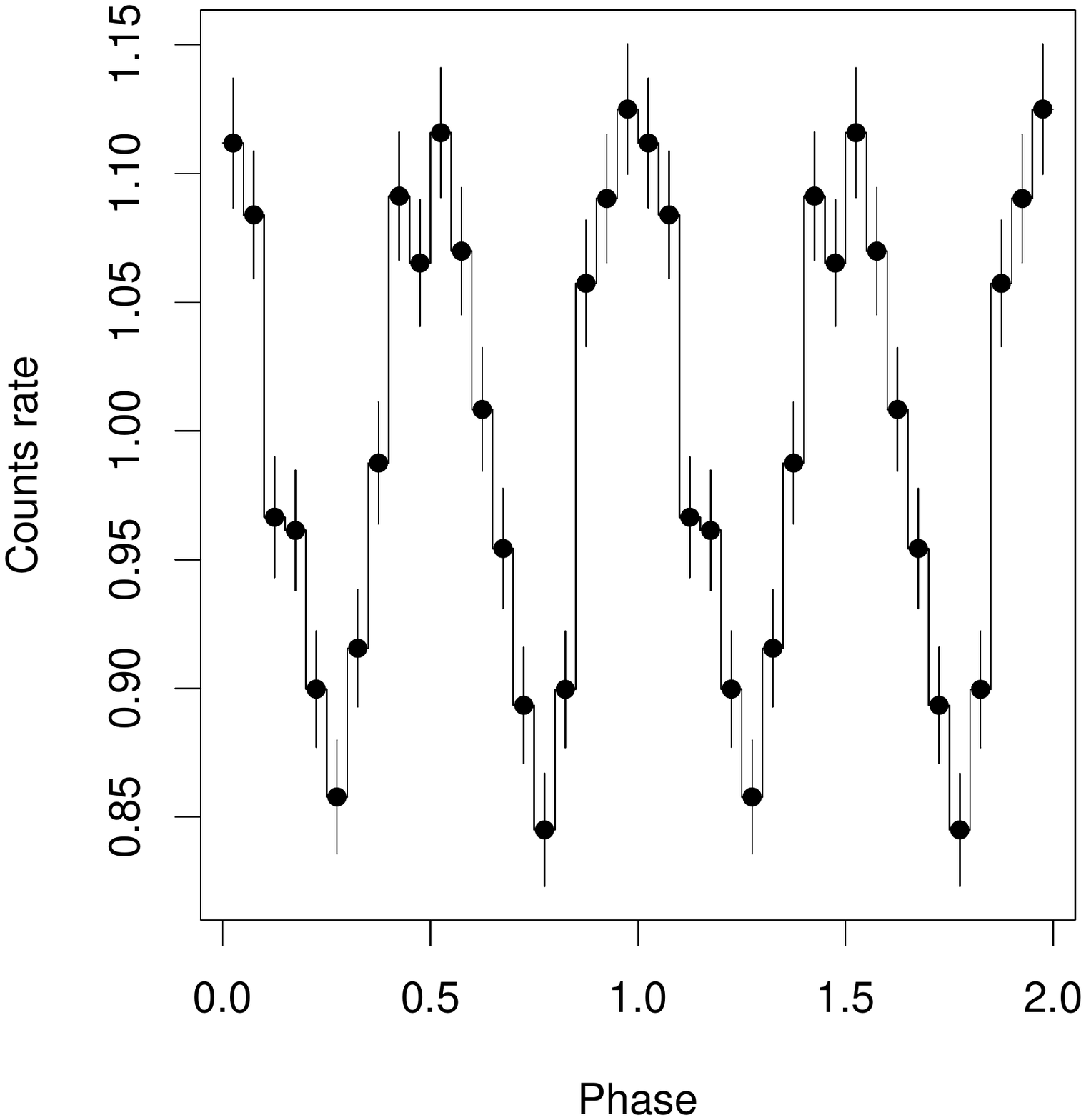}
    \includegraphics[bb=5 108 527 649,width=5.4cm]{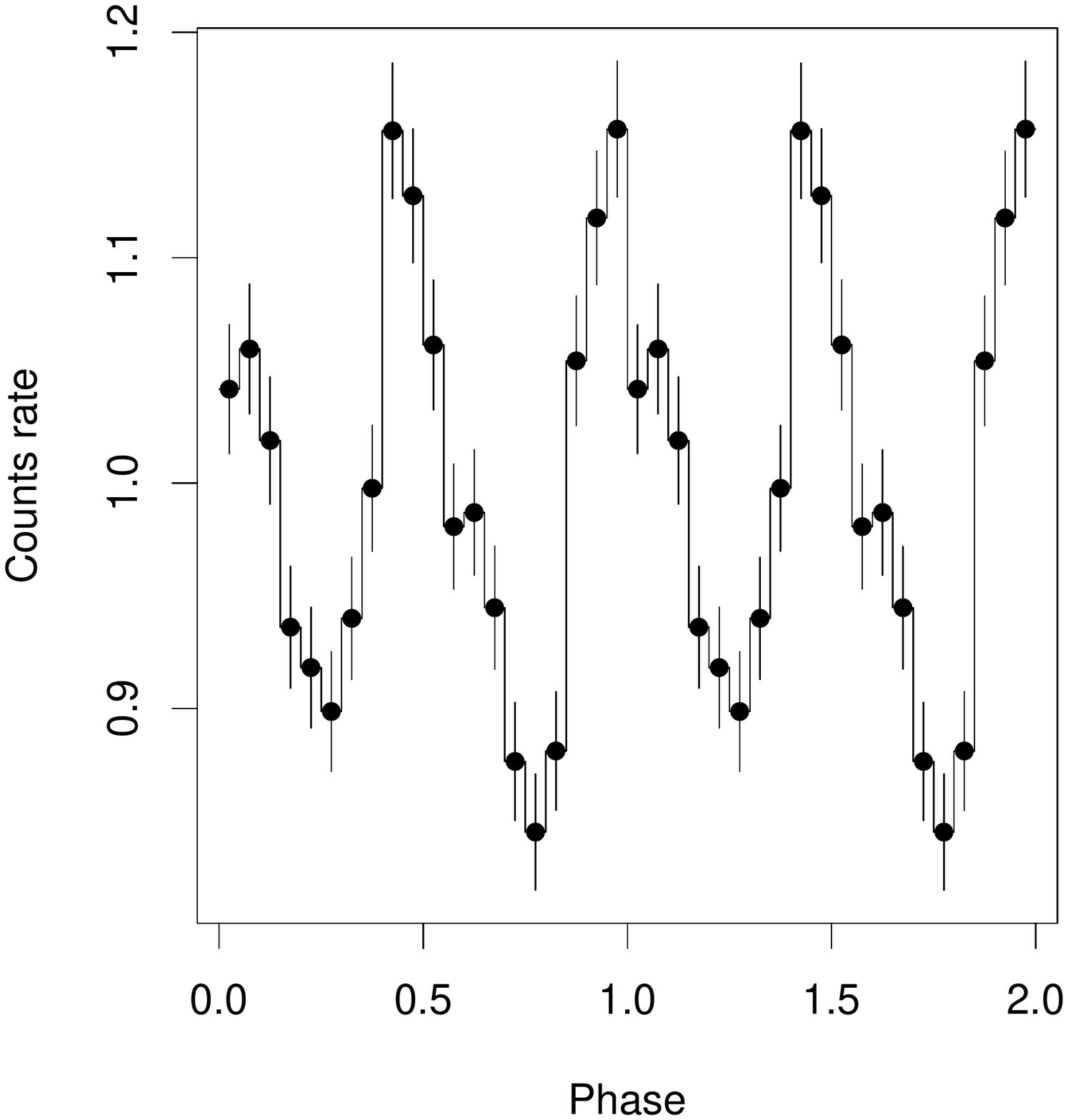}
    \includegraphics[bb=5 108 527 649,width=5.4cm]{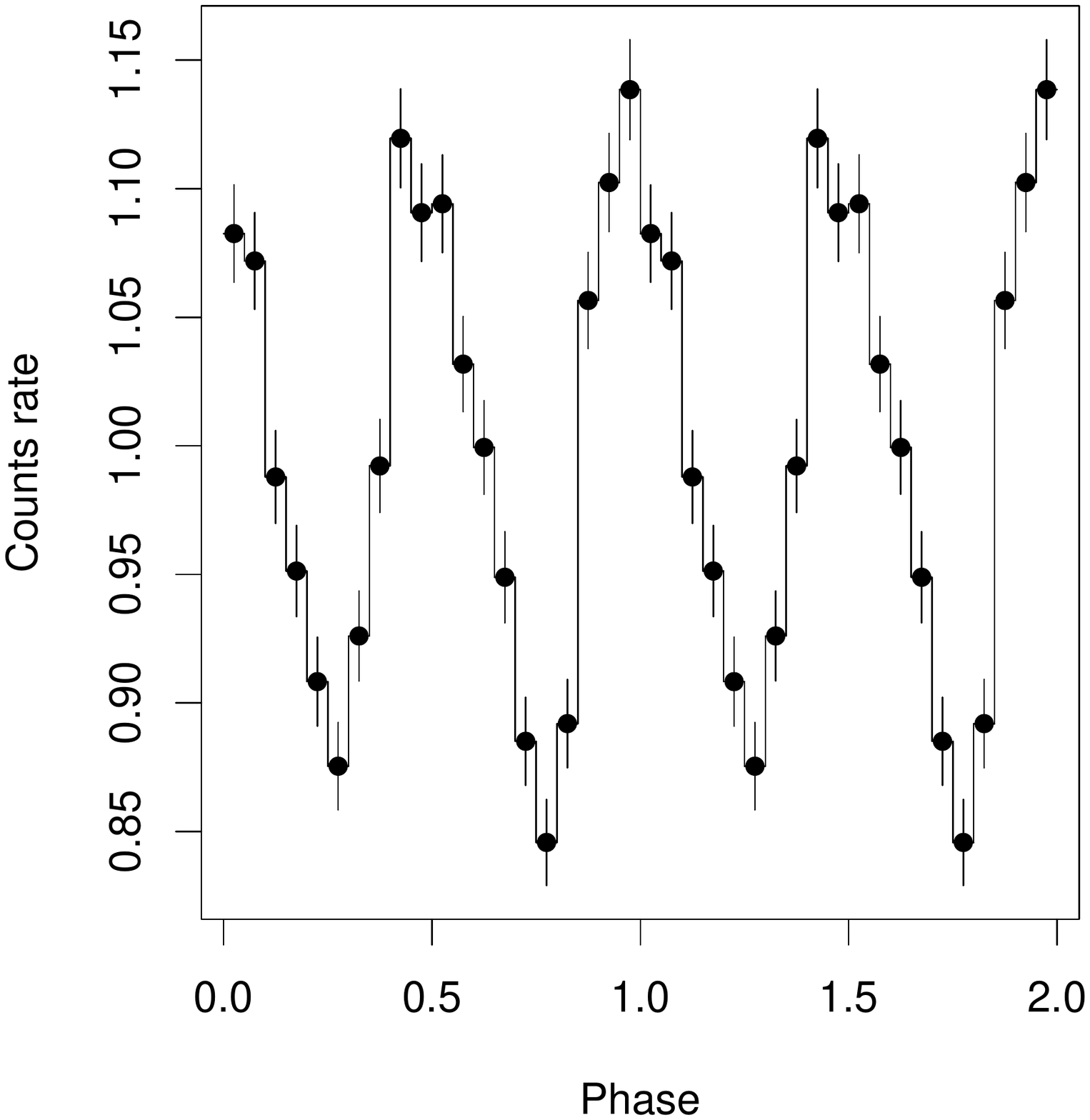}
  }
  \hbox{
    \includegraphics[bb=5 108 527 649,width=5.4cm]{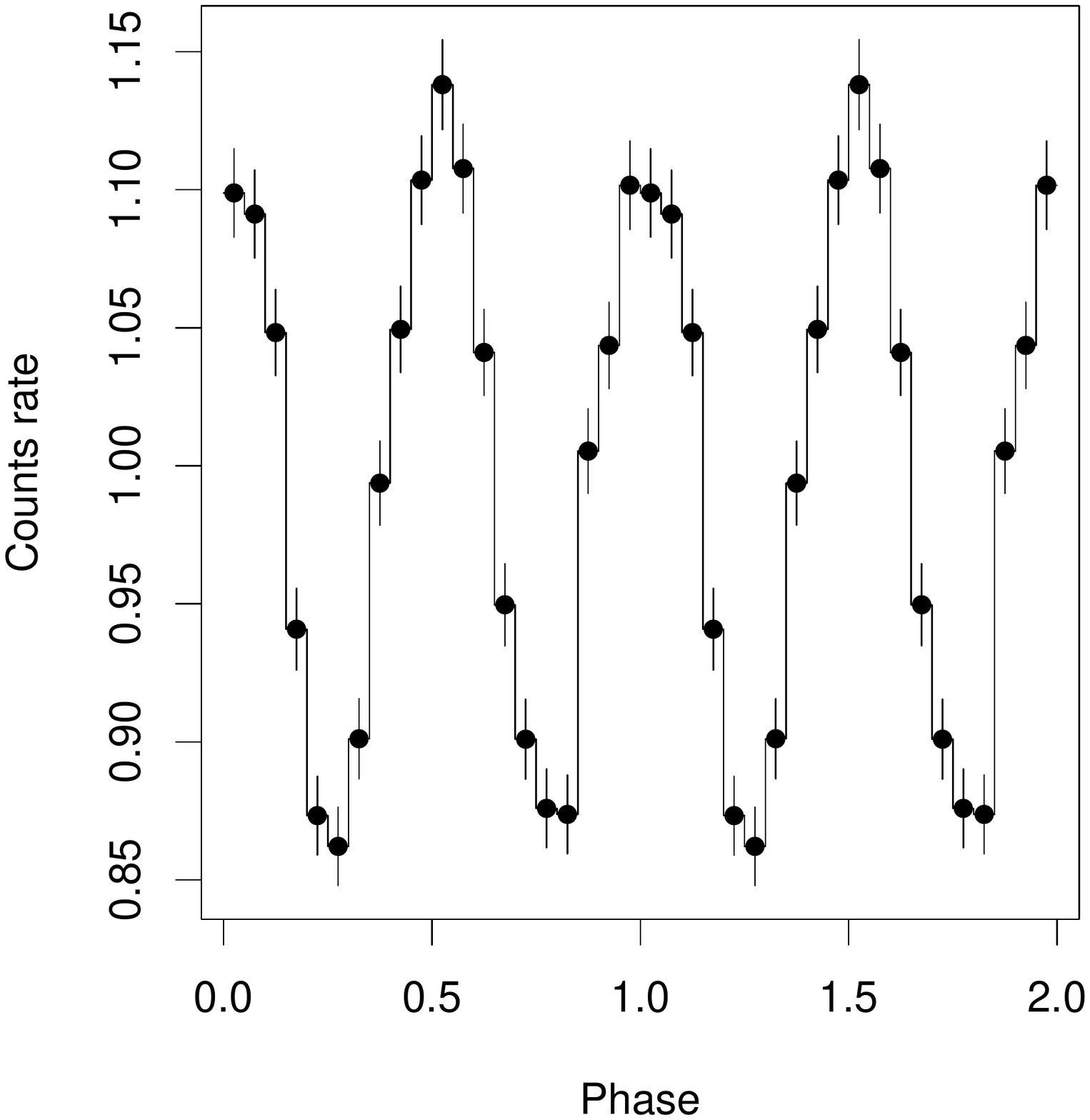}
    \includegraphics[bb=5 108 527 649,width=5.4cm]{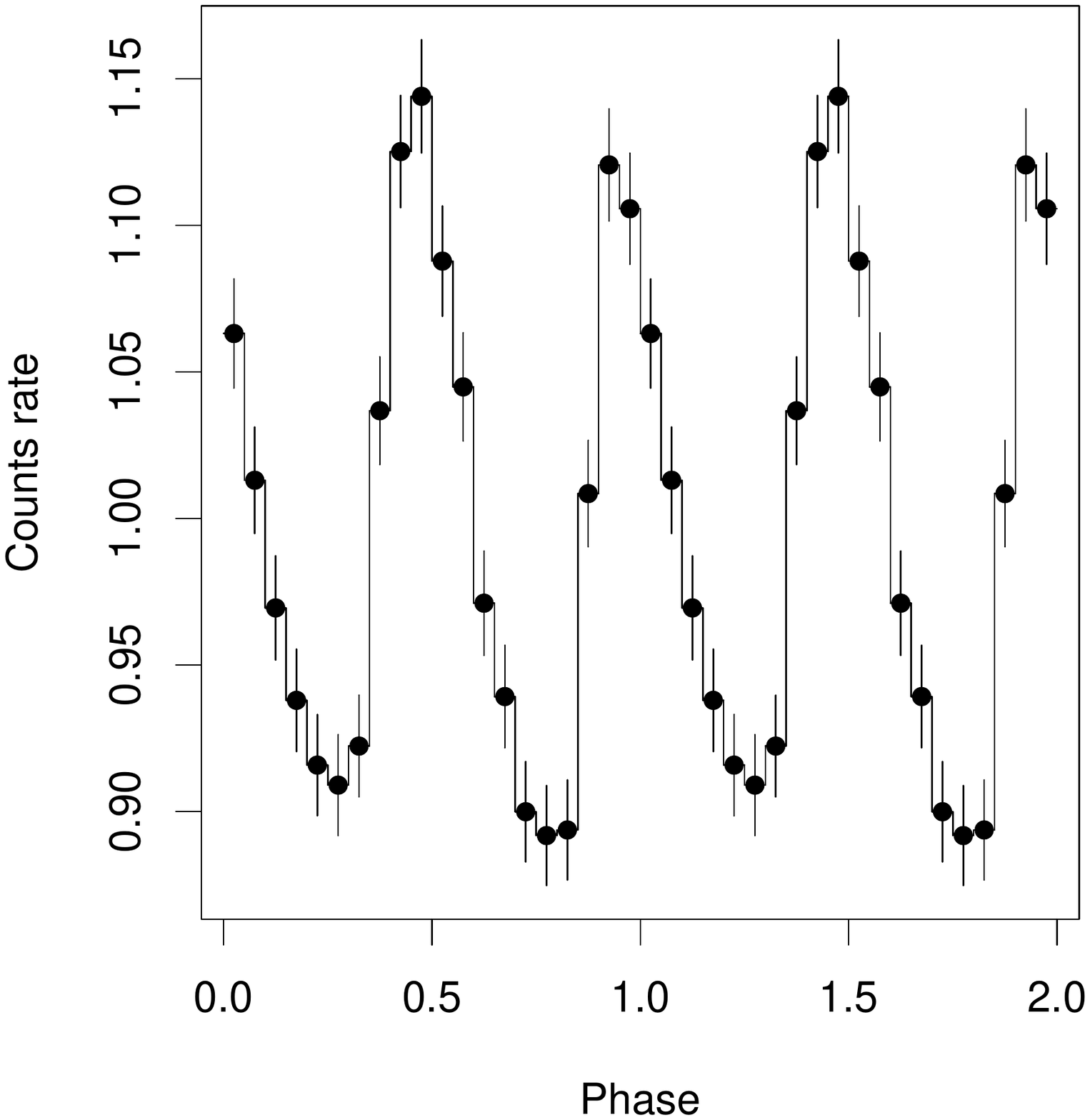}
    \includegraphics[bb=5 108 527 649,width=5.4cm]{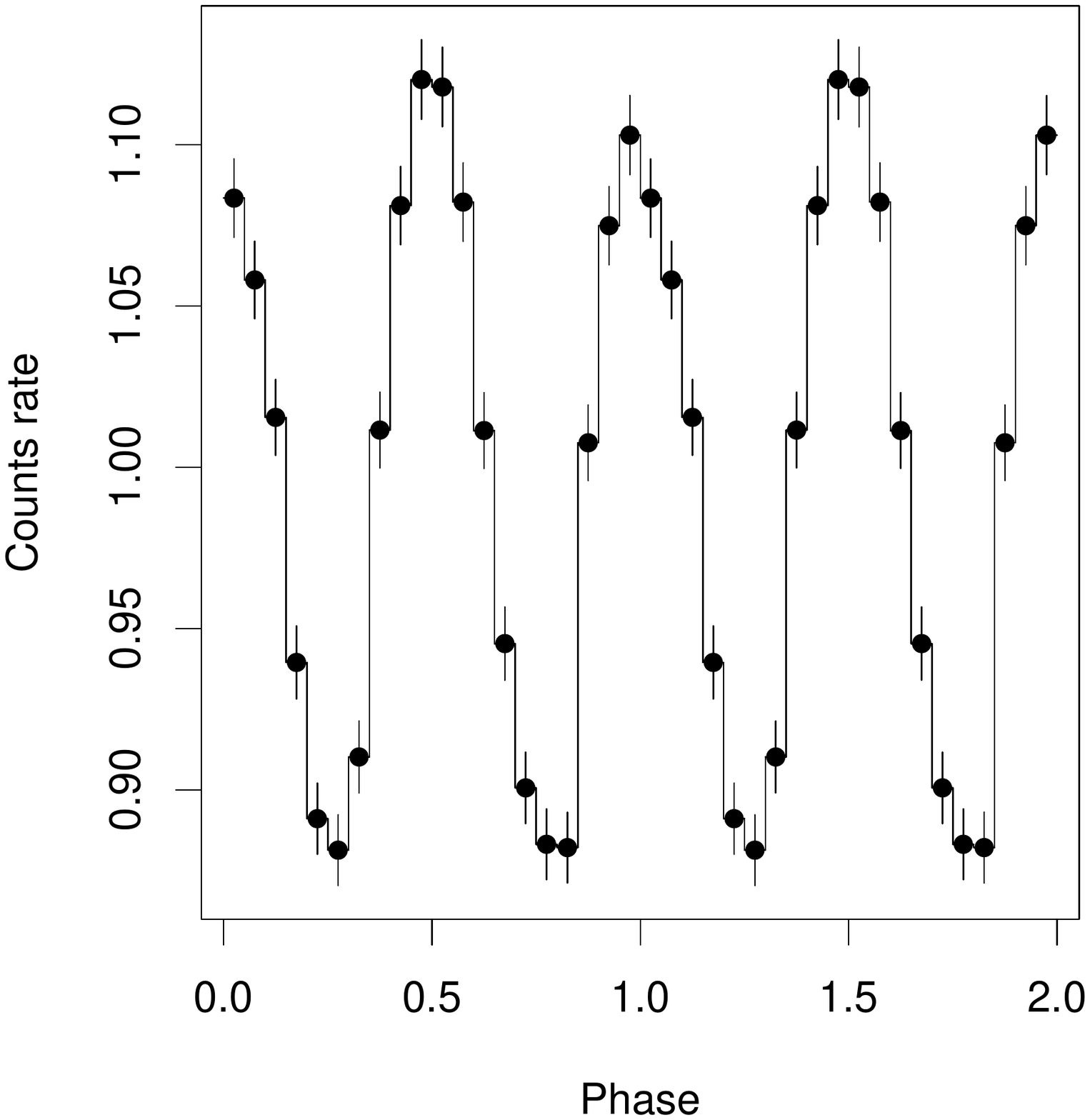}
  }
  \hbox{
    \includegraphics[bb=5 108 527 649,width=5.4cm]{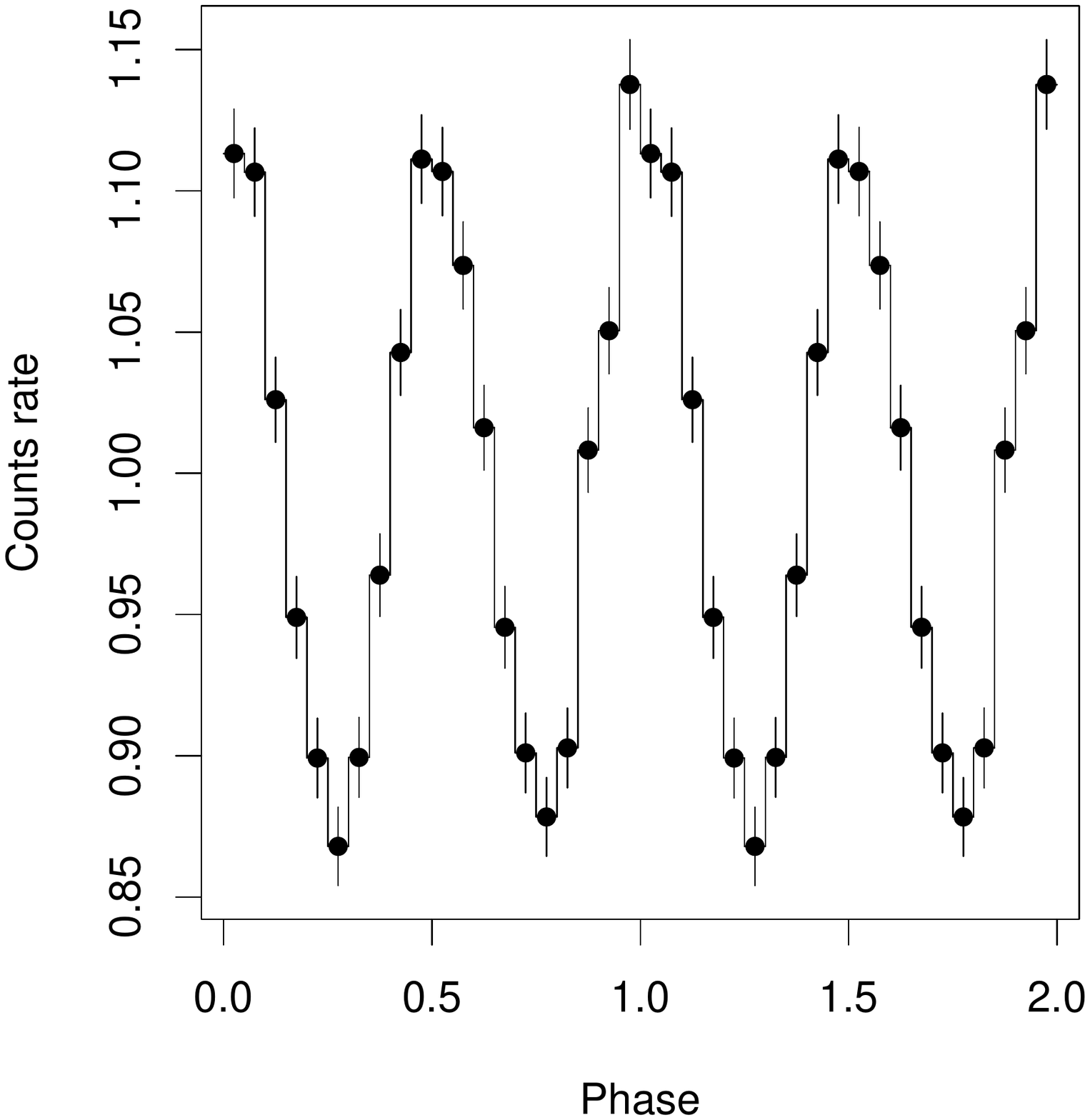}
    \includegraphics[bb=5 108 527 649,width=5.4cm]{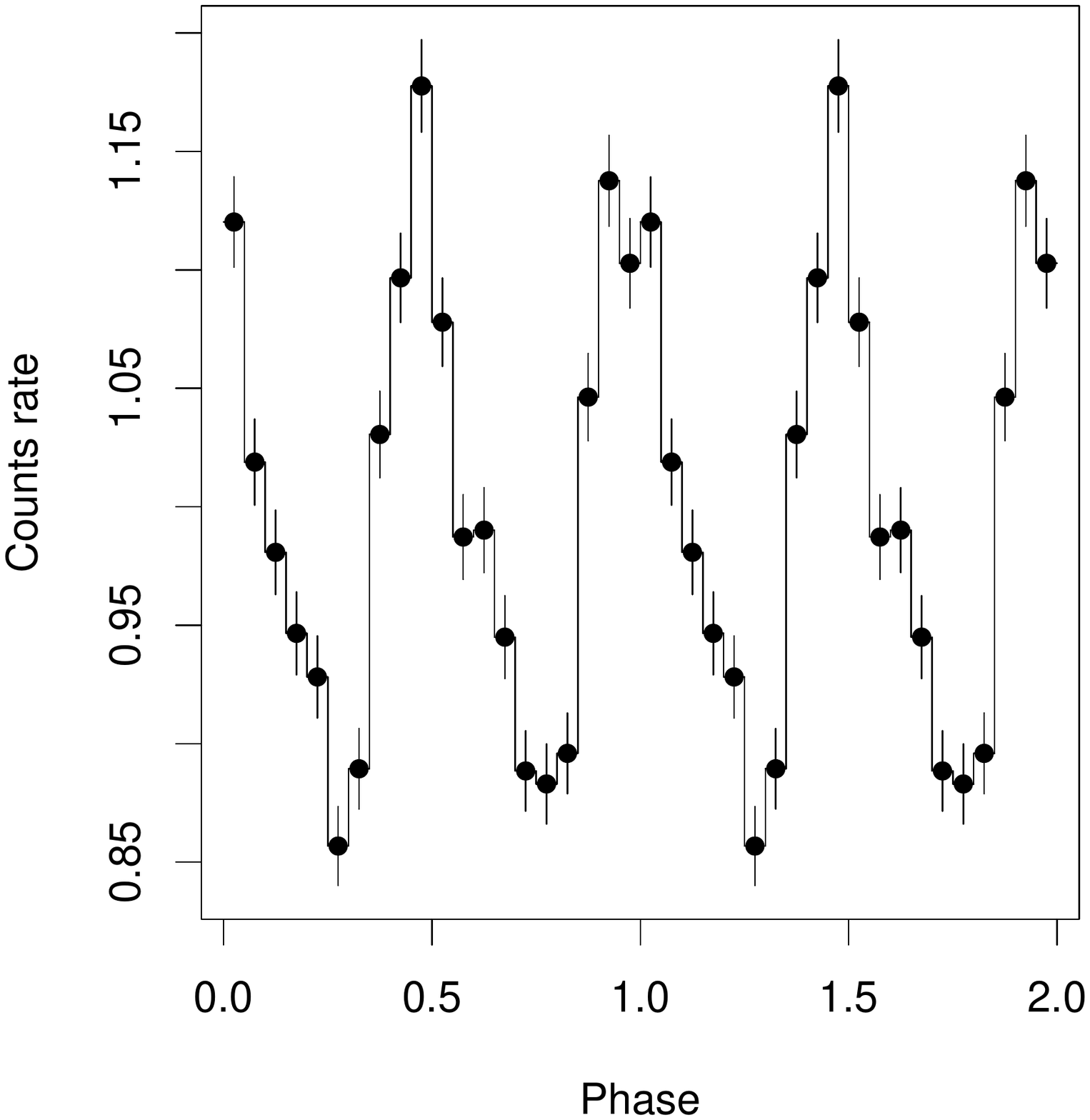}
    \includegraphics[bb=5 108 527 649,width=5.4cm]{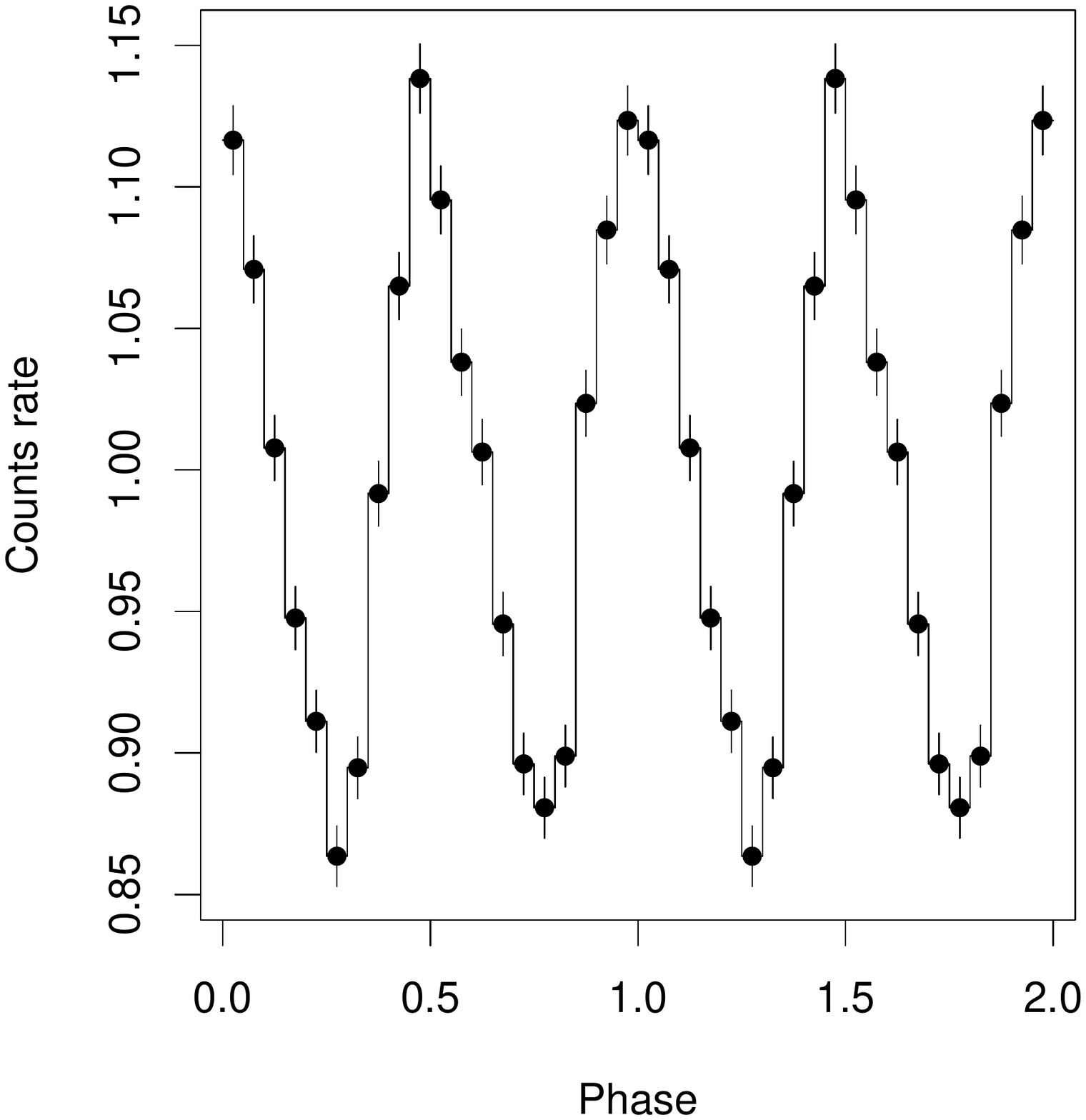}
  }
}
\vspace{-0.2cm}
\caption{For ObsIds 0650920101, 0670700201, 0670700301, and 0690070201 (see Fig.~\ref{resapp1}.)\label{resapp4}}
\end{figure*}

\begin{figure*}
\centering
\vspace{-0.2cm}
\vbox{
  \hbox{
    \includegraphics[width=8.0cm]{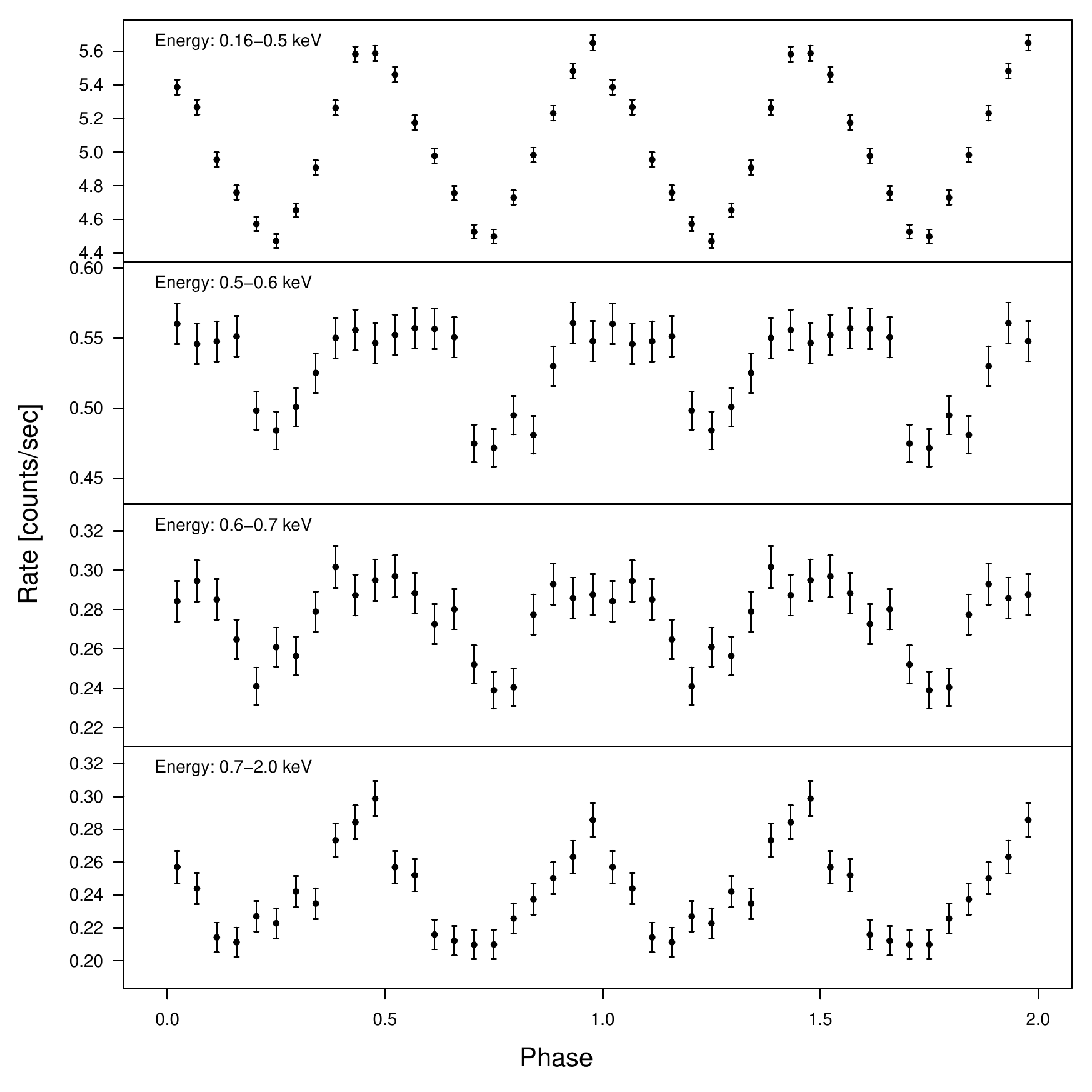}
    \includegraphics[width=8.0cm]{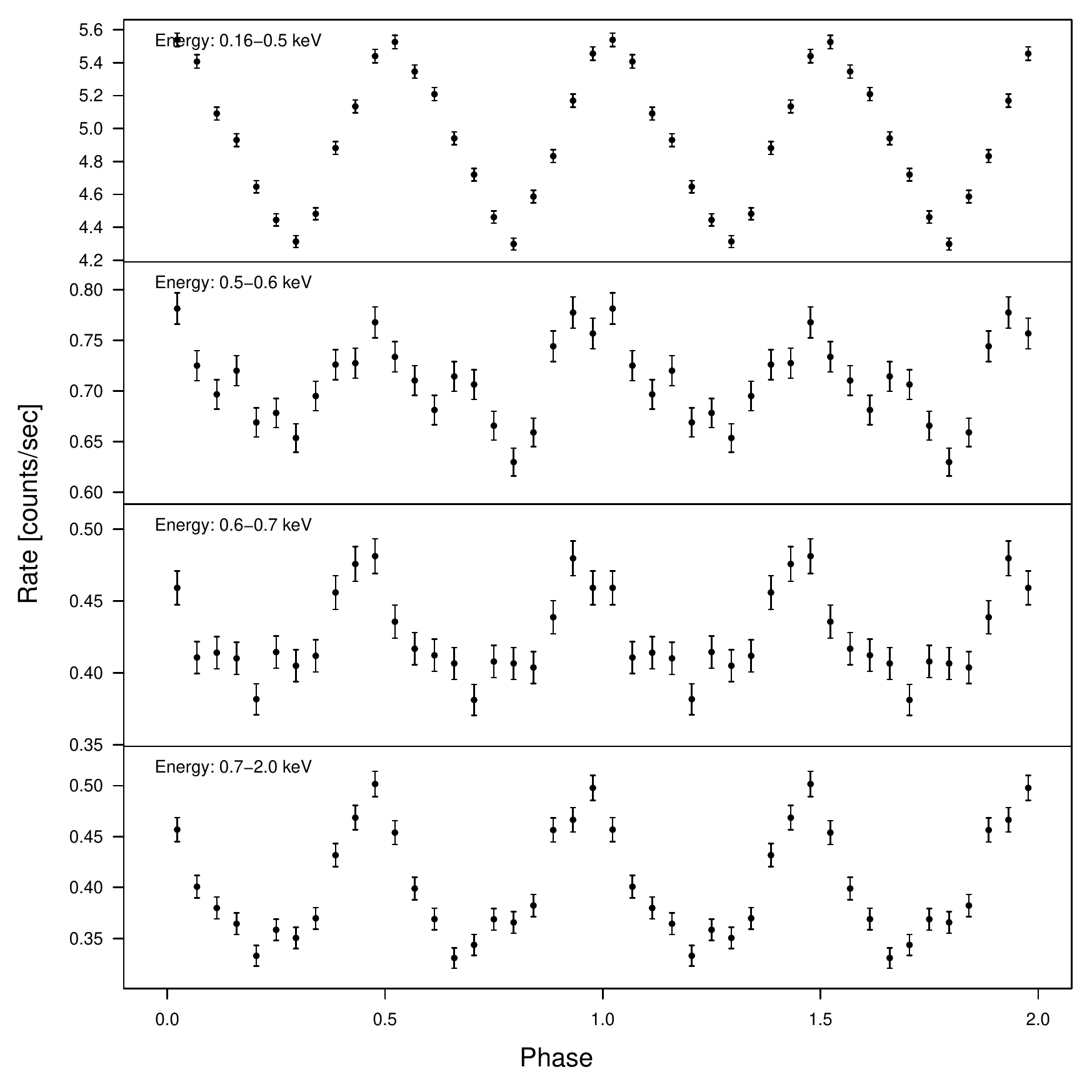}
  }
  \hbox{
    \includegraphics[width=8.0cm]{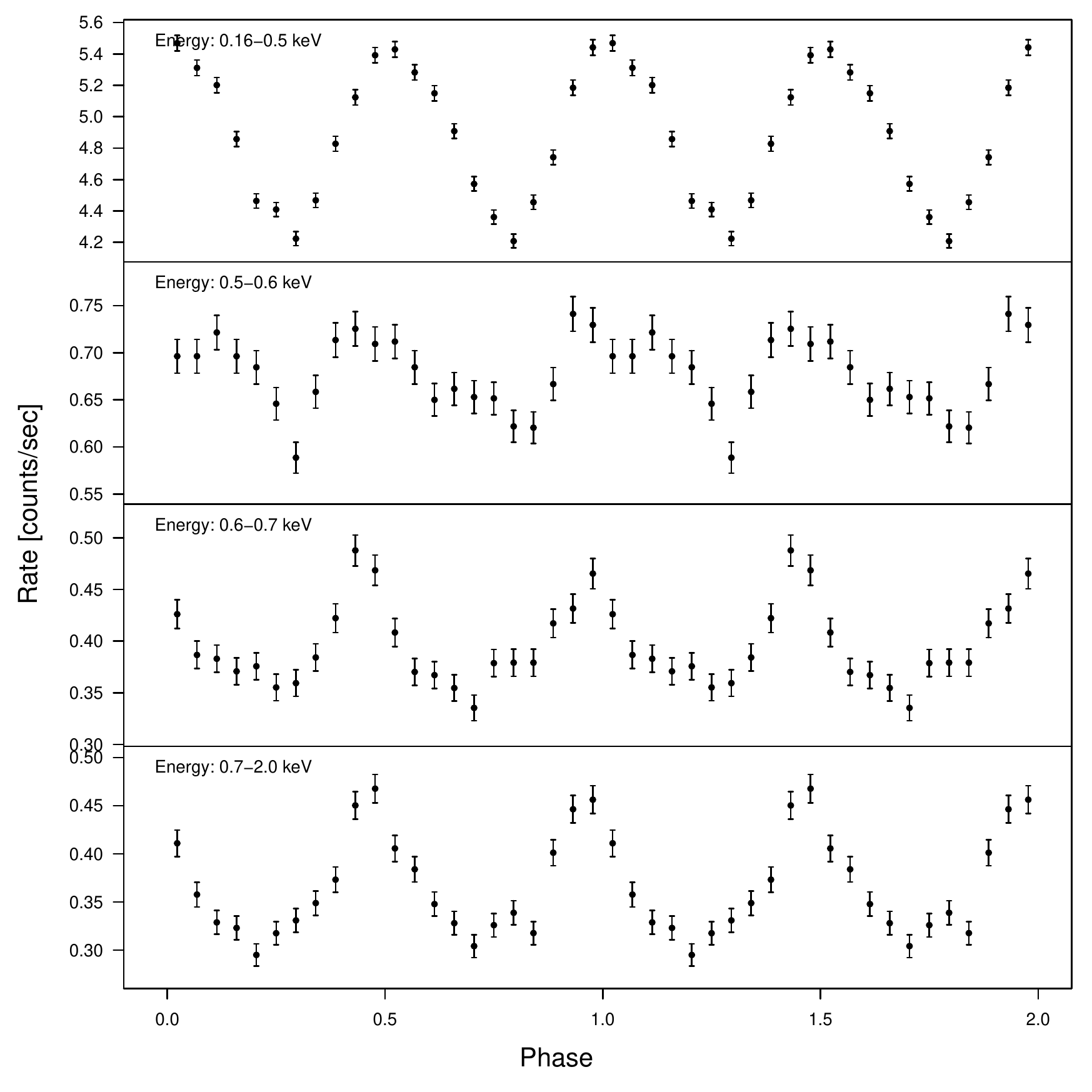}
    \includegraphics[width=8.0cm]{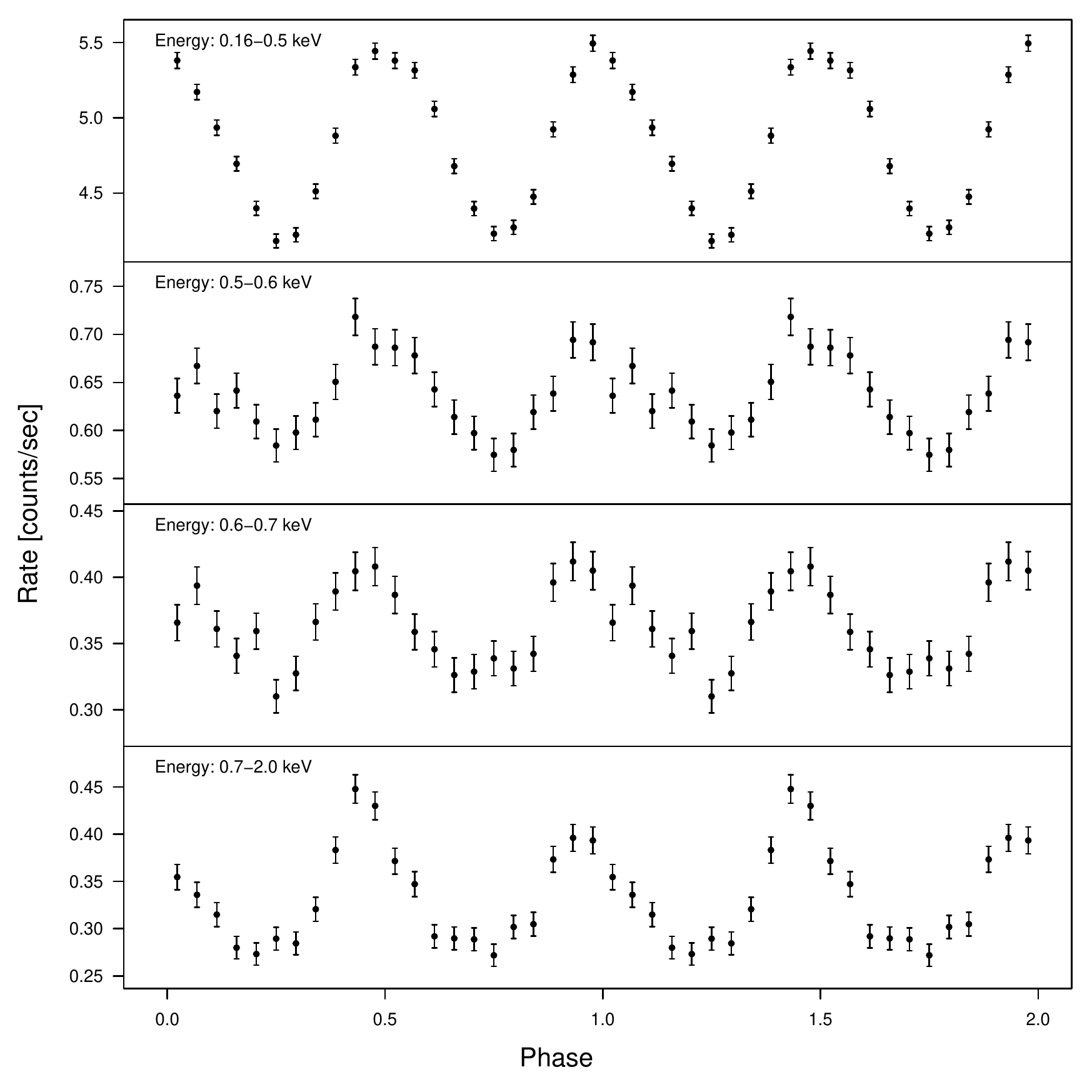}
  }
  \hbox{
    \includegraphics[width=8.0cm]{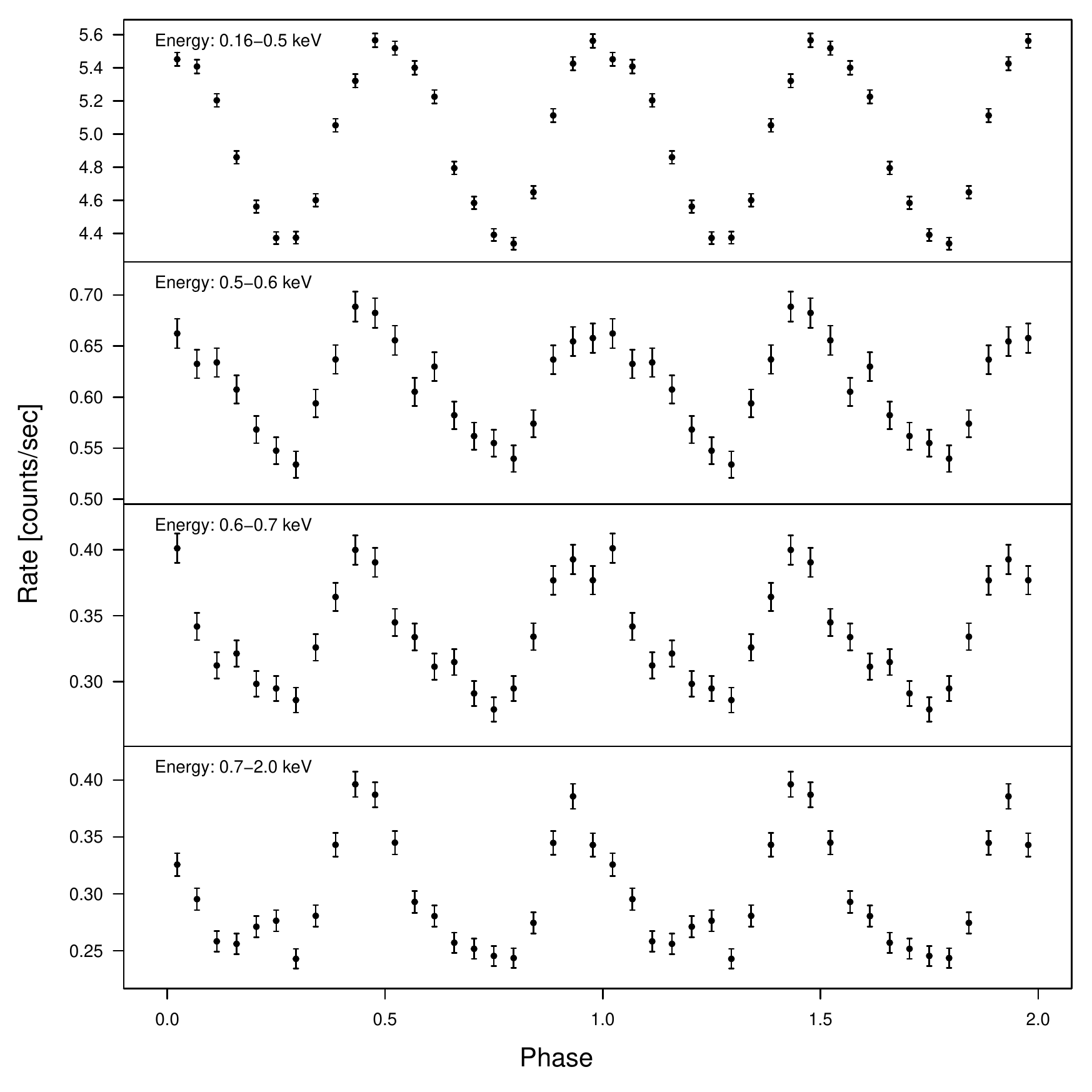}
    \includegraphics[width=8.0cm]{f10.pdf}
  }
}
\vspace{-0.2cm}
\caption{Combined phase-folded light curves of different data groups 
(GR I, II, III (left panel, from top to bottom), IV, V and all data sets (right panel) 
in different energy bands (0.16$-$0.50 keV, 0.50$-$0.60 keV,0.60$-$0.70 keV,0.70$-$2.0 keV).\label{fig:groups}}
\end{figure*}

\section{Phase-resolved spectra with MCMC}
\label{Mapp}

\begin{figure*}
\centering
\vbox{
  \hbox{
    \includegraphics[width=16.0cm]{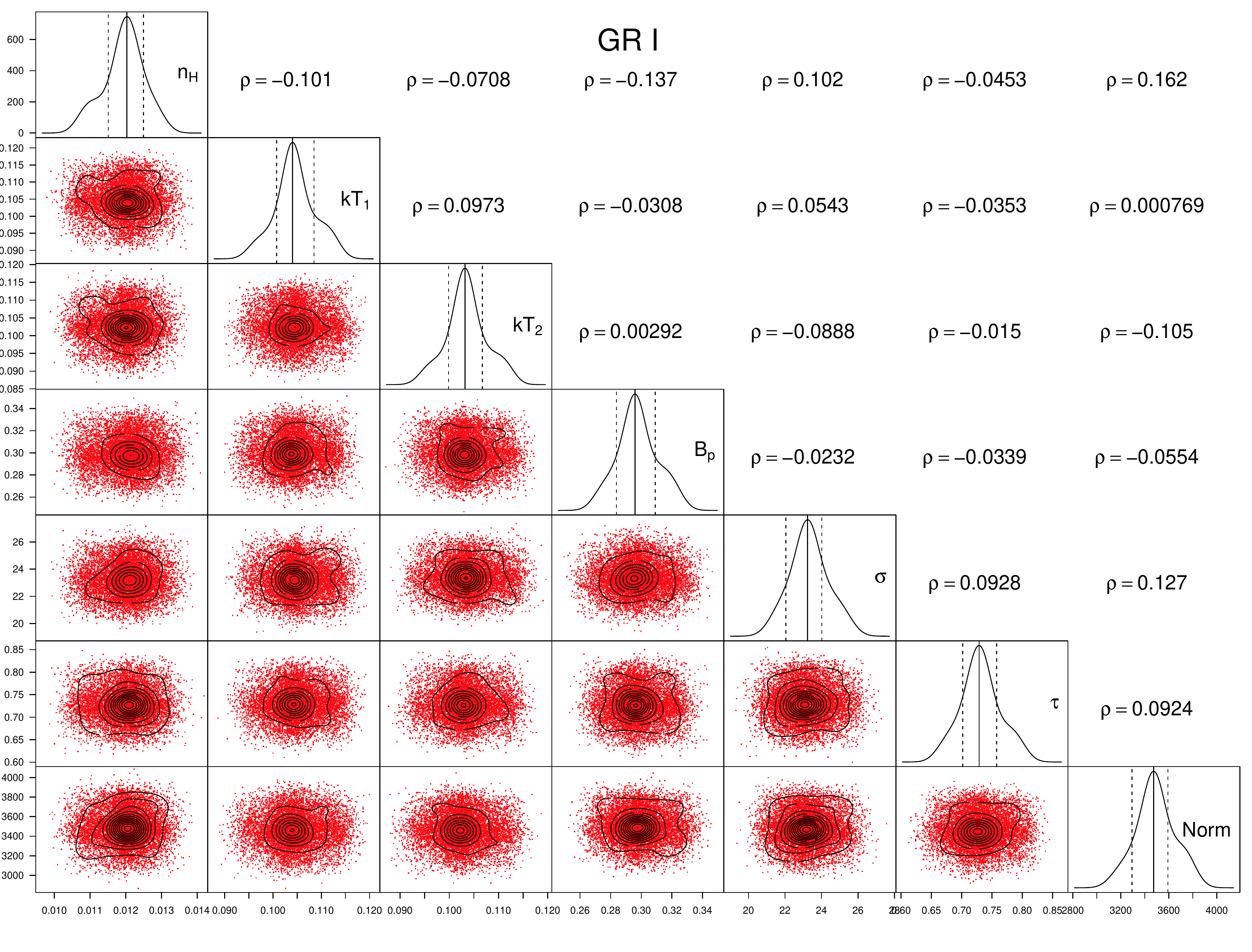}
  }
  \hbox{
    \includegraphics[width=16.0cm]{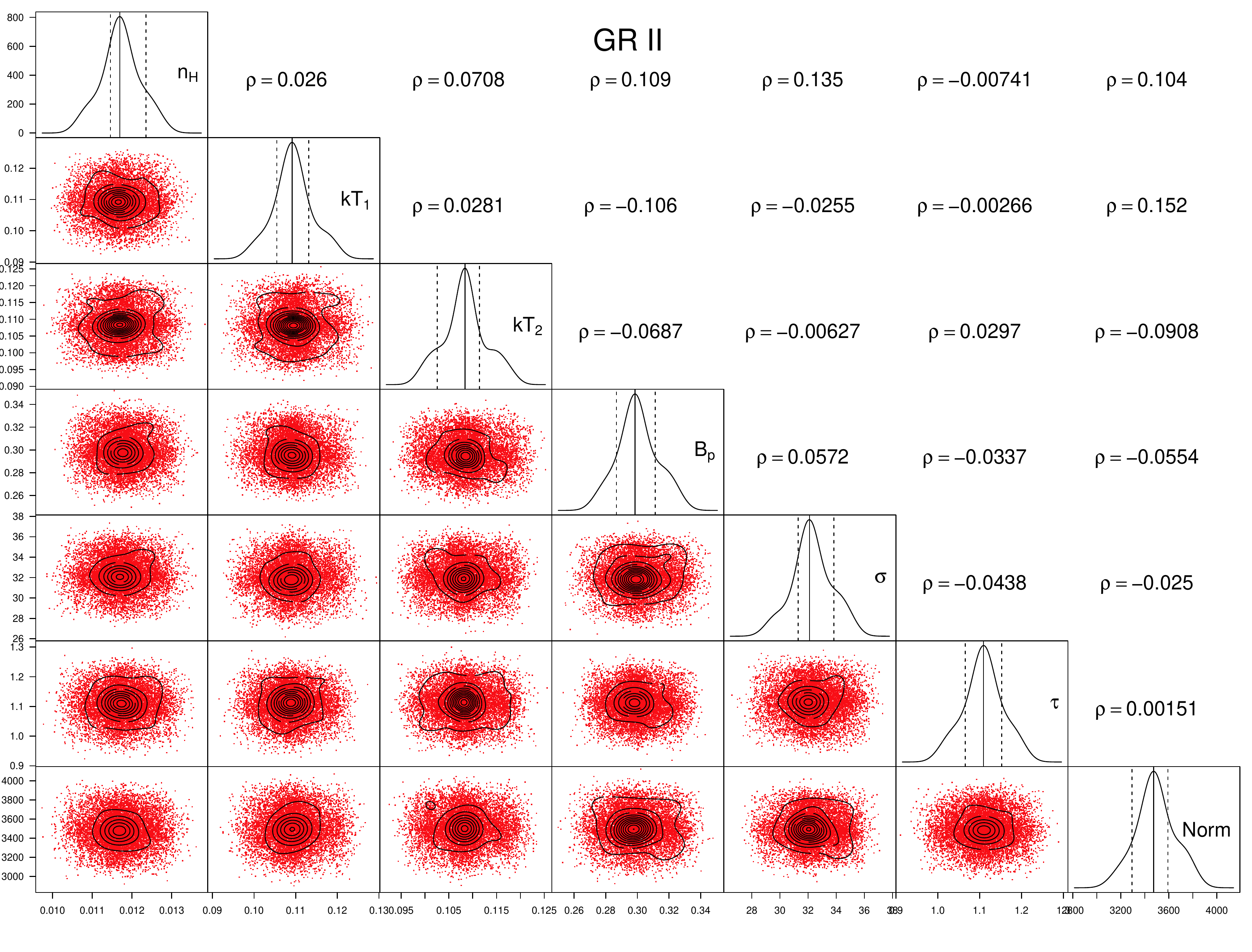}
  }
}
\vspace{-0.2cm}
\caption{Fited spectral model parameters (kept free for all data gropus) and MCMC verification results for
  data groups before the variation 
(GR I, upper panel),  and at the brightest stage (GR II, lower panel) of \xin.
In the diagonal are presented derived probability
density functions of the basic parameters of the fitted model. The lower
triangular part depicts samplings from the posteriors of two parameters, while
the upper one their mutual correlations.  
\label{fig:mcmc1}}
\end{figure*}

As already mentioned, in order to assess the degree of uniqueness and to estimate confidence intervals  
of the determined parameters for the model of an
emitting condensed iron surface with partially ionized hydrogen model
atmosphere, (see, Section~\ref{specphase}), we have additionally performed a
Markov Chain Monte Carlo (MCMC) fitting using the Goodman-Weare chain
generating algorithm (the default in {\it XSPEC}).
The results for the basic parameters of the fitted model (for two data groups I
and II) are presented as a matrix plot (see, Fig.~\ref{fig:mcmc1}). 

\end{appendix}


\begin{thebibliography}{51}
\expandafter\ifx\csname natexlab\endcsname\relax\def\natexlab#1{#1}\fi

\bibitem[{{Borghese} {et~al.}(2015){Borghese}, {Rea}, {Coti Zelati}, {Tiengo},
  \& {Turolla}}]{2015ApJ...807L..20B}
{Borghese}, A., {Rea}, N., {Coti Zelati}, F., {Tiengo}, A., \& {Turolla}, R.
  2015, \apjl, 807, L20

\bibitem[{{Buccheri} {et~al.}(1983){Buccheri}, {Bennett}, {Bignami}, {Bloemen},
  {Boriakoff}, {Caraveo}, {Hermsen}, {Kanbach}, {Manchester}, {Masnou},
  {Mayer-Hasselwander}, {Ozel}, {Paul}, {Sacco}, {Scarsi}, \&
  {Strong}}]{1983A&A...128..245B}
{Buccheri}, R., {Bennett}, K., {Bignami}, G.~F., {et~al.} 1983, A\&A, 128, 245

\bibitem[{{Connors}(1997)}]{1997daa..conf..251C}
{Connors}, A. 1997, in Data Analysis in Astronomy, ed. V.~{Di Gesu}, M.~J.~B.
  {Duff}, A.~{Heck}, M.~C. {Maccarone}, L.~{Scarsi}, \& H.~U. {Zimmerman},
  251--260

\bibitem[{{Cropper} {et~al.}(2004){Cropper}, {Haberl}, {Zane}, \&
  {Zavlin}}]{2004MNRAS.351.1099C}
{Cropper}, M., {Haberl}, F., {Zane}, S., \& {Zavlin}, V.~E. 2004, \mnras, 351,
  1099

\bibitem[{{Cropper} {et~al.}(2001){Cropper}, {Zane}, {Ramsay}, {Haberl}, \&
  {Motch}}]{2001A&A...365L.302C}
{Cropper}, M., {Zane}, S., {Ramsay}, G., {Haberl}, F., \& {Motch}, C. 2001,
  \aap, 365, L302

\bibitem[{{de Jager} \& {B{\"u}sching}(2010)}]{2010A&A...517L...9D}
{de Jager}, O.~C. \& {B{\"u}sching}, I. 2010, \aap, 517, L9

\bibitem[{{de Jager} {et~al.}(1989){de Jager}, {Raubenheimer}, \&
  {Swanepoel}}]{1989A&A...221..180D}
{de Jager}, O.~C., {Raubenheimer}, B.~C., \& {Swanepoel}, J.~W.~H. 1989, \aap,
  221, 180

\bibitem[{{de Vries} {et~al.}(2004){de Vries}, {Vink}, {M{\'e}ndez}, \&
  {Verbunt}}]{2004A&A...415L..31D}
{de Vries}, C.~P., {Vink}, J., {M{\'e}ndez}, M., \& {Verbunt}, F. 2004, \aap,
  415, L31

\bibitem[{{Gregory} \& {Loredo}(1992)}]{1992ApJ...398..146G}
{Gregory}, P.~C. \& {Loredo}, T.~J. 1992, \apj, 398, 146

\bibitem[{{Haberl}(2004)}]{2004AdSpR..33..638H}
{Haberl}, F. 2004, Advances in Space Research, 33, 638

\bibitem[{{Haberl}(2007)}]{2007Ap&SS.308..181H}
{Haberl}, F. 2007, \apss, 308, 181

\bibitem[{{Haberl} {et~al.}(2003){Haberl}, {Schwope}, {Hambaryan}, {Hasinger},
  \& {Motch}}]{2003A&A...403L..19H}
{Haberl}, F., {Schwope}, A.~D., {Hambaryan}, V., {Hasinger}, G., \& {Motch}, C.
  2003, \aap, 403, L19

\bibitem[{{Haberl} {et~al.}(2006){Haberl}, {Turolla}, {de Vries}, {Zane},
  {Vink}, {M{\'e}ndez}, \& {Verbunt}}]{2006A&A...451L..17H}
{Haberl}, F., {Turolla}, R., {de Vries}, C.~P., {et~al.} 2006, \aap, 451, L17

\bibitem[{{Haberl} {et~al.}(2004){Haberl}, {Zavlin}, {Tr{\"u}mper}, \&
  {Burwitz}}]{2004A&A...419.1077H}
{Haberl}, F., {Zavlin}, V.~E., {Tr{\"u}mper}, J., \& {Burwitz}, V. 2004, \aap,
  419, 1077

\bibitem[{{Hambaryan} {et~al.}(2002){Hambaryan}, {Hasinger}, {Schwope}, \&
  {Schulz}}]{2002A&A...381...98H}
{Hambaryan}, V., {Hasinger}, G., {Schwope}, A.~D., \& {Schulz}, N.~S. 2002,
  \aap, 381, 98

\bibitem[{{Hambaryan} \& {Neuh{\"a}user}(2016)}]{byur2016}
{Hambaryan}, V. \& {Neuh{\"a}user}, R. 2016, in Non-Stable Universe: Energetic
  Resources, Activity Phenomena and Evolutionary Processes: Byurakan 2016 (in
  press)

\bibitem[{{Hambaryan} {et~al.}(2009){Hambaryan}, {Neuh{\"a}user}, {Haberl},
  {Hohle}, \& {Schwope}}]{2009A&A...497L...9H}
{Hambaryan}, V., {Neuh{\"a}user}, R., {Haberl}, F., {Hohle}, M.~M., \&
  {Schwope}, A.~D. 2009, \aap, 497, L9

\bibitem[{{Hambaryan} {et~al.}(2014){Hambaryan}, {Neuh{\"a}user}, {Suleimanov},
  \& {Werner}}]{2014JPhCS.496a2015H}
{Hambaryan}, V., {Neuh{\"a}user}, R., {Suleimanov}, V., \& {Werner}, K. 2014,
  Journal of Physics Conference Series, 496, 012015

\bibitem[{{Hambaryan} {et~al.}(2011){Hambaryan}, {Suleimanov}, {Schwope},
  {Neuh{\"a}user}, {Werner}, \& {Potekhin}}]{2011A&A...534A..74H}
{Hambaryan}, V., {Suleimanov}, V., {Schwope}, A.~D., {et~al.} 2011, \aap, 534,
  A74

\bibitem[{{Hambaryan} {et~al.}(2015){Hambaryan}, {Wagner}, {Schmidt}, {Hohle},
  \& {Neuh{\"a}user}}]{2015AN....336..545H}
{Hambaryan}, V., {Wagner}, D., {Schmidt}, J.~G., {Hohle}, M.~M., \&
  {Neuh{\"a}user}, R. 2015, Astronomische Nachrichten, 336, 545

\bibitem[{{Heinke} {et~al.}(2006){Heinke}, {Rybicki}, {Narayan}, \&
  {Grindlay}}]{2006ApJ...644.1090H}
{Heinke}, C.~O., {Rybicki}, G.~B., {Narayan}, R., \& {Grindlay}, J.~E. 2006,
  \apj, 644, 1090

\bibitem[{{Ho} {et~al.}(2007){Ho}, {Kaplan}, {Chang}, {van Adelsberg}, \&
  {Potekhin}}]{2007MNRAS.375..821H}
{Ho}, W.~C.~G., {Kaplan}, D.~L., {Chang}, P., {van Adelsberg}, M., \&
  {Potekhin}, A.~Y. 2007, \mnras, 375, 821

\bibitem[{{Ho} {et~al.}(2008){Ho}, {Potekhin}, \&
  {Chabrier}}]{2008ApJS..178..102H}
{Ho}, W.~C.~G., {Potekhin}, A.~Y., \& {Chabrier}, G. 2008, \apjs, 178, 102

\bibitem[{{Hohle} {et~al.}(2012{\natexlab{a}}){Hohle}, {Haberl}, {Vink}, {de
  Vries}, \& {Neuh{\"a}user}}]{2012MNRAS.419.1525H}
{Hohle}, M.~M., {Haberl}, F., {Vink}, J., {de Vries}, C.~P., \&
  {Neuh{\"a}user}, R. 2012{\natexlab{a}}, \mnras, 419, 1525

\bibitem[{{Hohle} {et~al.}(2012{\natexlab{b}}){Hohle}, {Haberl}, {Vink}, {de
  Vries}, {Turolla}, {Zane}, \& {M{\'e}ndez}}]{2012MNRAS.423.1194H}
{Hohle}, M.~M., {Haberl}, F., {Vink}, J., {et~al.} 2012{\natexlab{b}}, \mnras,
  423, 1194

\bibitem[{{Hohle} {et~al.}(2009){Hohle}, {Haberl}, {Vink}, {Turolla},
  {Hambaryan}, {Zane}, {de Vries}, \& {M{\'e}ndez}}]{2009A&A...498..811H}
{Hohle}, M.~M., {Haberl}, F., {Vink}, J., {et~al.} 2009, \aap, 498, 811

\bibitem[{Mardia \& Jupp(2009)}]{mardia2009directional}
Mardia, K. \& Jupp, P. 2009, Directional Statistics, Wiley Series in
  Probability and Statistics (Wiley)

\bibitem[{Mardia {et~al.}(2008)Mardia, Hughes, Taylor, \&
  Singh}]{CJS:CJS5550360110}
Mardia, K.~V., Hughes, G., Taylor, C.~C., \& Singh, H. 2008, Canadian Journal
  of Statistics, 36, 99

\bibitem[{Mardia {et~al.}(2007)Mardia, Taylor, \& Subramaniam}]{BIOM:BIOM682}
Mardia, K.~V., Taylor, C.~C., \& Subramaniam, G.~K. 2007, Biometrics, 63, 505

\bibitem[{{Mendez} {et~al.}(2004){Mendez}, {de Vries}, {Vink}, \&
  {Verbunt}}]{2004cosp...35.2075M}
{Mendez}, M., {de Vries}, C.~P., {Vink}, J., \& {Verbunt}, F. 2004, in COSPAR
  Meeting, Vol.~35, 35th COSPAR Scientific Assembly, ed. J.-P. {Paill{\'e}},
  2075

\bibitem[{{Mereghetti}(2008)}]{2008A&ARv..15..225M}
{Mereghetti}, S. 2008, \aapr, 15, 225

\bibitem[{{Mori} \& {Ho}(2007)}]{2007MNRAS.377..905M}
{Mori}, K. \& {Ho}, W.~C.~G. 2007, \mnras, 377, 905

\bibitem[{{Motch} {et~al.}(2003){Motch}, {Zavlin}, \&
  {Haberl}}]{2003A&A...408..323M}
{Motch}, C., {Zavlin}, V.~E., \& {Haberl}, F. 2003, \aap, 408, 323

\bibitem[{{Ng} \& {Romani}(2007)}]{2007ApJ...660.1357N}
{Ng}, C.-Y. \& {Romani}, R.~W. 2007, \apj, 660, 1357

\bibitem[{{{\"O}gelman} \& {van den Heuvel}(1989)}]{1989ASIC..262.....O}
{{\"O}gelman}, H. \& {van den Heuvel}, E.~P.~J., eds. 1989, NATO Advanced
  Science Institutes (ASI) Series C, Vol. 262, {Timing Neutron Stars}

\bibitem[{{P{\'e}rez-Azor{\'{\i}}n} {et~al.}(2006){P{\'e}rez-Azor{\'{\i}}n},
  {Pons}, {Miralles}, \& {Miniutti}}]{2006A&A...459..175P}
{P{\'e}rez-Azor{\'{\i}}n}, J.~F., {Pons}, J.~A., {Miralles}, J.~A., \&
  {Miniutti}, G. 2006, \aap, 459, 175

\bibitem[{{Potekhin}(2010)}]{2010A&A...518A..24P}
{Potekhin}, A.~Y. 2010, \aap, 518, A24+

\bibitem[{{Poutanen} \& {Beloborodov}(2006)}]{2006MNRAS.373..836P}
{Poutanen}, J. \& {Beloborodov}, A.~M. 2006, \mnras, 373, 836

\bibitem[{{Schwope} {et~al.}(2005){Schwope}, {Hambaryan}, {Haberl}, \&
  {Motch}}]{2005A&A...441..597S}
{Schwope}, A.~D., {Hambaryan}, V., {Haberl}, F., \& {Motch}, C. 2005, \aap,
  441, 597

\bibitem[{{Schwope} {et~al.}(2007){Schwope}, {Hambaryan}, {Haberl}, \&
  {Motch}}]{2007Ap&SS.308..619S}
{Schwope}, A.~D., {Hambaryan}, V., {Haberl}, F., \& {Motch}, C. 2007, \apss,
  308, 619

\bibitem[{{Speagle} {et~al.}(2011){Speagle}, {Kaplan}, \& {van
  Kerkwijk}}]{2011ApJ...743..183S}
{Speagle}, J.~S., {Kaplan}, D.~L., \& {van Kerkwijk}, M.~H. 2011, \apj, 743,
  183

\bibitem[{{Str{\"u}der} {et~al.}(2001){Str{\"u}der}, {Briel}, {Dennerl},
  {Hartmann}, {Kendziorra}, {Meidinger}, {Pfeffermann}, {Reppin}, {Aschenbach},
  {Bornemann}, {Br{\"a}uninger}, {Burkert}, {Elender}, {Freyberg}, {Haberl},
  {Hartner}, {Heuschmann}, {Hippmann}, {Kastelic}, {Kemmer}, {Kettenring},
  {Kink}, {Krause}, {M{\"u}ller}, {Oppitz}, {Pietsch}, {Popp}, {Predehl},
  {Read}, {Stephan}, {St{\"o}tter}, {Tr{\"u}mper}, {Holl}, {Kemmer}, {Soltau},
  {St{\"o}tter}, {Weber}, {Weichert}, {von Zanthier}, {Carathanassis}, {Lutz},
  {Richter}, {Solc}, {B{\"o}ttcher}, {Kuster}, {Staubert}, {Abbey}, {Holland},
  {Turner}, {Balasini}, {Bignami}, {La Palombara}, {Villa}, {Buttler},
  {Gianini}, {Lain{\'e}}, {Lumb}, \& {Dhez}}]{2001A&A...365L..18S}
{Str{\"u}der}, L., {Briel}, U., {Dennerl}, K., {et~al.} 2001, \aap, 365, L18

\bibitem[{{Suleimanov} {et~al.}(2010){Suleimanov}, {Hambaryan}, {Potekhin},
  {van Adelsberg}, {Neuh{\"a}user}, \& {Werner}}]{2010A&A...522A.111S}
{Suleimanov}, V., {Hambaryan}, V., {Potekhin}, A.~Y., {et~al.} 2010, \aap, 522,
  A111+

\bibitem[{{Suleimanov} \& {Poutanen}(2006)}]{2006MNRAS.369.2036S}
{Suleimanov}, V. \& {Poutanen}, J. 2006, \mnras, 369, 2036

\bibitem[{{Suleimanov} {et~al.}(2011){Suleimanov}, {Poutanen}, {Revnivtsev}, \&
  {Werner}}]{2011ApJ...742..122S}
{Suleimanov}, V., {Poutanen}, J., {Revnivtsev}, M., \& {Werner}, K. 2011, \apj,
  742, 122

\bibitem[{{Tetzlaff} {et~al.}(2011){Tetzlaff}, {Eisenbeiss}, {Neuh{\"a}user},
  \& {Hohle}}]{2011MNRAS.417..617T}
{Tetzlaff}, N., {Eisenbeiss}, T., {Neuh{\"a}user}, R., \& {Hohle}, M.~M. 2011,
  \mnras, 417, 617

\bibitem[{{Turolla}(2009)}]{2009ASSL..357..141T}
{Turolla}, R. 2009, in Astrophysics and Space Science Library, Vol. 357,
  Neutron Stars and Pulsars, ed. {W.~Becker}, 141--164

\bibitem[{{van Kerkwijk} {et~al.}(2007){van Kerkwijk}, {Kaplan}, {Pavlov}, \&
  {Mori}}]{2007ApJ...659L.149V}
{van Kerkwijk}, M.~H., {Kaplan}, D.~L., {Pavlov}, G.~G., \& {Mori}, K. 2007,
  \apjl, 659, L149

\bibitem[{{Vigan{\`o}} {et~al.}(2014){Vigan{\`o}}, {Perna}, {Rea}, \&
  {Pons}}]{2014MNRAS.443...31V}
{Vigan{\`o}}, D., {Perna}, R., {Rea}, N., \& {Pons}, J.~A. 2014, \mnras, 443,
  31

\bibitem[{{Zane} \& {Turolla}(2006)}]{2006MNRAS.366..727Z}
{Zane}, S. \& {Turolla}, R. 2006, \mnras, 366, 727

\bibitem[{{Zavlin} {et~al.}(1995){Zavlin}, {Pavlov}, {Shibanov}, \&
  {Ventura}}]{1995A&A...297..441Z}
{Zavlin}, V.~E., {Pavlov}, G.~G., {Shibanov}, Y.~A., \& {Ventura}, J. 1995,
  \aap, 297, 441

\end{thebibliography}
\end{document}